\documentclass[aps,prd,nofootinbib,onecolumn,superscriptaddress,preprintnumbers,balancelastpage,longbibliography,nobibnotes]{revtex4-2}

\usepackage[dvipsnames]{xcolor}
\usepackage{graphicx,color,amsmath,amssymb,flushend,bm,mathrsfs,comment}
\definecolor{linkcolor}{rgb}{0.0, 0.28, 0.67}

\usepackage[
   colorlinks=true,
    urlcolor=linkcolor,
   anchorcolor=linkcolor,
    citecolor=linkcolor,
    filecolor=linkcolor,
    linkcolor=linkcolor,
    menucolor=linkcolor,
    linktocpage=true,
    pdfproducer=medialab,
    pdfa=true
]{hyperref}

\usepackage[capitalise]{cleveref}
\usepackage{tikz}
\usepackage{ulem}

\usepackage{amsmath}
\usepackage{physics}
\usepackage{simpler-wick}
\usepackage{amsfonts}
\usepackage{amssymb}
\usepackage[mathscr]{euscript}
\usepackage{setspace}
\usepackage{lipsum}
\usepackage{slashed}
\usepackage{cancel}
\usepackage{multirow}
\usepackage[utf8]{inputenc}
\usepackage{mathtools}

\usetikzlibrary{calc}
\usepackage{siunitx}
\DeclareSIUnit{\year}{yr}
\DeclareSIUnit{\parsec}{pc}

\usepackage[english]{babel}
\usepackage{csquotes}

\makeatletter
\newcommand*{\rom}[1]{\expandafter\@slowromancap\romannumeral #1@}
\makeatother
\newcommand{\bea}{\begin{eqnarray}\begin{aligned}}
\newcommand{\eea}{\end{aligned}\end{eqnarray}}
\newcommand{\chibar}{ {\overline{\chi}} }

\newcommand{\const}{\text{const}}

\newcommand{\mpl}{m_\text{pl}}

\newcommand{\alphaem}{\alpha_\text{em}}

\newcommand{\crit}{\text{crit}}

\newcommand{\rh}{\text{rh}}

\newcommand{\cm}{\text{cm}}

\newcommand{\EM}{\text{em}}
\newcommand{\hc}{\text{h.c.}}
\newcommand{\res}{\text{res}}

\newcommand{\Teq}{T_\text{eq}}
\newcommand{\eV}{\text{eV}}
\newcommand{\kev}{\text{keV}}
\newcommand{\mev}{\text{MeV}}
\newcommand{\gev}{\text{GeV}}
\newcommand{\tev}{\text{TeV}}

\newcommand{\dm}{\text{dm}}
\newcommand{\osc}{\text{osc}}
\newcommand{\local}{\text{local}}

\newcommand{\zn}{\mathbb{Z}_N}

\newcommand{\be}{\begin{equation}}
\newcommand{\ee}{\end{equation}}

\newcommand{\Sec}[1]{Sec.~\ref{sec:#1}}
\newcommand{\App}[1]{Appendix.~\ref{#1}}
\newcommand{\Fig}[1]{Fig.~\ref{#1}}
\newcommand{\Eq}[1]{Eq.~(\ref{#1})}
\newcommand{\bmtx}{\begin{pmatrix}}
\newcommand{\emtx}{\end{pmatrix}}

\newcommand{\F}{\Psi}
\newcommand{\barF}{\bar{\Psi}}

\newcommand{\Mol}{\text{M\o l}}

\DeclareMathSymbol{\mh}{\mathord}{operators}{`\-}

\newcommand{\cw}{\text{cw}}

\newcommand{\FI}{\text{FI}}

\newcommand{\LambdaKM}{\Lambda_\text{KM}}

\newcommand{\DG}{\text{D}}

\newcommand{\SM}{\text{SM}}
\newcommand{\DS}{\text{DS}}

\newcommand{\standard}{\text{std}}

\newcommand{\CMB}{\text{CMB}}
\newcommand{\BBN}{\text{BBN}}

\newcommand{\Mearth}{M_{\bigoplus}}
\newcommand{\Rearth}{R_{\bigoplus}}
\newcommand{\Qearth}{Q_{\gamma, \bigoplus}}

\newcommand{\UV}{\text{UV}}
\newcommand{\IR}{\text{IR}}

\newcommand{\CD}{\mathcal{C}_d}

\begin{document}

\preprint{DESY-23-013} 
\preprint{LAPTH-003/23}

\title{Cosmologically Varying Kinetic Mixing}

\author{Xucheng Gan}
\email{xg767@nyu.edu}

\affiliation{Center for Cosmology and Particle Physics, Department of Physics, New York University, New York, NY 10003, USA}

\author{Di Liu}
\email{di.liu@lapth.cnrs.fr}
\affiliation{Deutsches Elektronen-Synchrotron DESY, Notkestr. 85, 22607 Hamburg, Germany}
\affiliation{Laboratoire d'Annecy-le-Vieux de Physique Th\'eorique, CNRS -- USMB, BP 110 Annecy-le-Vieux, F-74941 Annecy, France}

\begin{abstract}
The portal connecting the invisible and visible sectors is one of the most natural explanations of the dark world. However, the early-time dark matter production via the portal faces extremely stringent late-time constraints. To solve such tension, we construct the scalar-controlled kinetic mixing varying with the ultralight CP-even scalar's cosmological evolution. To realize this and eliminate the constant mixing, we couple the ultralight scalar within $10^{-33}\text{eV} \lesssim m_0 \ll \text{eV}$ with the heavy doubly charged messengers and impose the $\mathbb{Z}_2$ symmetry under the dark charge conjugation. Via the varying mixing, the $\text{keV}-\text{MeV}$ dark photon dark matter is produced through the early-time freeze-in when the scalar is misaligned from the origin and free from the late-time exclusions when the scalar does the damped oscillation and dynamically sets the kinetic mixing. We also find that the scalar-photon coupling emerges from the underlying physics, which changes the cosmological history and provides the experimental targets based on the fine-structure constant variation and the equivalence principle violation. To ensure the scalar naturalness, we discretely re-establish the broken shift symmetry by embedding the minimal model into the $\mathbb{Z}_N$-protected model. When $N \sim 10$, the scalar's mass quantum correction can be suppressed much below $10^{-33}\text{eV}$. 
\end{abstract}

\maketitle

\tableofcontents

\section{Introduction}
\label{sec:introduction}

We know that dark matter exists, but we do not know the dark matter's particle nature. Even so, we can naturally imagine that the dark matter stays in the invisible sector, and the invisible and visible sectors are connected by the portal. Through the portal, the energy flow from the visible sector to the invisible sector, which is known as freeze-in~\cite{Hall:2009bx}, or vice versa, which is known as freeze-out. Hence, the dark matter reaches $\Omega_{\text{DM}}h^2 \simeq 0.12$, being compatible with the CMB anisotropy~\cite{Holdom:1985ag, Planck:2018vyg}. The kinetic mixing~\cite{Holdom:1985ag}, one of the three major portals~\cite{Holdom:1985ag, Falkowski:2009yz, Lindner:2010rr, GonzalezMacias:2015rxl, Batell:2017cmf, Batell:2017rol, Berlin:2018ztp, Silveira:1985rk, McDonald:1993ex, Burgess:2000yq, Patt:2006fw}, connects the photon and the dark photon as $\mathcal{L} \supset \epsilon F_{\mu \nu} F'^{\mu \nu}/2$. In the last few decades, the research on time-independent kinetic mixing has boosted on both the experimental and theoretical sides~\cite{Fabbrichesi:2020wbt, Caputo:2021eaa, Dienes:1996zr, Abel:2003ue, Goodsell:2009xc, Goodsell:2011wn, DelZotto:2016fju, Gherghetta:2019coi,Benakli:2020vng, Obied:2021zjc, Rizzo:2018ntg, Wojcik:2022rtk, Chiu:2022bni}, with only a few discussions on the spacetime-varying scenarios~\cite{Banerjee:2019asa, Baldes:2019tkl, Chakraborty:2020vec, Davoudiasl:2022ubh}. In the meantime, other varying constants are extensively discussed~\cite{Bekenstein:1982eu, Olive:2001vz, Dvali:2001dd, Chacko:2002mf, Fardon:2003eh, Fardon:2005wc, Weiner:2005ac, Ghalsasi:2016pcj, Berlin:2016bdv, Baker:2016xzo,  Baker:2017zwx, Baker:2018vos, Bian:2018mkl, Bian:2018bxr, Croon:2020ntf, Hashino:2021dvx, Guo:2022vxr, Baldes:2016gaf, Bruggisser:2017lhc,Ellis:2019flb,Berger:2019yxb, Ipek:2018lhm, Croon:2019ugf, Berger:2020maa, Howard:2021ohe, Fung:2021wbz, Fung:2021fcj, Elor:2021swj, Allali:2022yvx, Allali:2022sfm}. Moreover, extremely strong tensions exist between the early-time dark matter production through the portal and the late-time constraints on the portal, such as the dark photon dark matter freeze-in through the kinetic mixing portal~\cite{Pospelov:2008jk, Redondo:2008ec, An:2014twa}, the sterile neutrino dark matter freeze-in through the neutrino portal~\cite{Dodelson:1993je, Abazajian:2017tcc}, and the dark matter freeze-out through the Higgs portal~\cite{Escudero:2016gzx}. To solve such tension in the simplest way, we allow the portal evolve during the universe's expansion. However, there is no free lunch to evade the constraints without consequences. To be more specific, controlling the portal leaves significant imprints in our universe, which changes the early cosmological history and can be detected by experiments designed for general relativity testing and ultralight dark matter detection.

\begin{figure}[t]
\centering
\includegraphics[width=0.497\columnwidth]{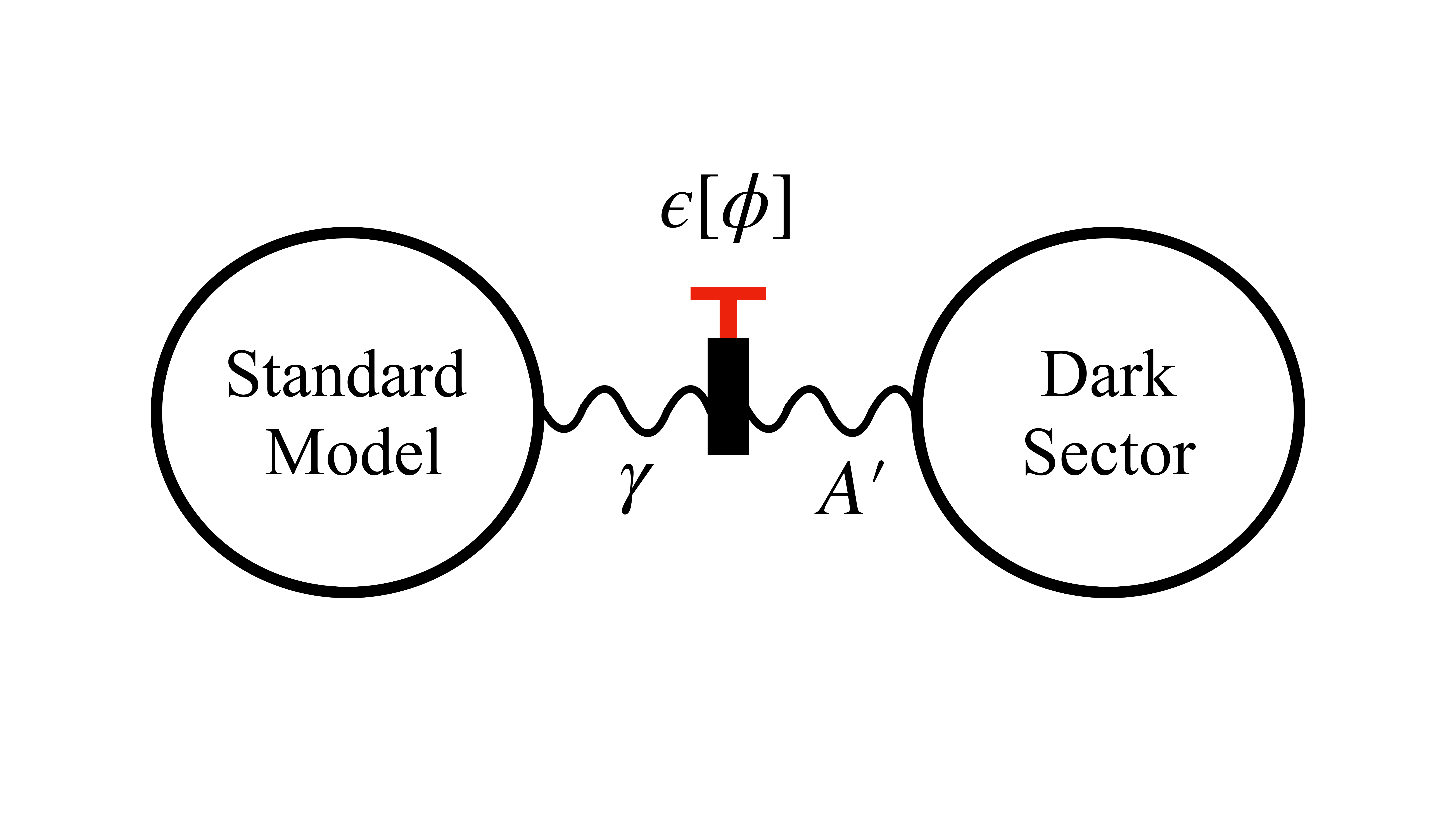}
\includegraphics[width=0.497\columnwidth]{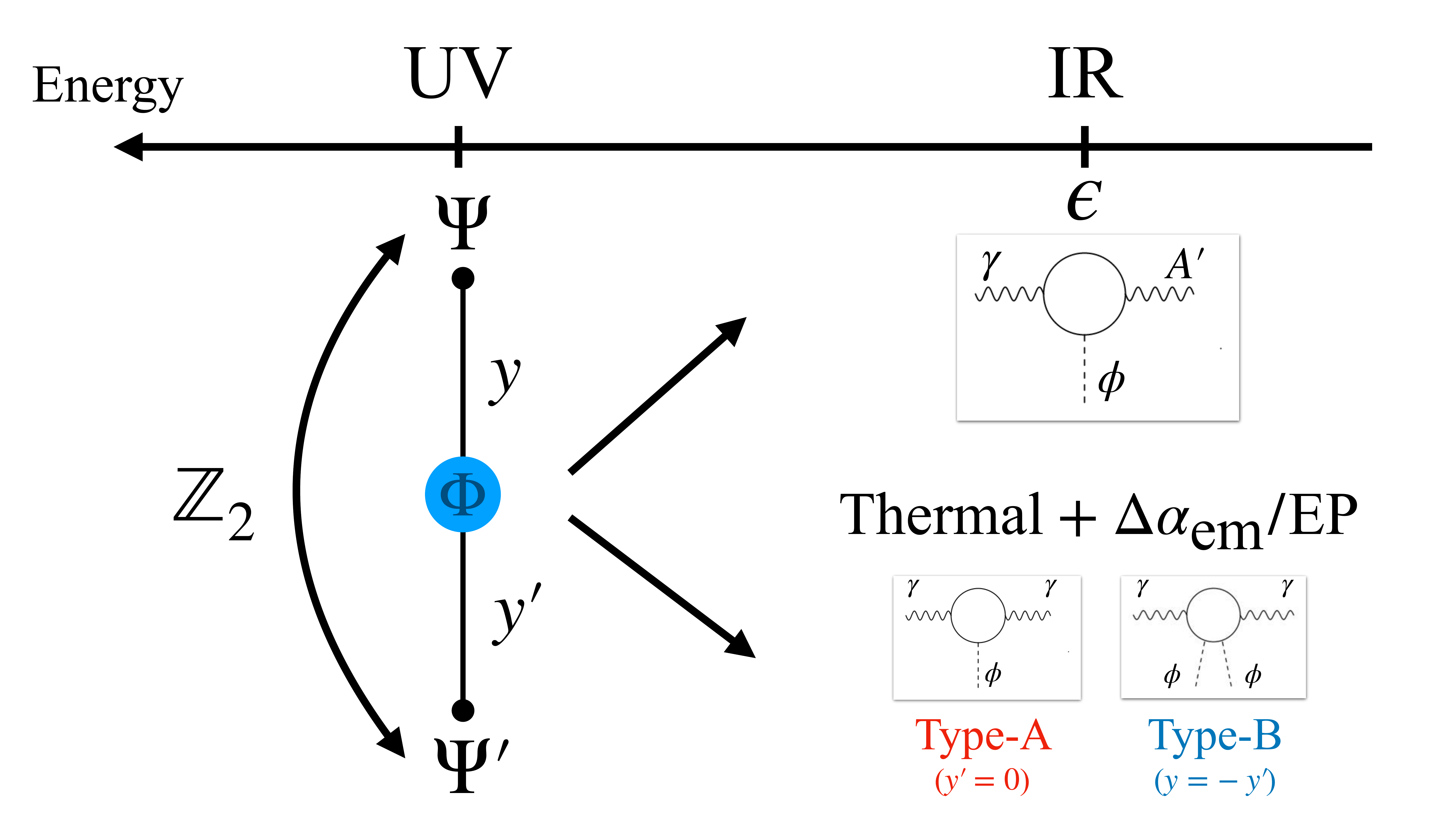}
\caption{{\bf Left}: The schematic diagram of the cosmologically varying kinetic mixing. The dark and standard model sectors are connected through the kinetic mixing controlled by the CP-even scalar $\phi$, the subcomponent of the dark matter in the late universe. Based on this model, the energy flows from the dark sector to the standard model sector in the early universe for the dark matter production with the portal opened. In the late universe, with the portal partially closed, the dark matter is safe from stringent constraints. 
{\bf Right}: The UV and IR theories. UV theory contains heavy messengers $\F$ and $\F^\prime$ carrying the same electromagnetic charge but the opposite dark charge. $\F$ and $\F^\prime$ are coupled with the scalar $\Phi$ via the Yukawa couplings $y$ and $y^\prime$. To eliminate the time-independent kinetic mixing, we impose the $\mathbb{Z}_2$-protected mass degeneracy between $\F$ and $\F^\prime$. In the IR theory, integrating out the heavy messengers induces the varying kinetic mixing and the scalar-photon couplings simultaneously. These scalar-photon interactions lead to nontrivial cosmological history caused by the thermal effect and the signals from the $\alphaem$ variation and the equivalence principle violation. Based on the relation of the scalar-messenger Yukawas, we classify the theory into two types with different phenomenologies: The type-A model~($y \neq 0, y'=0$) with the linear scalar-photon coupling and the type-B model~($y'=-y$) with the quadratic scalar-photon coupling. }
\label{fig:phiKM}
\end{figure}

In this work, we study the scalar-controlled kinetic mixing by meticulously exploring the top-down and bottom-up theories, cosmological history, $\kev-\mev$ dark photon dark matter production, and experimental signals from both the dark photon dark matter and the nonrelativistic ultralight scalar relic. To vary the kinetic mixing, we couple the ultralight scalar $\phi$, the CP-even degree of freedom predicted by the string theory~\cite{Wu:1986ac, Maeda:1987ku, Damour:1990tw, Damour:1994zq, Damour:1994ya, Damour:2002mi,Damour:2002nv}, to the heavy fermionic messengers doubly charged under the standard model $U(1)$ and dark $U(1)$. Here, the constant kinetic mixing is eliminated when the $\mathbb{Z}_2$ symmetry under the dark charge conjugation is imposed. Given this, in the low energy limit, the varying-mixing operator $\phi F F'$ emerges, along with the scalar-photon coupling, such as $\phi F^2$ or $\phi^2 F^2$.
Initially, $\phi$ has the early misalignment opening the portal for the dark photon dark matter production with the kinetic mixing $\epsilon_\FI \sim 10^{-12}$, which stems from the early-time $\mathbb{Z}_2$-breaking of the system. Afterward, $\phi$'s damped oscillation gradually and partially closes the portal, which stems from the late-time $\mathbb{Z}_2$-restoration. Through the evolution during the cosmological expansion, $\phi$ sets the benchmark kinetic mixing of the dark photon dark matter, which is free from stringent late-time constraints, such as stellar energy loss~\cite{Redondo:2008aa, Redondo:2013lna, An:2013yfc,  Hardy:2016kme}, direct detection~\cite{An:2014twa, Bloch:2016sjj, XENON:2018voc, XMASS:2018pvs, XENON:2019gfn, XENON:2020rca, XENONCollaboration:2022kmb}, and late-time decay bounds~\cite{Pospelov:2008jk, Redondo:2008ec, Essig:2013goa, Slatyer:2016qyl, Wadekar:2021qae}. At the same time, via the scalar-photon coupling, the ultralight scalar as the nonrelativistic relic in the mass range $10^{-33}\eV \lesssim m_0 \ll \eV$ changes the fine-structure constant, and the scalar as the mediator contributes to the extra force between two objects. Therefore, the experiments such as the equivalence principle~(EP) violation test~\cite{Smith:1999cr, Schlamminger:2007ht, Berge:2017ovy}, clock comparison~\cite{Arvanitaki:2014faa, VanTilburg:2015oza, Hees:2016gop, Barontini:2021mvu, collaboration2021frequency}, resonant-mass detector~\cite{Arvanitaki:2015iga}, PTA~\cite{Kaplan:2022lmz}, CMB~\cite{Stadnik:2015kia, Hart:2019dxi}, and BBN~\cite{Stadnik:2015kia, Sibiryakov:2020eir, Bouley:2022eer} can be used to test the scalar-controlled kinetic mixing, and the experimental targets are set by the dark photon freeze-in. If the signals from the dark photon dark matter and the ultralight scalar experiments appear consistently, we can confidently declare the verification of our model. In addition, given the scalar-photon coupling in the strong region, the scalar's high-temperature evolution is affected by the standard model plasma, which sets the early displacement, modifies the start of oscillation, and enhances the scalar's signal. To understand the whole setup classified under the exactness of the $\mathbb{Z}_2$ symmetry, one can refer to \Fig{fig:phiKM}.

To protect the CP-even scalar's naturalness caused by the heavy messengers and inspired by the former works on the discrete symmetries~\cite{Frieman:1995pm, Hook:2018jle, Hook:2019mrd, Das:2020arz, Dror:2020zru, Brzeminski:2020uhm, DiLuzio:2021pxd, Banerjee:2022wzk, Vileta:2022jou}, we embed the varying kinetic mixing into $N$ copies of the universes, where the $\mathbb{Z}_N$ symmetry rebuilds the global $U(1)$ shift symmetry in the discrete form. In such $\mathbb{Z}_N$-protected model, the scalar's lowest order mass term becomes $\Phi^{N}/\Lambda^{N-4}$, which reveals the exponential suppression of the quantum correction. For example, we only need $N \sim 10$ to suppress the scalar mass correction all the way down to $10^{-33}\eV$. Furthermore, to understand the additional cancellation from the exact $\mathbb{Z}_2$ symmetry and gain an accurate analytical result, we expand the $\mathbb{Z}_N$-invariant Coleman-Weinberg formally calculated to the leading order in \cite{Hook:2018jle, Brzeminski:2020uhm} to the arbitrary orders with two different methods, i.e., the Fourier transformation and the cosine sum rules. Other topics of the varying kinetic mixing from the supersymmetric Dirac gaugino model and the dark matter models via other cosmologically varying portals are preliminarily discussed in our work.

The remainder of this paper is organized as follows. In Sec.~\ref{sec:UV_model}, we build the minimal model for the scalar-controlled kinetic mixing and show that the scalar-photon coupling appears simultaneously. Based on whether the scalar-messenger couplings are $\mathbb{Z}_2$-invariant, we categorize the theory into the type-A model with the linear scalar-photon coupling and the type-B model with the quadratic scalar-photon coupling. In Sec.~\ref{sec:cos_his}, for the type-A and the type-B models, we do the systematic classification of the scalar evolution jointly affected by the thermal effect, bare potential effect, and cosmological expansion.  In Sec.~\ref{sec:dpdm_fi}, we discuss the dark photon dark matter freeze-in via the varying kinetic mixing. We also discuss the detection of the dark photon dark matter with the experimental targets set by the scalar mass and the experiments of the non-relativistic ultralight scalar relic with targets set by the dark photon dark matter freeze-in. In Sec.~\ref{sec:zn_varykm}, we build the $\mathbb{Z}_N$ model to protect the scalar's naturalness from the heavy messengers, discuss the extra cancellation from the exact $\mathbb{Z}_2$, and calculate the $\mathbb{Z}_N$-invariant Coleman-Weinberg potential. In Sec.~\ref{sec:dirac_gaugino}, we generate the varying kinetic mixing and the dark axion portal simultaneously from the Dirac gaugino model. In Sec.~\ref{sec:other_portal}, we preliminarily explore the dark matter models via other cosmologically varying portals and their experimental signals. Finally, in Sec.~\ref{sec:conclusion}, we summarize our results.  We also provide a series of appendices containing the details of the calculation, such as the analytical solutions of the high-temperature scalar evolution in \App{appx:analyt_sol}, the freeze-in of the dark photon dark matter in \App{appx:dpfi}, and the exact expansion of the $\mathbb{Z}_N$-invariant Coleman-Weinberg potential using the Fourier transformation and cosine sum rules in \App{appx:zn_vcw}.

\section{Minimal Model}
\label{sec:UV_model}
Extensive research has delved into the time-independent kinetic mixing between $U(1)_Y$ and $U(1)_d$, described by
\bea
\frac{\epsilon}{2} F^{\mu \nu} F'_{\mu \nu}.
\eea
Nevertheless, exploring time-varying kinetic mixing presents compelling motivations. From the perspective of the UV theory of the vector portal, the concept of the time-varying kinetic mixing offers a new UV realization, given that most of the current studies focus on the time-independent kinetic mixing~\cite{Dienes:1996zr, Abel:2003ue, Goodsell:2009xc, Goodsell:2011wn, DelZotto:2016fju, Gherghetta:2019coi,Benakli:2020vng, Obied:2021zjc, Rizzo:2018ntg, Wojcik:2022rtk, Chiu:2022bni}. From the perspective of dark matter models, the freeze-in of the $\kev-\mev$ dark photon dark matter~($\epsilon_\FI \sim 10^{-12}$) is excluded by the current constraints~\cite{Pospelov:2008jk, Redondo:2008ec, An:2014twa}. Making the kinetic mixing time-varying is one of the minimal solutions to open the dark photon parameter space and set a benchmark for dark photon detection. From the perspective of experiments, the varying kinetic mixing driven by the ultralight scalar is accompanied by scalar-photon coupling, which induces intriguing new signals due to the variation of the fine-structure constant and the violation of the equivalence principle. In this section, we will build the framework of the time-varying kinetic mixing through the minimal model and establish the connection between the varying kinetic mixing and the scalar-photon coupling.

To begin with, we recall the UV model of the time-independent kinetic mixing. This can be generated at the one-loop level by the fermions $\F$ and $\F'$ charged as $(e, e')$  and $(e, -e')$ under $U(1)_Y \times U(1)_d$~\cite{Holdom:1985ag}.
In the low energy limit, the constant kinetic mixing is
\be
\epsilon = \frac{e e'}{6 \pi^2} \log\frac{M}{M'},
\ee
where $M$ and $M^\prime$ are the masses of $\F$ and $\F^\prime$, respectively. To build the varying kinetic mixing, we eliminate the time-independent kinetic by imposing the mass degeneracy $M = M'$ and promote $\epsilon$ to a dynamical variable by imposing the $\mathbb{Z}_2$ symmetry under the dark charge conjugation, i.e., 
\bea
\label{eq:CD_0}
\CD: A' \rightarrow - A', \quad \epsilon \rightarrow -\epsilon.
\eea
Given this, the time-independent kinetic mixing is forbidden because $F F'$ is not invariant under $\CD$, whereas varying kinetic mixing is permitted. 

To realize this from the top-down theory, we introduce the Lagrangian 
%
\bea
\label{eq:simpuv}
\mathcal{L}_\UV \supset y \Phi \barF \F + y' \Phi \barF' \F' + \hc - M (\barF \F + \barF' \F') - \lambda \left( \left|\Phi\right|^2 - \frac{f^2}{2}\right)^2,
\eea
where $\Phi$ is a complex scalar, $y^{(\prime)}$ is the Yukawa coupling, $M$ is the heavy messenger mass, $\lambda$ is the self-coupling of the Complex scalar, and $f$ is the decay constant.  After the spontaneous breaking of the global $U(1)$ symmetry, we have $\Phi = i f e^{i \phi/f}/\sqrt{2}$, where $\phi$ is a CP-even real scalar.\footnote{More generically, we define $\Phi=\frac{f}{\sqrt{2}}e^{i(\phi/f+c)}$ with the arbitrary phase $c$, known as the Coleman-Callan-Wess-Zumino construction~\cite{Coleman:1969sm, Callan:1969sn}. Knowing that the transformation $\phi\rightarrow -\phi-\left(2c-\pi\right)f$ is equivalent to $\Phi\rightarrow -\Phi^\dagger$, under which the Yukawa interactions in \Eq{eq:simpuv} flip signs, we choose $c=\pi/2$ to fix the phase factor throughout the paper.} Given this, we define the dark charge conjugation in the UV theory as
\bea
\label{eq:CD_1} 
\CD: A \rightarrow A, \,\, A' \rightarrow - A',  \,\, \phi \rightarrow -\phi, \,\, \Psi \leftrightarrow \Psi'.
\eea
Considering either $y'=y=0$ or the more general case of $y'=-y$, both with $M'=M$, the Lagrangian is invariant under $\CD$. Given that $|y^{(\prime)}|\ll 1$  in our model, we regard $\CD$ as an approximate symmetry. 

We will see from the later discussion that as long as $\phi$ has a nonzero vacuum expectation value~(VEV), the dynamical kinetic mixing can be generated. Here, $\F^{(\prime)}$'s effective mass is
\bea
M^{(\prime)}(\phi) = M\left[ 1 + r^{(\prime)} \sin\frac{\phi}{f} \right],\,\,\,\, \text{where $r^{(\prime)} = \frac{\sqrt{2}\, y^{(\prime)} f}{M}$.}
\label{eq:M_phi}
\eea
%
In \Eq{eq:M_phi} and the rest part of this paper, we use \enquote{$(\prime)$} to denote the physical quantities in the SM sector~(without \enquote{$\prime$}) and the dark sector~(with \enquote{$\prime$}).  After integrating out $\Psi^{(\prime)}$, the Lagrangian of the effective kinetic mixing becomes
\be
\label{eq:Min_eps}
\mathcal{L}_\IR \supset \frac{\epsilon}{2}  F_{\mu \nu} F'^{\mu \nu}, \text{\quad  where $\epsilon = \frac{\sqrt{2}\, e e'\left( y - y'\right)}{ 6 \pi^2 M} f \sin\frac{\phi}{f}$}.
\ee
When $\phi \ll f$, the kinetic mixing in \Eq{eq:Min_eps} can be linearized as 
\bea
\label{eq:Min_eps_linear}
\epsilon \simeq \frac{\phi}{\LambdaKM}, \text{\quad where $\LambdaKM = \frac{6\pi^2 M}{\sqrt{2}\, e e' (y-y')}$.}
\eea
Therefore, in our model, the kinetic mixing varies with $\phi$'s cosmological evolution. 
Other than \Eq{eq:Min_eps}, another term invariant under $\CD$-transformation is the tadpole term expressed as
\bea
\mathcal{L}_\UV \supset -\frac{im_0^2 f}{\sqrt2} \Phi + \hc,
\eea
which arises as the potential\footnote{Providing that the system is invariant under $\CD$ in \Eq{eq:CD_1}, $V_0$ has to be the even function of $\phi$, while $\epsilon$ is the odd function of $\phi$. Given this, the phase difference between $\epsilon$ and $V_0$ is $(n+1/2)\pi$. Therefore, under the protection of the $\mathbb{Z}_2$ symmetry of $\mathcal{C}_d$, the potential's minima are necessarily aligned with the zeros of the kinetic mixing.}
\bea
\label{eq:V0}
V_{0} = m_0^2 f^2 \left[ 1 - \cos\left(\frac{\phi}{f}\right) \right].
\eea
During the cosmological evolution, $\phi$ initially stays at the nonzero displacement, so the portal is opened. As the universe expands, $\phi$ begins the damped oscillation around zero, so the portal is partially closed. In today's universe, $\phi$ consists of the nonrelativistic ultralight relic in the mass range $10^{-33}\eV \lesssim m_0 \ll \eV$, so the kinetic mixing has a nonzero residual. More fundamentally, this can be understood as the inverse $\mathbb{Z}_2$-breaking, as discussed in \cite{Weinberg:1974hy, Ramazanov:2021eya, Chang:2022psj, Ireland:2022quc}.

As we have shown in \Eq{eq:CD_1}, when $y' = -y$, the $\mathbb{Z}_2$ symmetry of $\CD$ is strictly preserved. When $y' \neq -y$, the $\mathbb{Z}_2$ symmetry is mildly broken but still approximate because $\abs{y^{(\prime)}} \ll 1$.\footnote{When $y' \neq -y$, the $\mathbb{Z}_2$ symmetry is mildly broken, which induces the two-loop constant kinetic mixing and the one-loop tadpole of $\phi$. Here, the small Yukawas highly suppress the two-loop constant mixing expressed as $\epsilon \sim (M/\LambdaKM)^2/e e'$. The cancellation of the $\phi$-tadpole may need fine-tuning, which can be realized in the $\mathbb{Z}_N$-protected model in Sec.~\ref{sec:zn_varykm}. Therefore, the $y' \neq -y$ case still has the approximate $\mathbb{Z}_2$ symmetry.} Based on this consideration, we classify our models into two types:
\bea
\label{eq:typeAB_Z2}
\text{Type-A:  $y\neq 0$, $y'=0$, Approximate $\mathbb{Z}_2$}. \quad \text{Type-B: $y' = - y$, Exact $\mathbb{Z}_2$.}
\eea

Due to $\phi \mh \Psi^{(\prime)}$ interaction, the scalar-photon coupling emerges from UV physics. At the one-loop level, the coupling between the CP-even scalar $\phi$ and the photon can be written as
\bea
\mathcal{L}_\IR \supset \frac{1}{4} \left(\frac{\Delta \alphaem}{\alphaem}\right) F_{\mu \nu} F^{\mu \nu},
\eea
where
\bea
\label{alpha_change}
\frac{\Delta \alphaem}{\alphaem}
= - \frac{e^2}{6 \pi^2} \left[ \log M(\phi) + \log M'(\phi)  \right] \supset \frac{\sqrt{2} e^2}{6 \pi^2} \left[- \frac{\left(y+y'\right) f}{M} \sin\left(\frac{\phi}{f}\right)+\frac{\left(y^2 + y'^2\right)  f^2}{\sqrt{2} M^2} \sin^2\left(\frac{\phi}{f}\right)\right]. 
\eea
Utilizing the classification in \Eq{eq:typeAB_Z2}, we have 
\be
\text{Type-A: $y\neq 0$, $y'=0$,\,\, $\frac{\Delta \alphaem}{\alphaem} \sim  \frac{\alphaem y \phi}{M}$,\quad \quad \quad Type-B: $y'=-y$,\,\, $\frac{\Delta \alphaem}{\alphaem} \sim \frac{\alphaem y^2 \phi^2}{M^2 }$},
\label{TypeAB_Def}
\ee
where the type-A and type-B models have linear and quadratic scalar-photon couplings, respectively. To compare with the experiments testing the fine-structure constant variation and the equivalence principle violation, we define the dimensionless constants $d_{\gamma,i} \,\,(i=1,2)$  through
\bea
\label{d1_d2_def}
\frac{\Delta \alphaem}{\alphaem} \coloneqq \left( \frac{\sqrt{4\pi} \phi}{\mpl} \right)^i \frac{d_{\gamma,i}}{i} \quad \quad \,\, (i=1,2),
\eea
where the indices ``$i=1$'' and ``$i=2$'' denote the type-A and type-B models, respectively. Comparing \Eq{d1_d2_def} with \Eq{TypeAB_Def}, we have 
\be
\label{d1_d2}
d_{\gamma,i} \sim \left(\frac{\mpl}{e' \LambdaKM}\right)^i \,\,\,\,\,(i=1,2).
\ee
Some literature uses the notation $\Delta \alphaem/\alphaem = \phi^i/\Lambda_{\gamma,i}^i$, from which we have $\Lambda_{\gamma,i} \sim e' \LambdaKM$.
In \Eq{d1_d2}, we find that with a fixed $\phi F F'$ operator inducing the effective kinetic mixing, smaller  $e'$ leads to larger $y^{(\prime)}$ such that the $\alphaem$-variation signals get stronger. Such small dark gauge coupling can be naturally generated in the large volume scenario in string compactification~\cite{Burgess:2008ri, Goodsell:2009xc, Cicoli:2011yh}.

To understand our model intuitively, one can refer to \Fig{fig:phiKM}. The left panel shows that when $\mathbb{Z}_2$ symmetry is imposed, the constant kinetic mixing is canceled, but the scalar-controlled kinetic mixing survives. In this case, as $\phi$ evolves during the cosmological evolution, the kinetic mixing becomes time-dependent. This provides a novel mechanism to generate a small but non-zero kinetic mixing. The right panel reveals that the scalar-photon couping is also generated as the byproduct when UV physics is considered. Based on the exactness of the $\mathbb{Z}_2$ symmetry, the theory can be classified as the type-A model with the linear scalar-photon couping and the type-B model with the quadratic scalar-photon couping. We will see from the later discussion that such scalar-photon couplings affect $\phi$'s evolution via the thermal effect from the SM plasma. They also change the fine-structure constant and violate the equivalence principle, which provides essential prospects for the experimental tests. 

\section{Cosmological History}
\label{sec:cos_his}

In this section, we discuss the ultralight scalar's cosmological evolution, which is affected by the scalar bare potential, thermal effect, and cosmological expansion. According to \cite{Brzeminski:2020uhm, Kapusta:2006pm}, the lowest order thermal contribution containing $\alphaem$ is at the two-loop level, and the free energy coming from that is
\bea
\label{2loop_F}
F_T \simeq  \frac{5 \pi \sum_i q_i^2}{72} \alphaem T^4 \times \mathcal{S}(T), \,\,\,\,\, \text{where}\,\,\, \mathcal{S}(T)=
\left\{
\begin{aligned}
& 1 & (T \gtrsim m_e)\\
& \frac{18}{5 \pi^3} \frac{m_e^2}{T^2} e^{-2m_e/T} & (T \ll m_e)
\end{aligned}.
\right.
\eea
In \Eq{2loop_F}, the suppression factor $\mathcal{S}(T)$ is $1$ when $e^{\pm}$ 
are relativistic but decreases exponentially after $e^{\pm}$ becomes non-relativistic. The factor $\sum_i q_i^2$~($q_i$ is the electric charge of the particle ``$i$'') quantifies the thermal contribution to the total free energy from the relativistic electric charged particles. As shown in \Eq{alpha_change}, when $\phi$ has nonzero VEV, the fine structure constant is modified, based on which $\phi$'s thermal potential can be obtained by replacing $\alphaem$ in \Eq{2loop_F} by $\alphaem(\phi)$. Combining \Eq{alpha_change} and \Eq{2loop_F}, we have~($r$ is defined in \Eq{eq:M_phi})
\bea
\label{TypeAB_VT}
\left\{
\begin{aligned}
\text{Type-A:} & \quad V_T \simeq -m_T^2 f^2 \sin\frac{\phi}{f}, & \text{where} &\,\,\, \frac{m_T}{\mathcal{S}^{1/2}} \sim \alphaem r^{1/2} \frac{T^2}{f}\\
\text{Type-B:} & \quad V_T \simeq \frac{1}{2} m_T^2 f^2 \sin^2 \frac{\phi}{f}, & \text{where} &\,\,\, \frac{m_T}{\mathcal{S}^{1/2}} \sim \alphaem r \frac{T^2}{f}
\end{aligned}.
\right. 
\eea

\begin{figure}[t]
\centering
\begin{tikzpicture}
\node at (-9,0){\includegraphics[width=0.43\columnwidth]{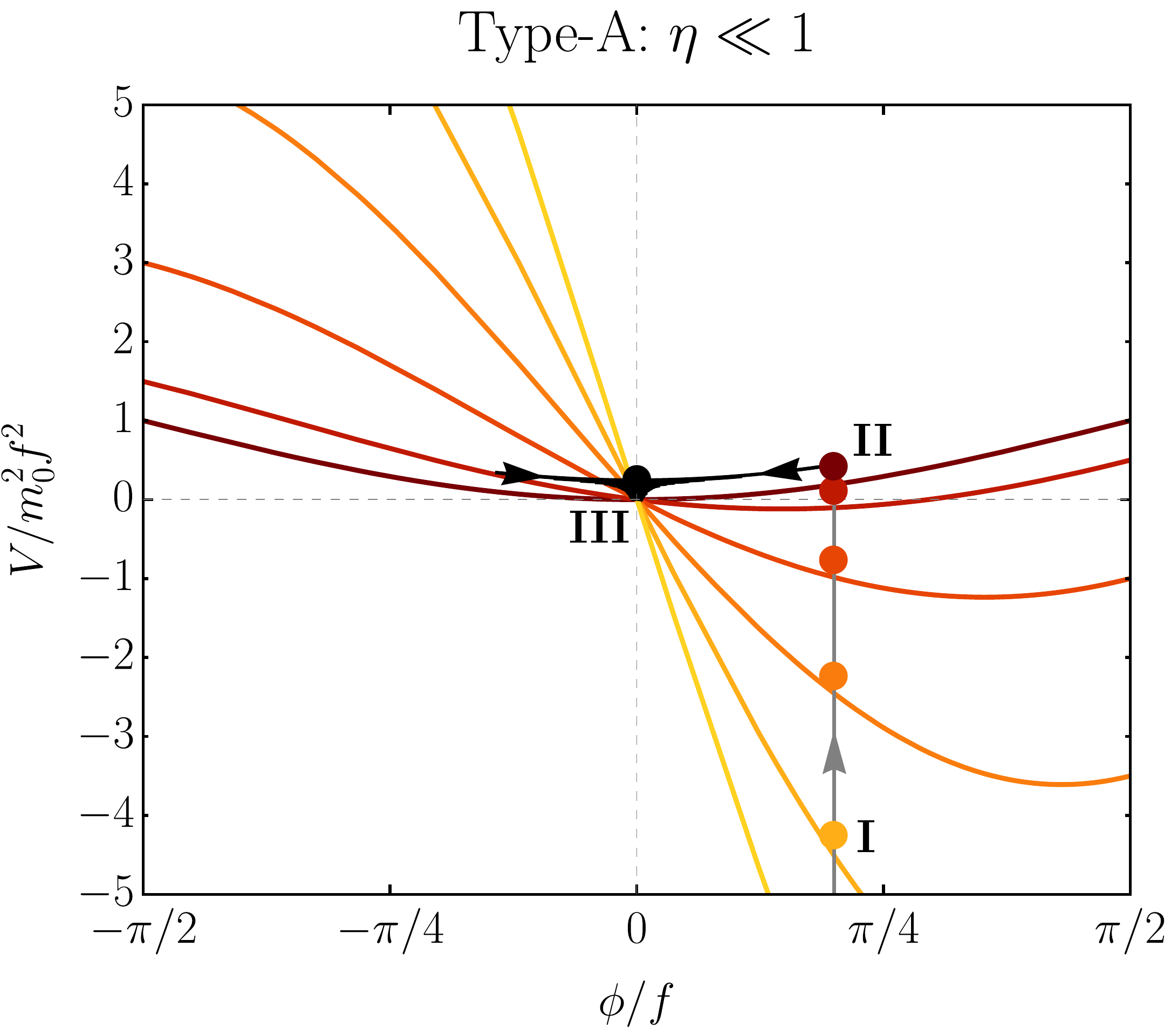}};
\node at (0, -0.025) {\includegraphics[width=0.429\columnwidth]{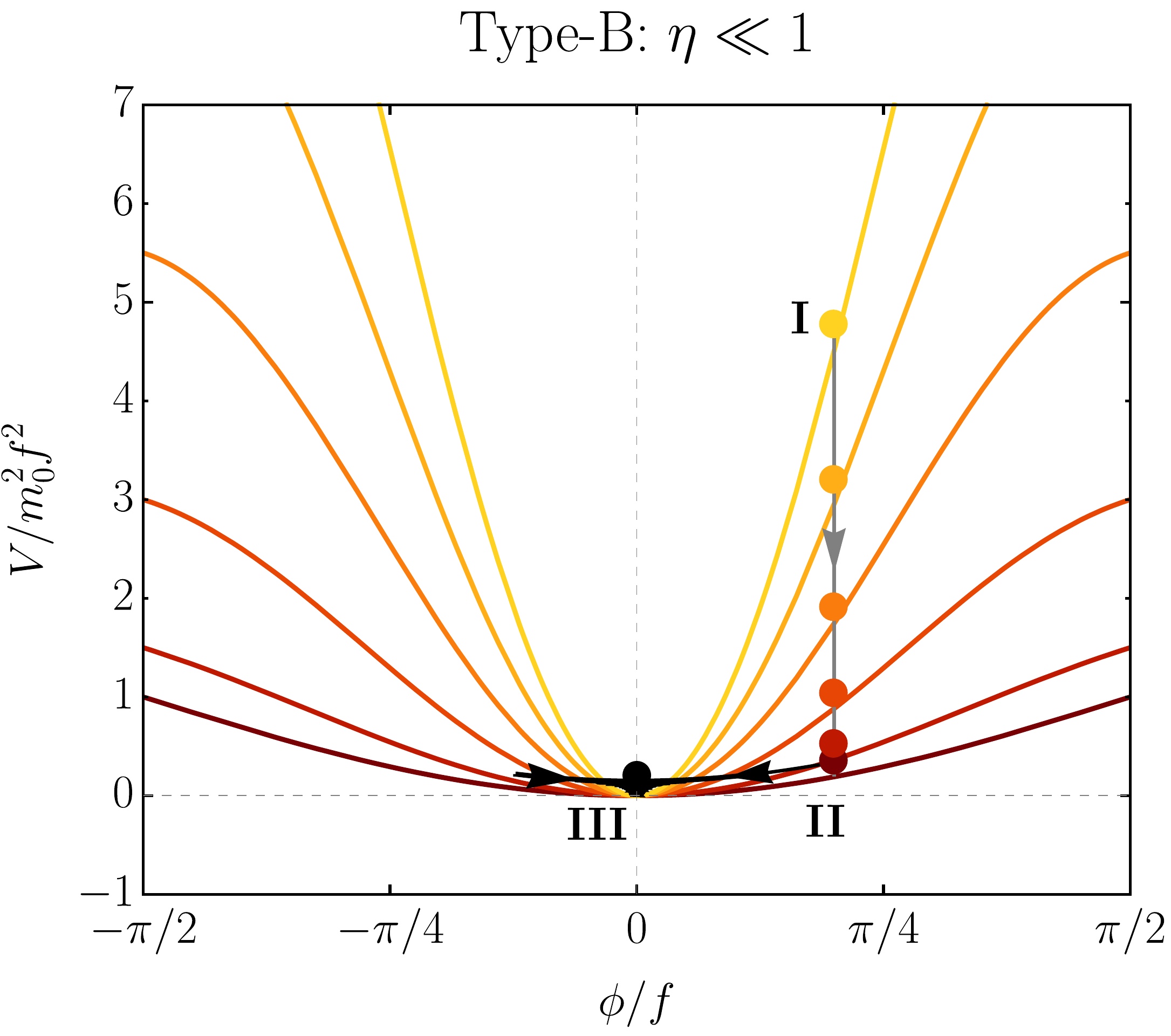}};
\node at (-9.12,-7.37){\includegraphics[width=0.44\columnwidth]{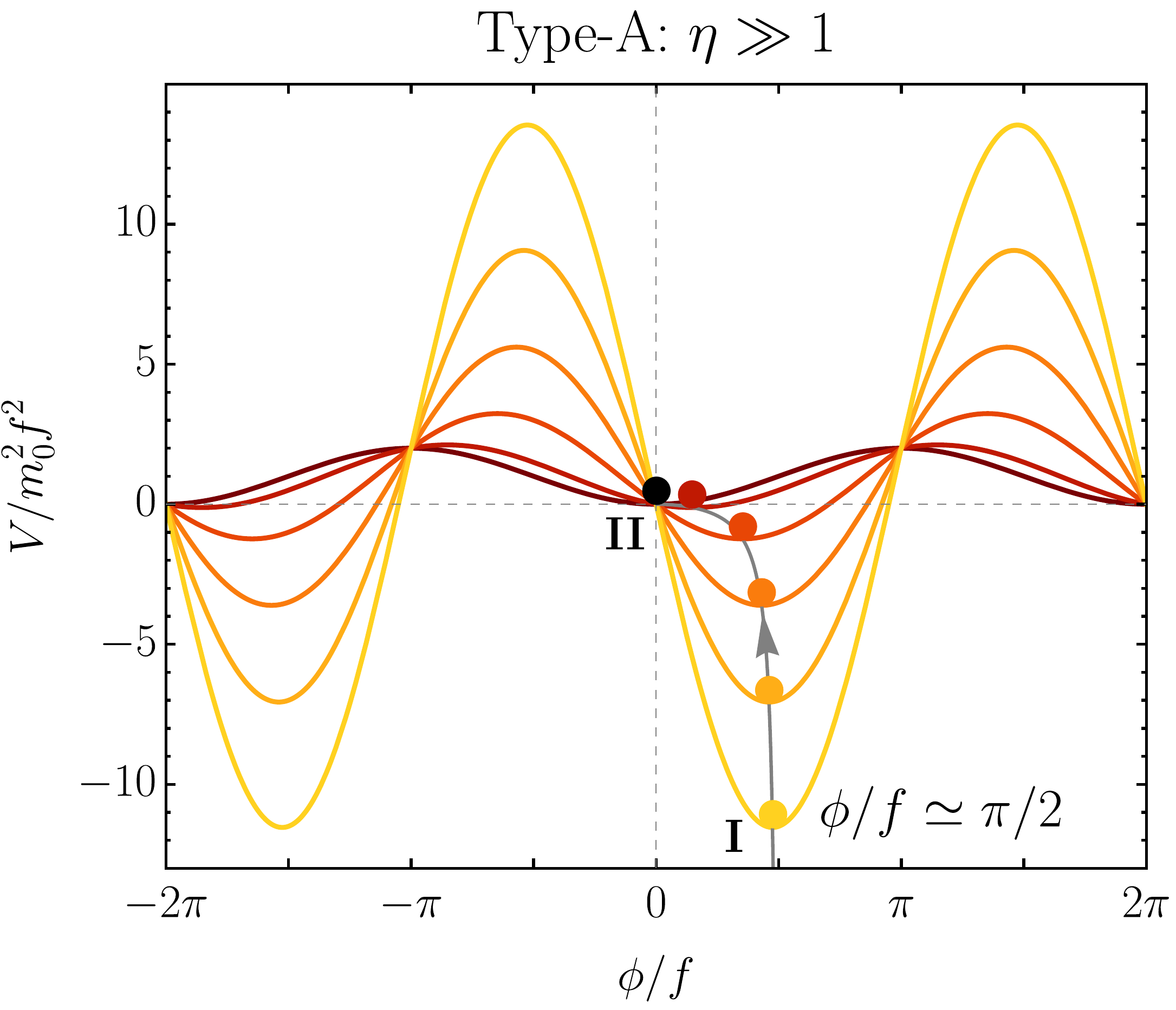}};
\node at (0.015,-7.3) {\includegraphics[width=0.423\columnwidth]{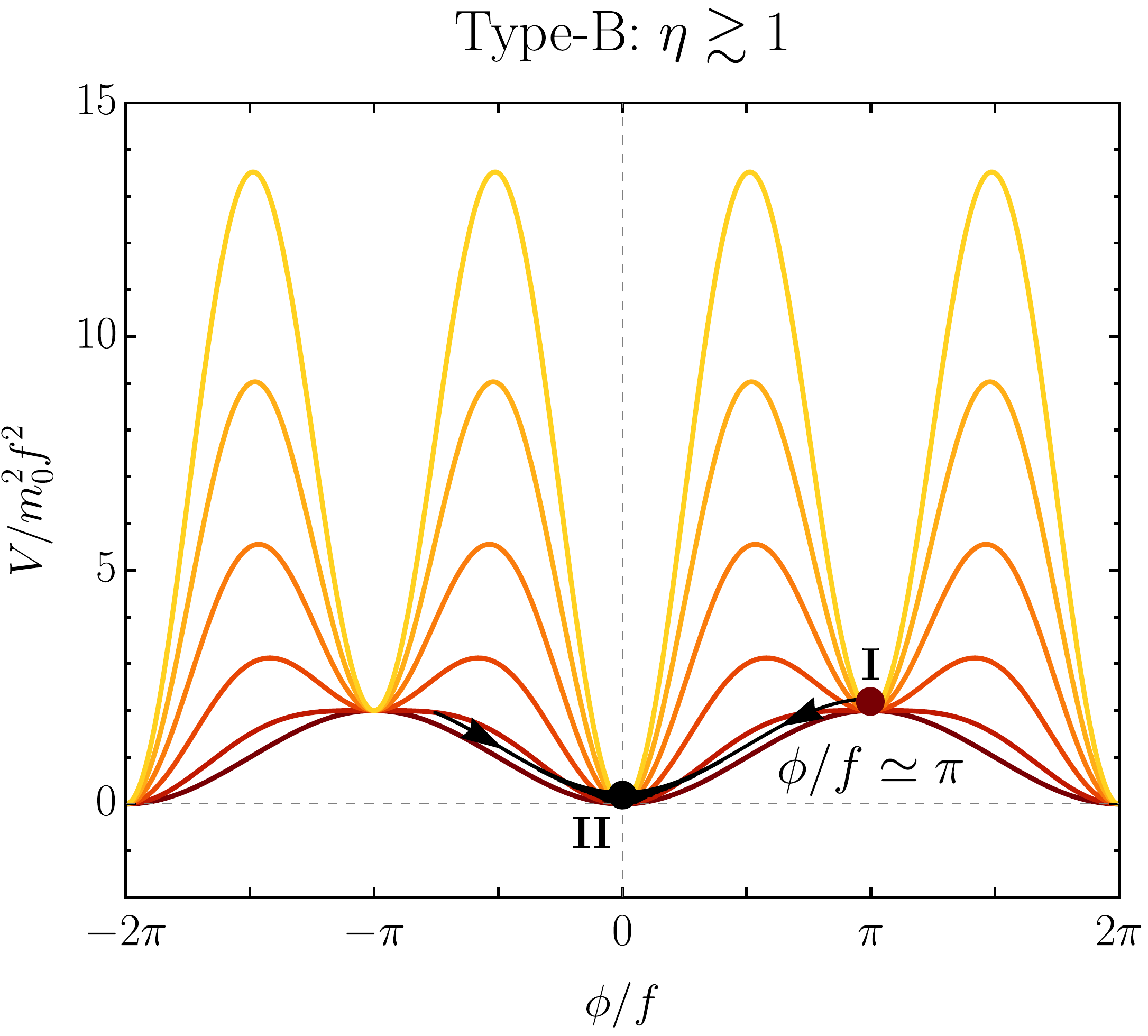}};
\end{tikzpicture}
\caption{Scalar evolution in the early universe. From yellow to dark red, $m_T/m_0 = 5,4,3,2,1,0$.  
{\bf Upper left and right}: Standard misalignment. Because $\eta \ll 1$, the thermal effects are negligible. At the early stage~($\text{\rom{1}} \rightarrow \text{\rom{2}}$), $\phi$ is frozen. When $3H \simeq m_0$~(\rom{2}), $\phi$ begins to oscillate with $\abs{\phi} \propto T^{3/2}$~($\text{\rom{2}} \rightarrow \text{\rom{3}}$) and converges to zero~(\rom{3}). {\bf Lower left}: Type-A model with $\eta \gg 1$. When the temperature is high, $\phi$ oscillates and converges to $\pi f/2$. When $T \gtrsim m_e$, the variation of the thermal potential is adiabatic, so $\phi$ tracks the potential minimum $\phi_{\min} = f \arctan(m_T^2/m_0^2)$~($\text{\rom{1}} \rightarrow \text{\rom{2}}$). When $T \ll m_e$, the adiabatic condition is violated, $\phi$ cannot follow $\phi_{\min}$ and starts to oscillate obeying $\abs{\phi} \propto T^{3/2}$, which is not shown here.
{\bf Lower right}: For the type-B model with $\eta \gtrsim 1$ and $\phi$ initially trapped inside the wrong vacuum, $\phi$ obeys $\abs{\phi - \pi f} \propto T^{1/2}$ at high temperatures. As the universe cools down, $\phi$ follows $\abs{\phi} \propto T^{3/2}$~($\text{\rom{1}} \rightarrow \text{\rom{2}}$). }
\label{fig:V_AB}
\end{figure}

Here, $T_\rh$ is much smaller than $M$, so $\F$'s contribution to the thermal potential is exponentially suppressed. In addition, because there is no dark electron, the thermal effect from the dark sector is also negligible. \Fig{fig:V_AB} shows how the scalar potential and the scalar field evolve with the temperature under different circumstances. 
From yellow to dark red, $m_T/m_0=5,4,3,2,1,0$. 
For the thermal potential, the local minimums of the type-A and type-B models are~($n\in \mathbb{N}$)
\be
\label{eq:TypeAB_VT_min}
\text{Type-A:} \,\,\, \frac{\phi_{\min}}{f} = \arctan\left(\frac{m_T^2}{m_0^2}\right) + 2n\pi , \,\,\,\,\, \text{Type-B:} \,\,\, \frac{\phi_{\min}}{f} = 
\left\{
\begin{aligned}
& n \pi  & \,\,\, (m_T > m_0)\\
& 2 n \pi  & \,\,\, (m_T \leq m_0)
\end{aligned}
\right. .
\ee
In the following discussion,  we focus on the range $- 2 \pi f \leq \phi \leq 2 \pi f$ without loss of generality. For the type-A model, because $m_T \gg m_0$ in the early epoch, $\phi_{\min} \simeq \pi f/2$. When $m_T \ll m_0$, $\phi_{\min}$ continuously shifts to zero. For the type-B model, when $m_T > m_0$, within the $2\pi f$ periodicity there are three local minimums, i.e., $\phi_{\min} = 0, \pm \pi f$. When $m_T \leq m_0$, only $\phi_{\min} =0 $ is the true minimum. During the entire process, $\phi$ keeps the classical motion without the tunneling from $\sim\pi f$ to $0$ because $f$'s largeness makes the tunneling rate much smaller than $H$~\cite{Coleman:1977py, Callan:1977pt, Linde:1981zj, Lee:1985uv, Duncan:1992ai, Intriligator:2006dd, Pastras:2011zr}.

Because $m_T, H \propto T^2$ when $T\gtrsim m_e$, we define a dimensionless quantity
\be
\label{eq:eta_def}
\eta \coloneqq \frac{2 m_T}{\mathcal{S}^{1/2} H} \sim \alphaem \frac{\mpl}{f} \times \left\{ 
\begin{aligned}
& r^{1/2} & \text{(Type-A)}\\
& r & \text{(Type-B)}
\end{aligned}\,\,\,
\right. 
\ee
to classify $\phi$'s evolution. The motion of $\phi$ is underdamped if $\eta > 1$, because the thermal effect dominates over the universe's expansion; In contrast, if $\eta < 1$, $\phi$'s motion is overdamped under the Hubble friction. $\eta=1$ denotes the critical case. To relate $\eta$ with experimental observables quantified by $d_{\gamma,i}~(i=1,2)$, we write $\eta$ as
\be
\label{eta_from_de}
\text{Type-A: $\eta \sim \left(\frac{\alphaem d_{\gamma,1} \mpl}{f} \right)^{\frac{1}{2}}$, \,\, Type-B: $\eta \sim \left( \alphaem d_{\gamma,2} \right)^{\frac{1}{2}}$}.
\ee

\begin{figure}[t]
\centering
\begin{tikzpicture}
\node at (-9.5,0){\includegraphics[width=0.45\columnwidth]{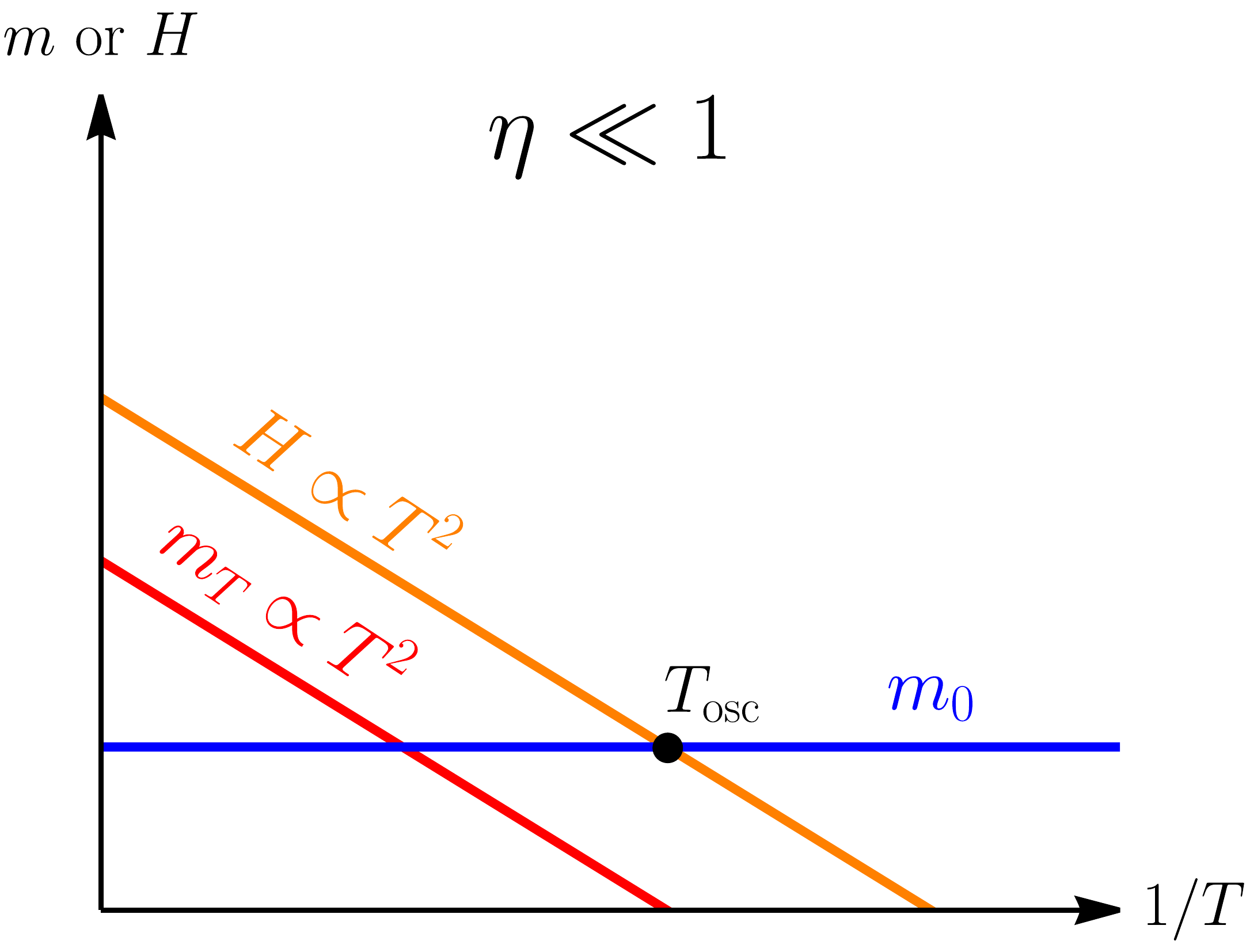}};
\node at (-0.2, 0.02) {\includegraphics[width=0.45\columnwidth]{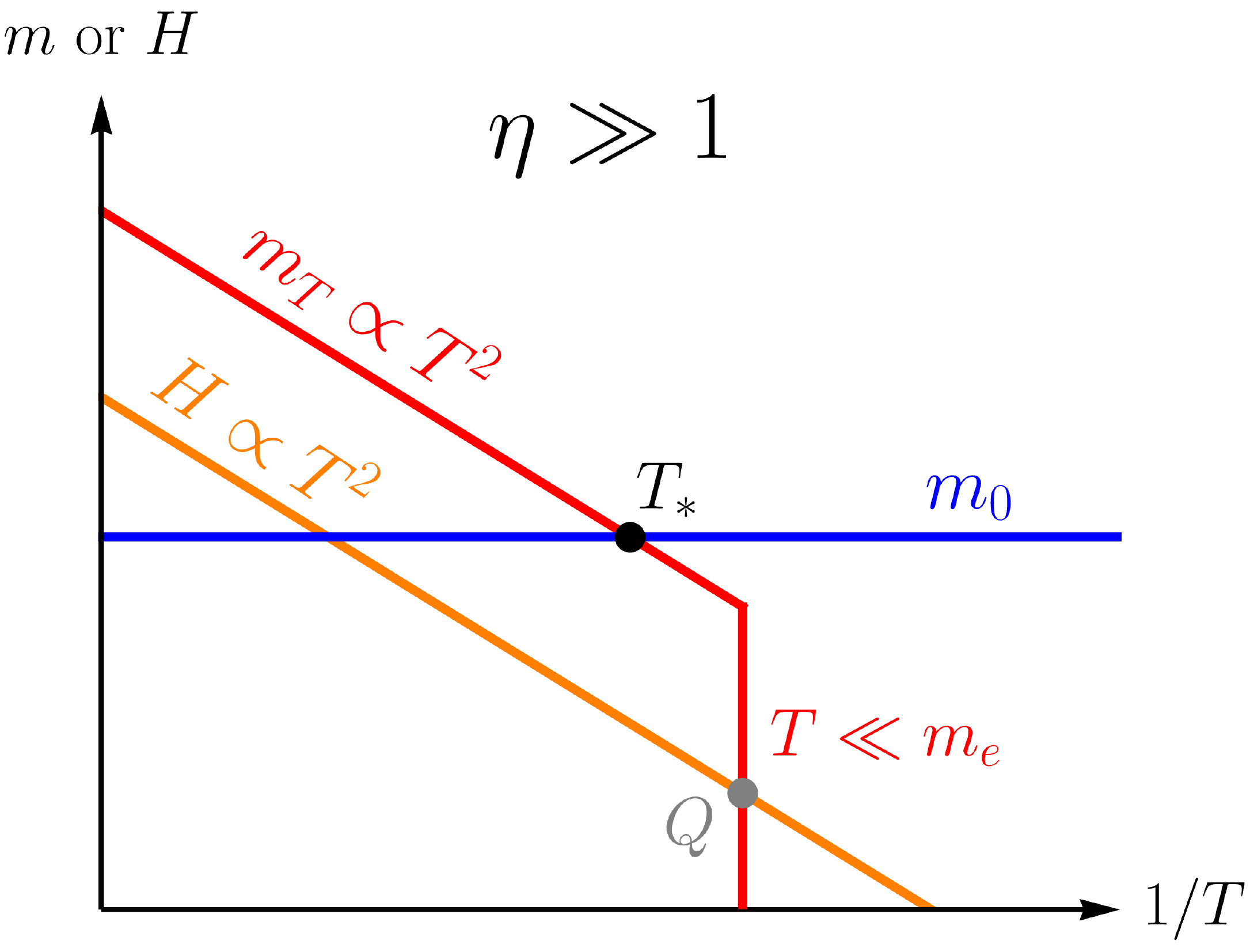}};
\end{tikzpicture}
\caption{The cosmological evolution of $m_T$~(red), $m_0$~(blue) and $H$~(orange) shown in the $\log$ scale. The $m_T$ and $H$ are parallel with each other when $T \gtrsim m_e$, because $H, m_T \propto T^2$. {\bf Left}:\, $\eta \ll 1$. The thermal effect is negligible, therefore $\phi$'s evolution obeys the standard misalignment. $\phi$ stays constant and then starts oscillation at $T_{\osc} \sim (m_0 \mpl)^{1/2}$ obeying $\abs{\phi} \propto T^{3/2}$.{\bf Right}:\, $\eta \gg 1$. The thermal effect is dominant. At $T_*$, $m_T = m_0$. At ``$Q$'', $H \sim m_T$. When $T\gg T_*$, $\phi$ converges to the thermal minimum obeying $\abs{\phi - \phi_{\min}} \propto T^{1/2}$. For the type-A model, when $T \gtrsim m_e$, $\phi$ follows $\phi_{\min} = f \arctan(m_T^2/m_0^2)$. When $T \ll m_e$, $\phi$ cannot follow $\phi_{\min}$ but begins to oscillate. For the type-B model, we focus on the case with $\phi$ being inside the wrong initial vacuum. When $T \gg T_*$, $\phi$ is trapped inside $\phi_{\min} = \pi f$. When $T \lesssim T_*$, $\phi$ begins oscillating obeying $\abs{\phi} \propto T^{3/2}$. If $T_* \ll T|_{3H=m_0}$, $\phi$'s oscillation is postponed. This means $\phi$ does the trapped misalignment.}
\label{fig:morHEvo}
\end{figure}

Because we compare $m_T$ and $m_0$ to determine whether the thermal effect dominates over the bare potential effect, we define the temperature $T_*$ at which $m_T = m_0$. Combining \Eq{TypeAB_VT} and \Eq{eq:eta_def}, we have 
\bea
\label{eq:T_thermal}
m_T=m_0:  \quad T_* \sim \max\left[0.1 m_e, \left(\frac{m_0 \,\mpl}{\eta}\right)^{1/2} \right].
\eea
%
For $m_T$ to be smaller than $m_0$, we only need one of the following two conditions to be satisfied: 1.\,$T \ll m_e$. Here $m_T$ is exponentially suppressed. 2.\,The unsuppressed part of $\phi$'s thermal mass, i.e., $m_T/\mathcal{S}^{1/2}$, is smaller than $m_0$. 

Based upon the comparison between the thermal and bare potential effects, we classify the scalar evolution as
\bea
\label{eq:misalignment_classify}
\text{\bf Standard Misalignment:}\,\, \eta \ll 1, \quad \quad \text{\bf Thermal Misalignment:}\,\, \eta \gtrsim 1.
\eea
When $\eta \ll 1$, the thermal effect is negligible. When $\eta\gtrsim 1$, the thermal effect is important, and one should consider it alongside the bare potential and the universe expansion.

\subsection{Standard Misalignment: $\eta \ll 1$}\label{subsec:eta<<1}
In this case,  when $3H \gtrsim m_0$, $\phi$ obeys
\be
\label{eq:approx_delphi_eta<<1}
\phi - \phi_{\min} \propto T^{\frac{\eta^2}{4}},
\ee
which means $\phi$ is frozen in the early universe. To have an overall picture, one could refer to the left panel of \Fig{fig:morHEvo}: The $H$-line is higher than the $m_T$-line during the whole process, meaning that the thermal effect is negligible. Therefore, one only needs to focus on the $H$-line and $m_0$-line. After the crossing of $H$-line and $m_0$-line, the scalar begins the damped oscillation whose amplitude obeys $\abs{\phi} \propto T^{3/2}$. The upper left and upper right panels of \Fig{fig:V_AB} show how $\phi$ moves for the type-A and type-B models separately when $\eta\ll 1$: From \rom{1} to \rom{2}, $\phi$ remains the same. This means that the initial field displacement determines $\phi$'s starting amplitude $\abs{\phi}_\osc$. Here, we focus on the model with the natural initial condition, i.e., $\abs{\phi}_\osc/{f} \sim \mathcal{O}(1)$. Afterwards, $\phi$ begins to oscillate at the temperature
\be
\label{eq:T_phi_eta<<1}
T_{\osc} = T|_{3H=m_0} \sim \text{few} \times 10^{-1} \eV \times \left( \frac{m_0}{10^{-28}\eV}\right)^{\beta_T}, \quad \text{where} \,\, \beta_T = 
\left \{ 
\begin{aligned}
\frac{1}{2} & \quad \,\,\,   (m_0 \gtrsim 10^{-28}\eV) \\
\frac{2}{3} & \quad \,\,\,   ( 10^{-33} \eV \lesssim m_0 \lesssim 10^{-28}\eV)
\end{aligned}
\right. .
\ee
$\beta_T$ in \Eq{eq:T_phi_eta<<1} and later mentioned $\beta_\phi$ in \Eq{phi_evo_eta<<1} are determined by $H$'s power law of $T$, which is $T^{2}$ in the radiation-dominated universe and $T^{3/2}$ in the matter-dominated universe. For $m_0 \simeq 10^{-28}\eV$, the scalar starts oscillation at the matter-radiation equality~($T \sim \eV$). 

Because the inflation smears out the field anisotropy, $\phi$ does spatially homogeneous oscillation, which is
\be
\label{phi_evo_eta<<1}
\phi(t) \simeq \abs{\phi} \cos\left(m_0 t\right), \,\, \text{where $\abs{\phi} = \frac{\sqrt{2 \rho_\phi}}{m_0} \propto T^{3/2}$.}
\ee
From~\cite{Preskill:1982cy, Arias:2012az} one knows that $\phi$ satisfies $\rho_\phi \propto T^3$ and $w_\phi = p_\phi/\rho_\phi \simeq 0$, so $\phi$ is part of the dark matter with the fraction $\mathcal{F} = \Omega_\phi/\Omega_\dm$. Without loss of generality, we choose $\mathcal{F}=10^{-3}$ as a benchmark value, such that $\phi$'s parameter space is not excluded by Lyman-$\alpha$ forest~\cite{Irsic:2017yje, Kobayashi:2017jcf, Armengaud:2017nkf, Zhang:2017chj, Nori:2018pka, Rogers:2020ltq}, CMB/LSS~\cite{Hlozek:2014lca, Lague:2021frh}, galaxy rotational curves~\cite{Bernal:2017oih, Robles:2018fur, Bar:2018acw, Bar:2019bqz, Bar:2021kti}, and ultra-faint dwarf galaxies~\cite{Dalal:2022rmp}.
After substituting \Eq{eq:T_phi_eta<<1} into \Eq{phi_evo_eta<<1}, one can get the starting oscillation amplitude
\be
\label{eq:phi_osc_eta<<1}
\abs{\phi}_{\osc} \sim \text{few} \times 10^{16}\gev \times \left( \frac{\mathcal{F}}{10^{-3}}\right)^{1/2} \left(\frac{10^{-28}\eV}{ m_0}\right)^{\beta_\phi}, \quad \,\,\text{where}\,\,\beta_\phi =  \left \{
\begin{aligned}
\frac{1}{4} & \quad \,\,\,   (m_0 \gtrsim 10^{-28}\eV)\\
0 & \quad \,\,\, ( 10^{-33} \eV \lesssim m_0 \lesssim 10^{-28}\eV)
\end{aligned}
\right.
\ee
In the case where $m_0 \gg 10^{-25}\,\eV$, $\phi$'s de Broglie wavelength is much smaller than the scale of the Milky Way halo, so $\phi$ behaves like a point-like particle similar to other cold DM particles. For this reason, $\phi$'s local density is $\rho_{\phi,\local} \simeq \mathcal{F} \rho_\local$, where $\rho_\local \simeq 0.4 \,\gev/\cm^3$ is DM's local density near earth. By effectively adding an enhancement factor
\be
\label{eq:Enhance_Factor}
\mathcal{E} = \left(\frac{\rho_\local}{\rho_{\text{average}}} \right)^{1/2} \simeq 6 \times 10^2
\ee
where $\rho_{\text{average}} = \rho_c \Omega_\text{DM} \simeq 1.3 \,\kev/\cm^{3} $ on $\abs{\phi}$ in \Eq{phi_evo_eta<<1}, we get $\phi$'s amplitude today near the earth, which is $\abs{\phi}_0 \simeq (2 \mathcal{F} \rho_\local)^{1/2}/m_0$. Here, $\abs{\phi}_0$ denotes  $\phi$'s local oscillation amplitude today. If $m_0 \ll 10^{-25}\eV$, oppositely, $\phi$ cannot be trapped inside the Milky Way halo's gravitational potential well. In this case, today's $\phi$ field is homogeneous, so the enhancement factor in \Eq{eq:Enhance_Factor} should not be included, from which we have $\abs{\phi}_0 = (2 \mathcal{F} \rho_\text{average})^{1/2}/m_0$. $\phi$'s oscillation amplitude in the middle mass range needs numerical simulation, which is left for future exploration.

\begin{figure}[t]
\centering
\begin{tikzpicture}
\node at (-9.2,0){\includegraphics[width=0.472\columnwidth]{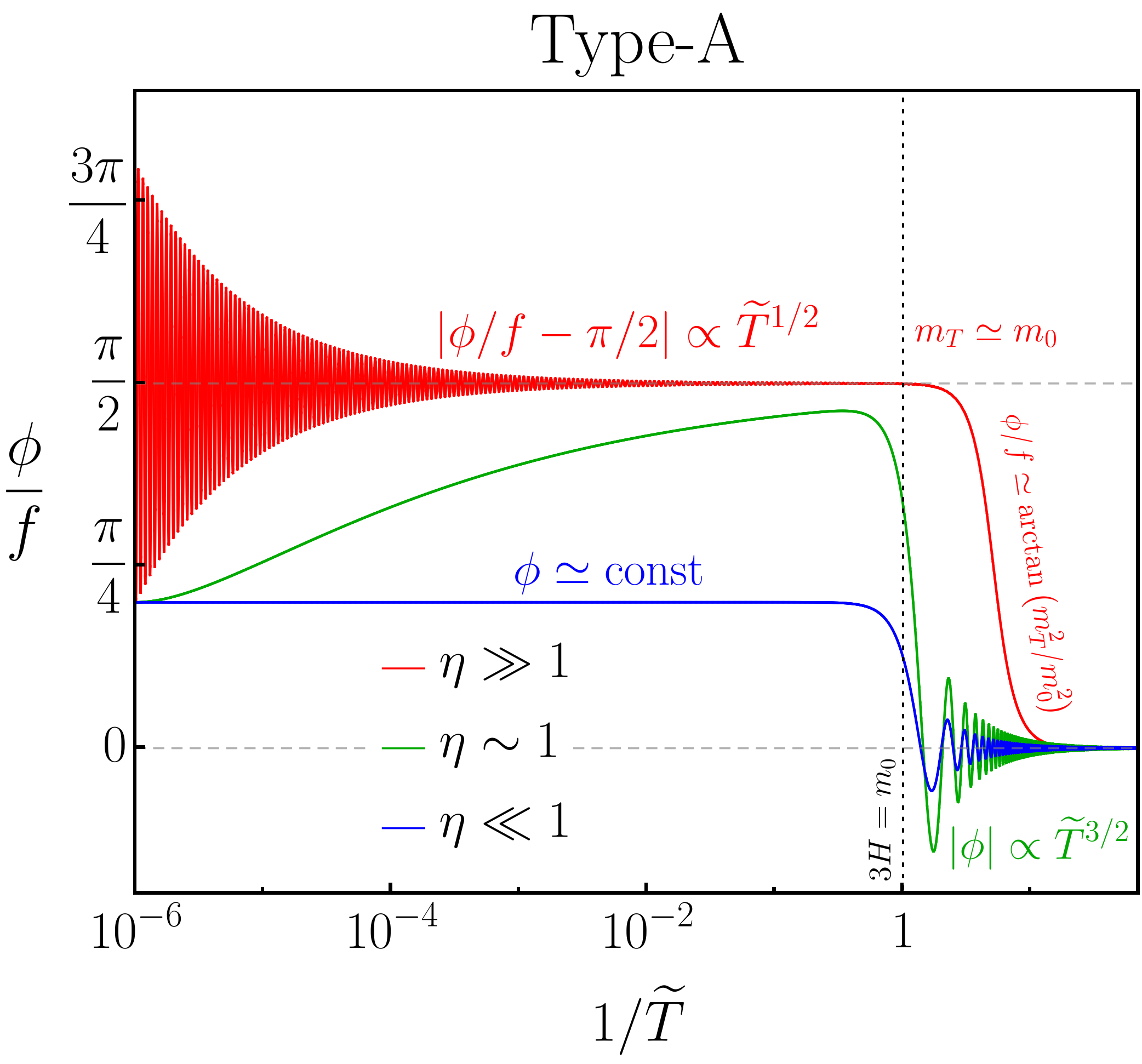}};
\node at (0, -0.01) {\includegraphics[width=0.48\columnwidth]{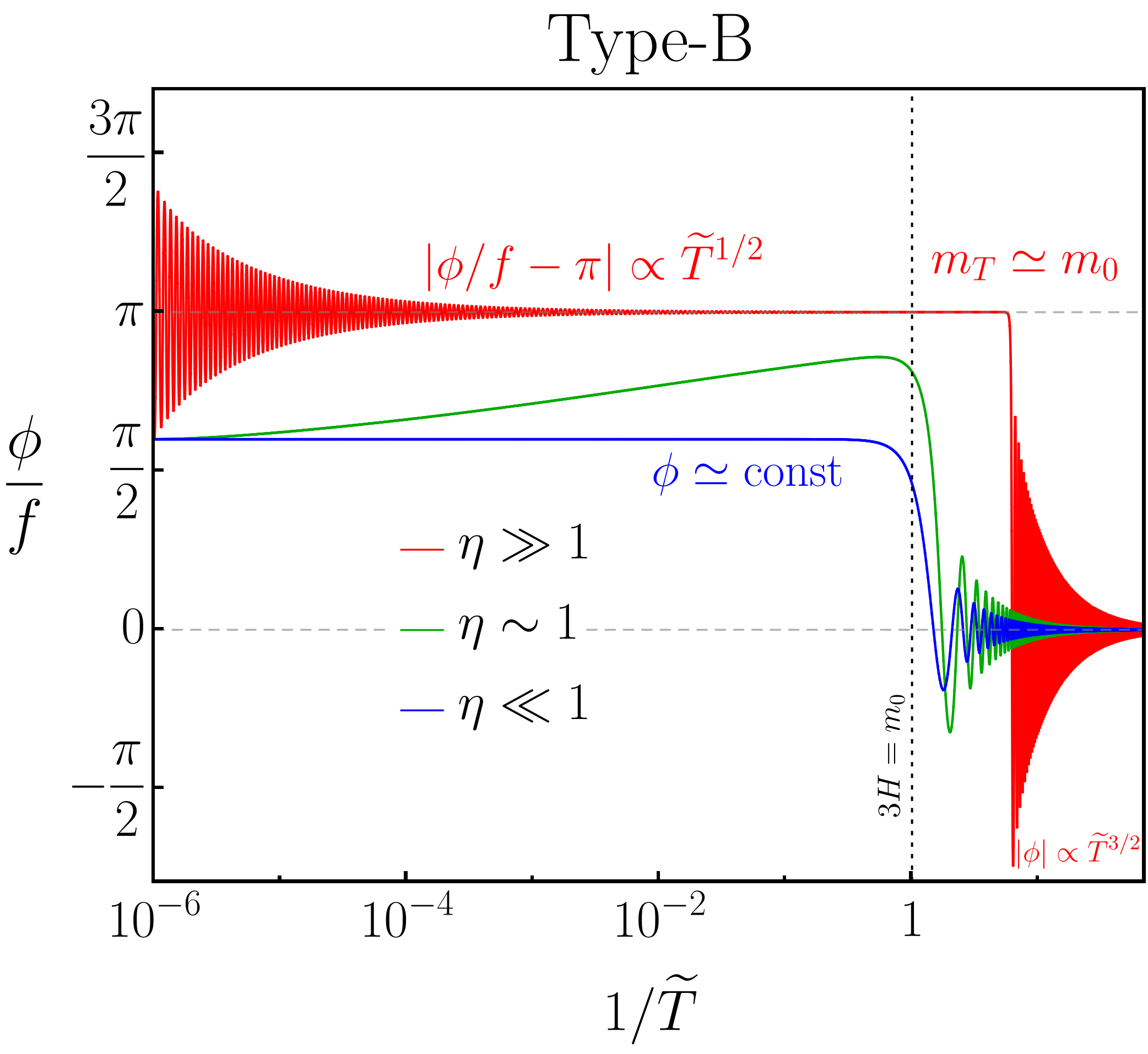}};
\end{tikzpicture}
\caption{$\phi$'s cosmological evolution. We define $\widetilde{T} \coloneqq T/T|_{3H=m_0}$. The red, green, and blue lines represent the $\eta \gg 1$, $\eta \sim 1$, and $\eta \ll 1$ cases, respectively. Both green and blue lines start oscillation at $\widetilde{T} \sim 1$, but they are different in the sources of the displacement: The green line's  displacement is thermally determined to be $\abs{\phi}_\osc \simeq \abs{\phi_{\min}}$, whereas the blue lines' oscillation amplitude depends on the initial conditions. 
{\bf Left}: The type-A model. For the red line, when $m_T \gg m_0$, $\phi$ obeys $\abs{\phi- \pi f/2} \propto \widetilde{T}^{1/2}$. As the temperature decreases, $\phi\simeq f \arctan(m_T^2/m_0^2)$ until $T \sim m_e$. After that, $\phi$ cannot follow the potential minimum but oscillates obeying $\abs{\phi} \propto \widetilde{T}^{3/2}$.
For the green line, when $\widetilde{T}\gtrsim 1$, $\phi$ gradually slides to the thermal minimum $\pi f/2$ following \Eq{eq:approx_delphi_eta}. When $\widetilde{T} \sim 1$, $\phi$ obeys $\abs{\phi} \propto \widetilde{T}^{3/2}$. The blue line represents the standard misalignment where the thermal effect is negligible. When $\widetilde{T} <1$, $\abs{\phi} \propto \widetilde{T}^{3/2}$. 
{\bf Right}: The type-B model. The red line represents the situation where $\eta\gg 1$ and $\pi f/2 \lesssim \phi_\rh \lesssim 3 \pi f/2$. When $m_T \gtrsim m_0$, $\phi$ obeys $\abs{\phi- \pi f} \propto \widetilde{T}^{1/2}$. When $m_T \lesssim m_0$, $V''|_{\phi = \pi f} \lesssim 0$, so $\phi$ rolls down from $\pi f$ and oscillates following $\abs{\phi} \propto \widetilde{T}^{3/2}$. The red line starts oscillation later than $\widetilde{T} \simeq 1$, which is classified as the trapped misalignment. The green line represents the situation where $\eta \sim 1$ and $\pi f/2 \lesssim \phi_\rh \lesssim 3 \pi f/2$. When $\widetilde{T} \gtrsim 1$, $\phi$ gradually slides to the thermal minimum $\pi f$ obeying \Eq{eq:approx_delphi_eta}. When $\widetilde{T} \lesssim 1$, $\abs{\phi} \propto \widetilde{T}^{3/2}$. 
}
\label{fig:phi_evo}
\end{figure}

\subsection{Thermal Misalignment: $\eta \gtrsim 1$}\label{subsec:eta>=1}

In this case, the thermal effect from the SM plasma plays a decisive role in $\phi$'s evolution in the early universe. To have a full picture, one can refer to the right panel of \Fig{fig:morHEvo}: At the early stage, since $m_T$ and $H$ are both proportional to $T^2$, their lines are approximately parallel in the plot with the $\log$ scale. 
In the plot, we label the cross point of $m_T$-line and $m_0$-line with ``$T_*$'', which is already defined in \Eq{eq:T_thermal}. When $T \lesssim T_*$, the effect from the bare potential dominates over the effect from the thermal potential. When $T \ll m_e$, the $m_T$-line drops fast following $e^{-m_e/T}$ and crosses the $H$-line. We label the cross point of the $m_T$-line and the $H$-line as ``$Q$'', and we have
\bea
\label{eq:TQ_HQ}
T_Q \sim m_e/\log \eta, \quad H_Q \sim 10^{-16}/\log^2 \eta \,\,\eV. 
\eea
As we will see in the rest of this section, comparing $m_0$ and $H_Q$ is vital in determining the temperature at which $\phi$ starts the late-time oscillation.

Let us first discuss $\phi$'s movement in the stage $T \gg T_*$, during which the bare potential can be omitted in the high-temperature environment. In \App{appx:analyt_sol}, we solve the scalar evolution for this situation. Because $\phi-\phi_{\min}$ is the linear combination of $T^{\frac{1\pm\sqrt{1-\eta^2}}{2}}$, to describe the thermal convergence more quantitatively, we need the specific value of $\eta$. We recall that the potential minimums are $\phi_{\min}=\pi f/2$ for the type-A model and $\phi_{\min}=0, \pm \pi f$ for the type-B model.  Given the initial condition $\dot{\phi}_\rh \simeq 0$, we approximately have~
\be
\label{eq:approx_delphi_eta}
\phi - \phi_{\min} \propto 
\left\{
\begin{aligned}
& T^{\frac{1-\sqrt{1-\eta^2}}{2}} & \quad  (\eta \leq 1)\\
& T^{\frac{1}{2}} \cos\big[ \sqrt{\eta^2-1} \log\left(T/T_\rh\right)/2 \big]  & \quad (\eta > 1)
\end{aligned}
\right. ,
\ee
where $\eta = 1$ is the critical value determining how $\phi$ evolves towards the local minimum. 

If $\eta \leq 1$~(but not too small), $\phi$ gradually slides to $\phi_{\min}$, as shown in the green lines in the left and right panels of \Fig{fig:phi_evo}. Incidentally, in the limit $\eta \ll 1$, we come back to \Eq{eq:approx_delphi_eta<<1} where $\phi$ is nearly frozen in the early universe, which is related to the blue lines in both panels of \Fig{fig:phi_evo}. Here, the $T^{\frac{1+\sqrt{1-\eta^2}}{2}}$ term is negligible because it describes the movement with non-zero $\dot{\phi}_\rh$. If $\eta > 1$, $\phi$'s movement towards $\phi_{\min}$ is oscillating because $T^{\frac{1 \pm i\sqrt{\eta^2-1}}{2}} = T^{1/2} e^{\pm i \sqrt{\eta^2-1} \log T/2}$. Obeying the power law $T^{1/2}$, $\phi$'s oscillation amplitude decreases as the temperature goes down. This can be explained more intuitively: When $T \gtrsim m_e$, the adiabatic condition of the WKB approximation is satisfied~($\dot{m}_T/m_T^2 \sim 1/\eta \lesssim 1$), so there is no particle creation or depletion for $\phi$, or equivalently speaking, its number density is conserved. For this reason, $\phi$'s oscillation amplitude obeys $\abs{\phi - \phi_{\min}} \simeq \sqrt{2 n_\phi/m_T} \propto T^{\frac{1}{2}}$. \Eq{eq:approx_delphi_eta} reveals an interesting phenomenon: As long as there is a hierarchy between $T_\rh$ and $m_e$, which is natural in most of the inflation models, $\phi$ converges to the local minimum of the thermal potential. For example, in $\eta\gtrsim 1$ case, given that $T_\rh \sim 10\,\gev$, when the universe temperature drops to $T\sim m_e$, the field deviation from the local minimum becomes $\abs{\phi-\phi_{\min}}/\phi_{\min} \sim 10^{-2}$. Higher $T_\rh$ leads to even smaller field displacement from the thermal minimum. Such determination of scalar's misalignment through the thermal effect is named the thermal misalignment in several recent works~\cite{Brzeminski:2020uhm, Batell:2021ofv, Batell:2022qvr, Croon:2022gwq}. Other mechanisms setting the scalar's nonzero initial displacement can be found in ~\cite{Co:2018mho, Takahashi:2019pqf, Huang:2020etx}.  As shown in \Eq{eq:misalignment_classify}, we use the term thermal misalignment throughout our paper for the cases satisfying $\eta \gtrsim 1$ to distinguish them from the standard misalignment in which the thermal effect does not play a role. We use such a definition because when this condition is satisfied, the thermal effect from the SM bath washes out $\phi$'s sensitivity of the initial condition and dynamically sets the field displacement. 

When $T \lesssim T_*$, the bare potential becomes important in $\phi$'s evolution. In the $m_0 \lesssim H_Q$ case, the ultralight scalar's evolution is simply the combination of the damped oscillation in the early-time wrong vacuum and the damped oscillation in the late-time true vacuum with
\bea
T_\osc \simeq T|_{3H=m_0}, \quad  \abs{\phi}_\osc \simeq \abs{\phi_{\min}},
\eea
where $\phi_{\min}= \pi f/2$ for the type-A model and $\phi_{\min} = \pm \pi f$ for the type-B model with $\phi$ initially in the wrong vacuum. We do not put more words on that because this kind of thermal misalignment can be treated as the standard misalignment with the thermally determined initial condition. Now, we shift our focus to the $m_0 \gtrsim H_Q$ case in the rest of this section.  Since the total potentials of the type-A and type-B models have different shapes depending on the exactness of the $\mathbb{Z}_2$ symmetry, and the scalar evolution depends on the numerical value of $\eta$, let us describe the scalar evolution case by case: 
\begin{itemize}
\item {\bf Type-A, $\eta \gg 1$}.

When $T \gg T_*$, $\phi$ does the damped oscillation, which converges to $\pi f/2$. Afterward, when $T \sim T_*$, or equivalently speaking, $m_T \sim m_0$, the potential minimum begins to shift from $\pi f/2$ to $0$, obeying $\phi= f \arctan(m_T^2/m_0^2)$ as shown in \Eq{eq:TypeAB_VT_min}. When the universe temperature is much higher than $m_e$, the adiabatic condition is satisfied because $\dot{m}_T/m_T^2 \sim 1/\eta \ll 1$. We show the movement of $\phi$ in the lower left panel of \Fig{fig:V_AB} and the red line in the left panel of \Fig{fig:phi_evo}. However, as the temperature drops below $m_e$, because $\dot{m}_T/m_T^2 \sim e^{m_e/T}/\eta \gg 1$, $\phi$ is not able to respond to the sudden variation of the potential minimum anymore and begins the oscillation. 
Here, the oscillation temperature and the starting amplitude can be written as
\be
\label{phiTe_arctan}
T_\osc \sim T_Q , \quad \abs{\phi}_\osc  \sim f \,(T_Q/T_*)^4,
\ee
where $T_Q\sim m_e/\log \eta$ as defined in \Eq{eq:TQ_HQ}. Given $f \ll \mpl$, the hierarchy between $T_*$ and $T_Q$ strongly suppresses $\phi$'s late-time oscillation amplitude, leading to $\phi$'s minuscule relic abundance. Even though this phase does not appear in the following context of the dark photon dark matter freeze-in in \text{Sec}.~\ref{sec:dpdm_fi},  we still discuss this phase in this section for completeness.

\item {\bf Type-A, $\eta \sim 1$}.

In the early time when $T \gg T|_{3H=m_0}$, $\phi$ slides to the thermal minimum $\pi f/2$. When $T \sim T|_{3H=m_0}$, the scalar begins the late-time damped oscillation obeying $\abs{\phi} \propto T^{3/2}$ with the starting temperature and amplitude 
\bea
T_\osc  \sim T|_{3H = m_0}, \quad  \abs{\phi}_\osc \simeq \pi f/2. 
\eea
In \Fig{fig:V_AB}, we do not list such a case because, even though categorized as the thermal misalignment, it can be decomposed into the standard misalignment~(upper left panel of \Fig{fig:V_AB}) plus the determined initial amplitude. In the left panel of \Fig{fig:phi_evo}, the green line describes such a situation: When $\widetilde{T}\gtrsim 1$, the green line gradually slides to the $\pi f/2$. When $\widetilde{T}\sim 1$, the green line begins the damped oscillation with the amplitude scaled as $\abs{\phi} \propto \widetilde{T}^{3/2}$.

\item {\bf Type-B,  $\eta \gtrsim 1$}.

Unlike the type-A model, there is no continuous shift of the potential minimum during the cosmological evolution. Therefore, we only need to focus on the moment when the local minimum flips. Here, we mainly focus on the case in which $\phi$'s initial condition satisfies $\pi f/2 \lesssim \phi_\rh \lesssim 3\pi f/2$. In this case, because of the thermal effect when $T \gg T_*$, $\phi$ converges to the local minimum $\pi f$ inside the wrong vacuum. When $T \lesssim T_*$, at the point $\phi = \pi f$, the second order derivative becomes nonpositive, i.e., $V''|_{\phi = \pi f} \lesssim 0$, thereafter $\phi$ begins the oscillation around zero. From \Eq{eq:T_thermal}, we have
\bea
\label{eq:Tosc_phiosc_TypeB_eta<<1}
T_\osc \simeq T_*, \quad \abs{\phi}_\osc \simeq \pi f. 
\eea
According to \Eq{eq:T_thermal}, we recall that $T_* \sim T|_{3H=m_0}/\eta^{1/2} \sim (m_0 \mpl/\eta)^{1/2}$ when $m_0 \gtrsim \eta H_Q$, and $T_* \sim 0.1 m_e$ when $H_Q \lesssim  m_0 \lesssim \eta H_Q$. 

For the $\eta \gg 1$ case, one could look at the red line in the right panel of \Fig{fig:phi_evo}. Here, $\phi$ oscillates around the thermal minimum $\pi f$ with the power law $\abs{\phi - \pi f} \propto T^{1/2}$ when $T \gg T_*$. Afterward, when $T \simeq T_*$, alternatively speaking, $m_T \simeq m_0$, $\phi$ begins the oscillation following $\abs{\phi} \propto T^{3/2}$. The plot shows the apparent postponement of the scalar oscillation for the red line compared with the other two lines. Because the CP-even scalar $\phi$'s oscillation is postponed, the evolution of $\phi$ can be classified as the trapped misalignment, which is formally investigated in axion models~\cite{Nakagawa:2020zjr, DiLuzio:2021gos}. In this case, $\phi$'s relic abundance is enhanced given the same $\abs{\phi}_\osc$ or $f$. We can also think about such characteristics inversely: For $\phi$ to reach the same abundance quantified by $\mathcal{F}$, one only needs smaller $\abs{\phi}_\osc$ or $f$. To be more quantitative, one can write the misalignment at the beginning of the oscillation as
\bea
\label{eq:typeB_phiosc_rescale}
\abs{\phi}_\osc \simeq \abs{\phi}_{\osc, \, \standard} \left(T_*/T|_{3H=m_0}\right)^{3/2},
\eea
where $\abs{\phi}_{\osc,\standard}$ denotes the starting amplitude for the standard misalignment as shown in \Eq{eq:phi_osc_eta<<1}. From \Eq{eq:typeB_phiosc_rescale}, one can see that the necessary early misalignment $\abs{\phi}_\osc$ is rescaled by a factor of $(T_*/T|_{3H=m_0})^{3/2}$, which shows that $\abs{\phi}_\osc$ is much smaller compared with $\abs{\phi}_{\osc,\standard}$ when $\mathcal{F}$ is determined.

For the $\eta \sim 1$ case, one could refer to the green line in the right panel of \Fig{fig:phi_evo}. In the early stage, i.e., $T \gtrsim T|_{3H=m_0}$, $\phi$ slowly moves to $\pi f$ obeying $\abs{\phi - \pi f} \propto T^{\frac{1-\sqrt{1-\eta^2}}{2}}$. When $T \simeq T|_{3H=m_0}$, $\phi$ begins the damped oscillation with the power law $\abs{\phi} \propto T^{3/2}$. From the discussion above, we can find that the $\eta \sim 1$ case has the thermal determination of the initial condition but does not have the postponement of the oscillation. Therefore, $\phi$ is in the phase of the thermal misalignment but not in the phase of the trapped misalignment. 

Finally, let us briefly discuss the case where $- \pi f/2\lesssim \phi_\rh \lesssim \pi f/2$. Here we have that $\phi$ oscillates obeying $\abs{\phi} \propto T^{1/2}$ when $T \gg T_*$, and then oscillates obeying $\abs{\phi} \propto T^{3/2}$ when $T\lesssim T_*$. Because the late-time amplitude is suppressed by a factor of $(T_\rh/T_*)^{1/2}$, given the nonnegligible fraction of $\phi$ among the dark matter~(For example, $\mathcal{F} \sim 10^{-3}$), $f$ depends on $T_\rh$ and may be larger than $\mpl$. Therefore, this paper focuses on the case where $\phi$ is initially localized inside the wrong vacuum. 
\end{itemize}

\section{Dark Photon Dark Matter} \label{sec:dpdm_fi}

In this section, we discuss the freeze-in of the $\kev-\mev$ dark photon dark matter via varying kinetic mixing. As shown in \cite{Pospelov:2008jk, Redondo:2008ec, An:2014twa}, the dark photon freeze-in through the time-independent kinetic mixing~($\epsilon_\FI \sim10^{-12}$) is excluded by the dark matter direct detection, stellar energy loss, the CMB energy injection, and the galactic photon spectrum. Alternative dark photon dark matter production mechanisms include misalignment~\cite{Nelson:2011sf, Arias:2012az, Alonso-Alvarez:2019ixv, Nakayama:2019rhg},
gravitational production~\cite{Graham:2015rva, Ema:2019yrd, Ahmed:2020fhc, Kolb:2020fwh, Wang:2022ojc, Redi:2022zkt}, the radiation of the cosmic string network~\cite{Long:2019lwl, Kitajima:2022lre}, and axion tachyonic instability~\cite{Agrawal:2018vin, Dror:2018pdh, Co:2018lka, Bastero-Gil:2018uel, Co:2021rhi}. Realizing that all the aforementioned mechanisms do not rely on kinetic mixing, which is indispensable for dark photon detection, we provide a minimal extension of the dark photon dark matter freeze-in where the ultralight scalar's evolution dynamically sets the kinetic mixing's experimental benchmarks. In addition, we want to stress the significance of detecting the ultralight scalar $\phi$ in the whole mass range, i.e., $10^{-33}\eV \lesssim m_0 \ll \eV$: Even though $\phi$ cannot be $100\%$ dark matter when $m_0 \lesssim 10^{-17}\eV$ according to the fuzzy dark matter bounds~\cite{Irsic:2017yje, Kobayashi:2017jcf, Armengaud:2017nkf, Zhang:2017chj, Nori:2018pka, Rogers:2020ltq, Dalal:2022rmp} and the superradiance constraints~\cite{Arvanitaki:2014wva, Stott:2018opm, Davoudiasl:2019nlo, Unal:2020jiy}, tiny amount of $\phi$'s relic can still open the gate for the main component dark matter's production.

Here, we briefly introduce our setup. The dark photon dark matter is produced through the operator $\mathcal{L} \supset \phi F_{\mu \nu} F'^{\mu \nu}/2 \LambdaKM$ whose effective kinetic mixing is supported by $\phi$'s non-zero VEV in the early universe. Thereafter, when $T \simeq T_\osc$, $\phi$ begins the damped oscillation following $\abs{\phi} \propto  \left(T/T_\osc\right)^{3/2}$. Because most of the constraints are imposed when $T \ll T_\osc$, the dark photon's parameter space can be vastly extended, and the ratio $T_0/T_\osc$ determines today's local kinetic mixing for future dark matter detection. In addition, because the $\phi F^2$ or $\phi^2 F^2$ operator is induced simultaneously from the UV theory as discussed in \text{Sec}.~\ref{sec:UV_model}, testing the fine-structure constant variation and the equivalence principle violation through the ground-based experiments
~\cite{Smith:1999cr, Schlamminger:2007ht, VanTilburg:2015oza, Hees:2016gop, Hees:2018fpg, Barontini:2021mvu, collaboration2021frequency, Banerjee:2022sqg, Baggio:2005xp, Arvanitaki:2015iga, Badurina:2019hst, MAGIS-100:2021etm}, satellite-based experiments
~\cite{Berge:2017ovy, Tsai:2021lly, AEDGE:2019nxb, Brzeminski:2022sde}, astrophysics
~\cite{Kaplan:2022lmz, Hamaide:2022rwi}, and cosmology
~\cite{Stadnik:2015kia, Hart:2019dxi, Sibiryakov:2020eir, Bouley:2022eer, Hamaide:2022rwi} open a new window for the dark matter experiments.

\subsection{Dark Photon Production}

\begin{figure}[t]
\centering
\includegraphics[width=0.6\columnwidth]{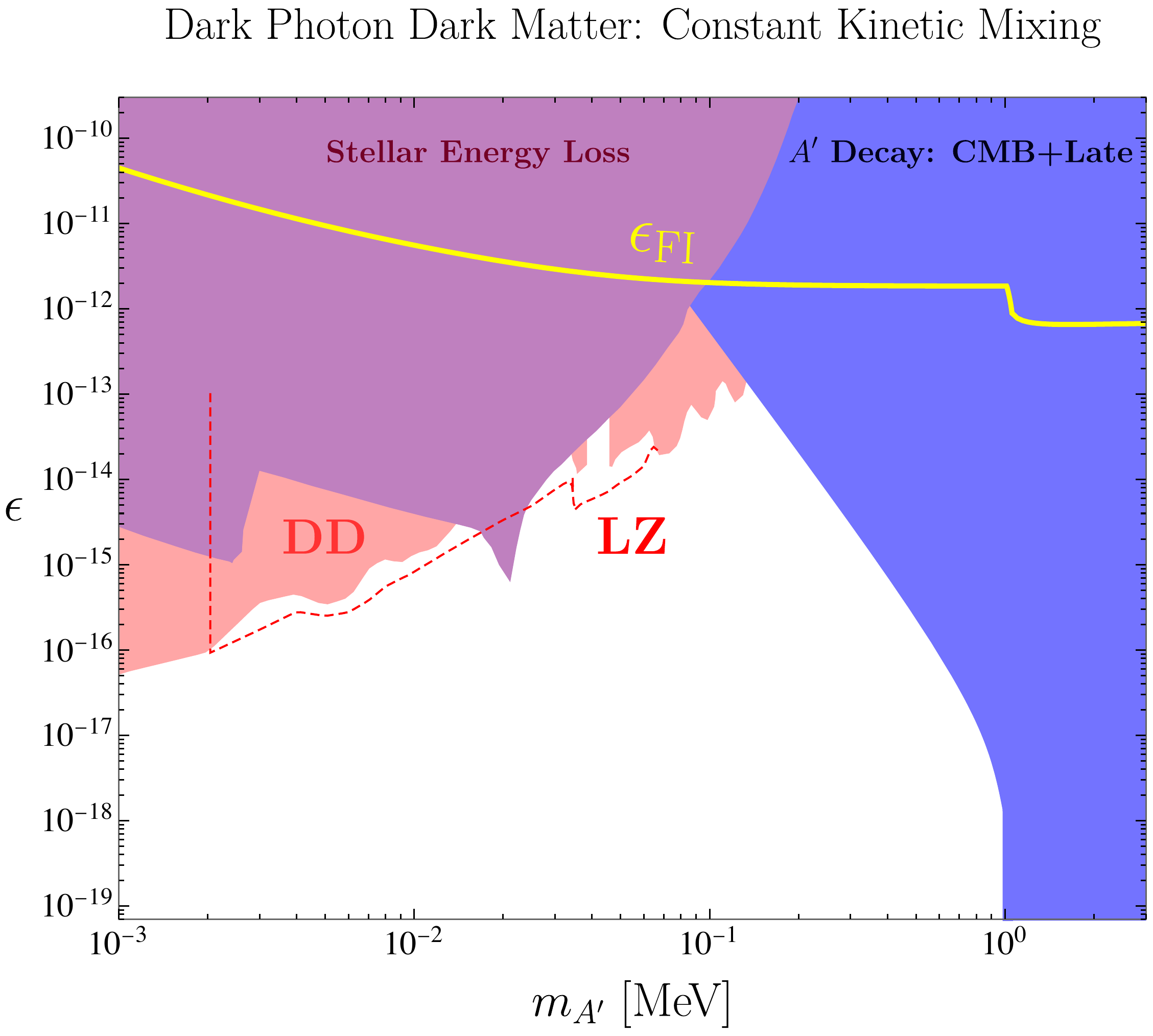}
\caption{The parameter space of the dark photon dark matter with the time-independent kinetic mixing $\epsilon$. The yellow line labeled with ``$\epsilon_\FI$'' is the dark photon freeze-in line with $\Omega_{A'} h^2 = 0.12$. The purple region represents the stellar energy loss constraints of red giants, horizontal branches, and the sun~\cite{Redondo:2008aa, Pospelov:2008jk, Redondo:2008ec, Redondo:2013lna, An:2013yfc, An:2014twa, Hardy:2016kme}. The shaded red region denotes the dark matter direct detection~(DD) constraints~\cite{An:2014twa, Bloch:2016sjj, XENON:2018voc, XMASS:2018pvs, XENON:2019gfn, XENON:2020rca, XENONCollaboration:2022kmb}. 
The dashed red line is the projection of the LZ experiment~\cite{LZ:2021xov}. The blue region denotes the dark photon dark matter decay constraints from CMB and late time, which are comparable with each other when $\epsilon$ is a constant~\cite{Pospelov:2008jk, Redondo:2008ec, An:2014twa, McDermott:2017qcg}.  
For the dark photon, its dominant channel is $A'\rightarrow 3 \gamma$ when $m_{A'}<2 m_e$, or $A' \rightarrow e^- e^+$ when $m_{A'} \geq 2 m_e$. From the plot, one can see that dark photon dark matter freeze-in via the constant $\epsilon$ is ruled out. }
\label{fig:DP_bound}
\end{figure}
In this section, we do the back-of-envelope calculation of the dark photon freeze-in through the kinetic mixing. To simplify the discussion, we focus on the region where the dark photon is produced before the scalar oscillation. Detailed calculations can be found in \App{appx:dpfi}.

When $m_{A'} < 2m_e$, the Boltzmann equation is
\bea
\label{eq:Boltz_AToAp}
\dot{n}_{A'} + 3H n_{A'} \simeq n_\gamma \langle \Gamma_{\gamma \rightarrow A'} \rangle, ~~ \text{where $\langle \Gamma_{\gamma \rightarrow A'} \rangle \sim \frac{\epsilon^2 m_{A'}^4}{T} \delta(m_\gamma^2 - m_{A'}^2)$}.
\eea
$n_{\gamma}$ and $n_{A'}$ are the photon and dark photon number densities, $m_\gamma$ is the plasmon mass, and $\langle \Gamma_{\gamma \rightarrow A'} \rangle$ is the thermally-averaged $\gamma \rightarrow A'$ transition rate. From \Eq{eq:Boltz_AToAp}, we know $\gamma \rightarrow A'$ oscillation happens when $m_\gamma \simeq m_{A'}$. Because $\Gamma_{A'} \ll H$ in the experimentally allowed region, the dark photon decay does not affect its abundance. Plugging $m_\gamma^2 \simeq 2 \pi \alphaem T^2/3$ into \Eq{eq:Boltz_AToAp}, we have
\bea
\label{eq:T_Omega_m_Ap<2me}
T_{\gamma \rightarrow A'} \sim 8 m_{A'}, \quad \, \Omega_{A', \, \gamma \rightarrow A'} \sim \epsilon^2 \alphaem^{3/2}  \,\frac{\mpl}{\Teq},
\eea
where $T_{\gamma \rightarrow A'}$ is the resonant temperature, $\Teq$ is matter-radiation equality temperature, and $\Omega_{A', \, \gamma \rightarrow A'}$ is the corresponding dark photon abundance. 
Given \Eq{eq:T_Omega_m_Ap<2me} and $\Omega_{A'}h^2\simeq 0.12$, we have
\bea
\label{eq:eps_FI}
\epsilon_\FI \sim 10^{-12},
\eea 
which describes the horizontal behavior of the yellow line in \Fig{fig:DP_bound}. In the region $m_{A'} \ll m_e$, $\gamma \rightarrow A'$ happens when $T \ll m_e$, $m_\gamma^2$ is exponentially suppressed, so $\epsilon_\FI$ needs to be larger as compensation. From \App{appx:dpfi}, we have 
$\epsilon_\FI \propto m_{A'}^{-3/2}$, which explains the yellow line's slope when $m_{A'}$ is small. 

When $m_{A'} \geq 2 m_e$, $e^- e^+ \rightarrow A'$ dominates over $\gamma \rightarrow A'$. The Boltzmann equation is
\bea
\label{eq:Boltz_EEToAp}
\dot{n}_{A'} + 3 H n_{A'} \simeq n_{e^-} n_{e^+} \langle \sigma_{e^- e^+ \rightarrow A'} \rangle, \quad \, \text{where $n_{e^-} n_{e^+} \langle \sigma_{e^- e^+ \rightarrow A'} \rangle \sim \epsilon^2 \alphaem m_{A'}^{5/2} T^{3/2} e^{-m_{A'}/T}$}.
\eea
Inside \Eq{eq:Boltz_EEToAp}'s right-hand side, the factor $e^{-m_{A'}/T}$ suppresses the dark photon production when $T \ll m_{A'}$. From \Eq{eq:Boltz_EEToAp}, we know the dark photon's production temperature and the relic abundance are
\bea
\label{eq:T_Omega_m_Ap>2me}
T_{e^- e^+ \rightarrow A'} \sim m_{A'}, \quad \, \quad \Omega_{A', \,e^- e^+ \rightarrow A'} \sim \epsilon^2 \alphaem \frac{\mpl}{\Teq}.
\eea
From \Eq{eq:T_Omega_m_Ap>2me}, we find that   $\Omega_{A', \,e^- e^+ \rightarrow A'}$ is similar to  $\Omega_{A', \, \gamma \rightarrow A'}$ but different by an $\alphaem^{1/2}$ factor. Therefore, $\epsilon_\FI$ in $m_{A'} \geq 2m_e$ is slightly smaller than $\epsilon_\FI$ in $m_{A'} < 2m_e$, which explains the lowering of the yellow line when $m_{A'}\geq 2m_e$.  


Being different from the time-independent kinetic mixing, our model has extra UV freeze-in channels, such as $\gamma \rightarrow A' \phi$ and $e^- e^+ \rightarrow A' \phi$. However, these channels are subdominant. Here, we have
\bea
\label{eq:NonVEV_vs_VEV}
\frac{\Omega_{A', \gamma \rightarrow A' \phi}}{\Omega_{A', \gamma \rightarrow A'}}  \sim \frac{m_{A'} T_\rh}{\abs{\phi}_\osc^2} \ll 1,
\eea
where $\Omega_{A', \gamma \rightarrow A' \phi}$ is the dark photon abundance from $\gamma \rightarrow A' \phi$. For the ultralight $\phi$, because a smaller mass leads to a larger scalar amplitude, it is easy to realize $|\phi|_\osc^2 \gg m_{A'} T_\rh$. Therefore, most of the dark photon comes from $\gamma \rightarrow A'$, because $\gamma \rightarrow A' \phi$ are highly suppressed by the large $\LambdaKM$, while $\gamma \rightarrow A'$ is compensated with large $\abs{\phi}_\osc$. Similar discussions can be applied to other UV freeze-in channels, such as $e^- e^+ \rightarrow A' \phi$.

\subsection{Signatures and Constraints}\label{subsec:DPFI_signal}

For our model, the experiments can be categorized into two types: 1.~The dark photon dark matter detection. This relies on the varying kinetic mixing between the visible and dark sectors. 
2.~The ultralight scalar detection. This is based on the scalar-photon couplings, which vary the fine-structure constant and violate the equivalence principle. 

\subsubsection{ Detection of the Dark Photon Dark Matter }
\begin{figure}[t]
\centering
\includegraphics[width=0.49\columnwidth]{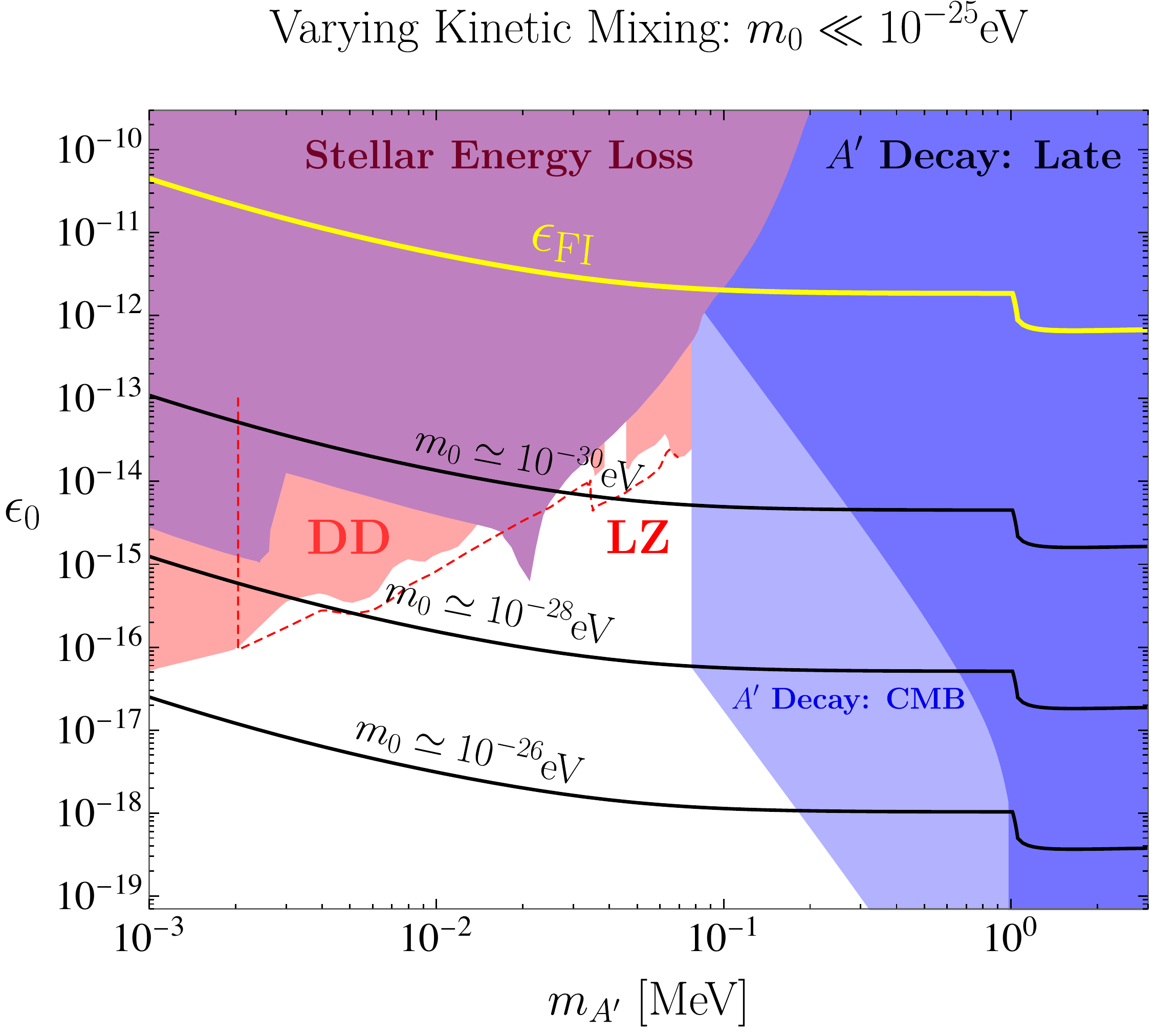}
\hfill
\includegraphics[width=0.49\columnwidth]{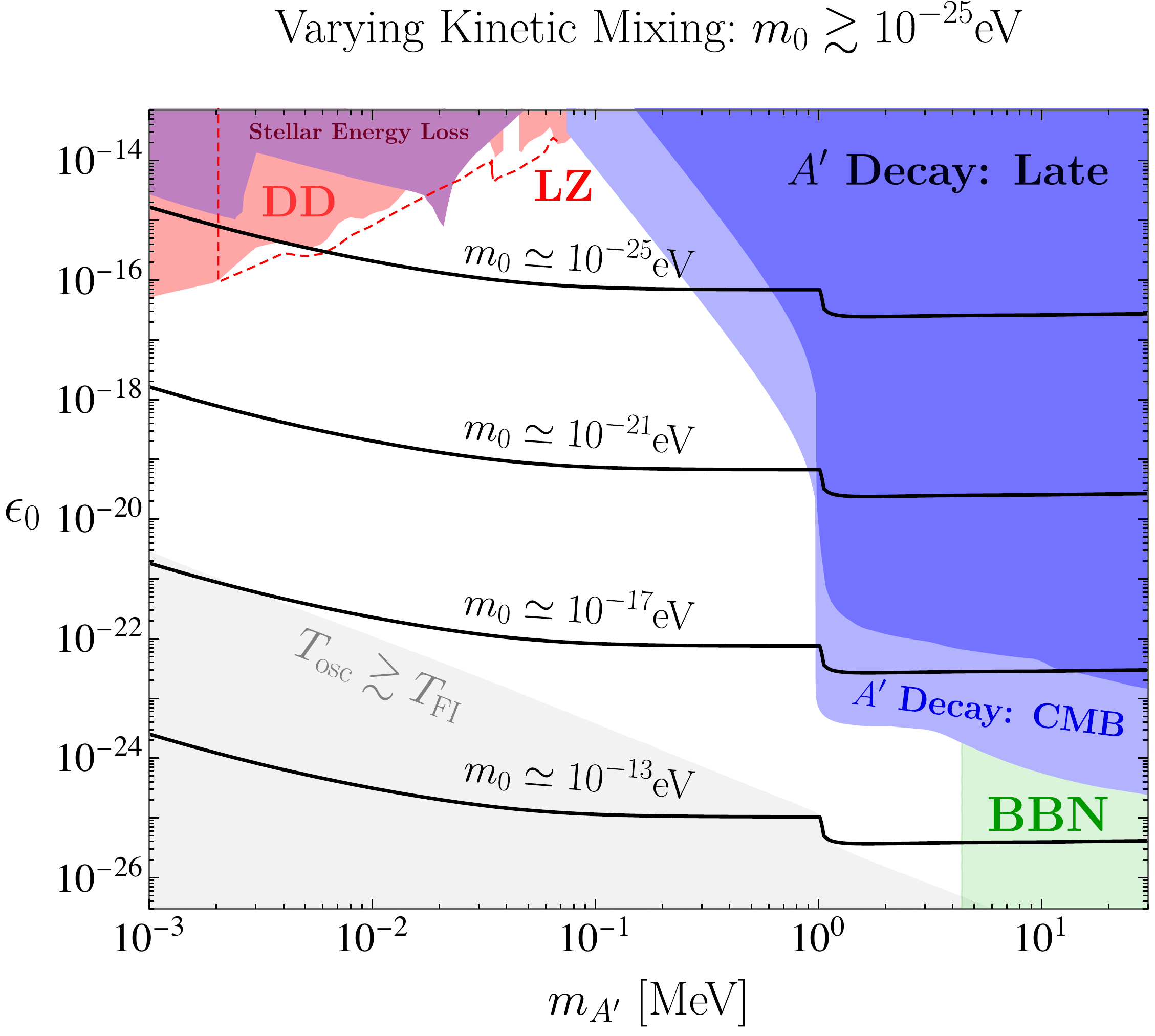}
\caption{The parameter space of the dark photon dark matter frozen-in through the time-varying mixing.  $\epsilon_0$ is today's local kinetic mixing near the earth. The direct detection~(DD) constraints are shown as the red-shaded region. The dashed red line denotes the projection of LZ. The stellar energy loss constraints are shown as the purple-shaded region. The constraints from the dark photon decay during late time, CMB and BBN are shown as the dark blue, light blue, and green regions, respectively. {\bf Left}: $m_0 \ll 10^{-25}\eV$. We choose $m_0 \simeq 10^{-30}, 10^{-28}, 10^{-26}\,\eV$ as the benchmark values for the freeze-in lines of the dark photon dark matter with varying mixing. They are shown as the black lines. 
{\bf Right}: $m_0 \gtrsim 10^{-25}\eV$. We choose $m_0 \simeq 10^{-25}, 10^{-21}, 10^{-17}, 10^{-13}\,\eV$ as the benchmark values for the dark photon freeze-in. The light gray region denotes the parameter space where $T_\osc \gtrsim T_\FI$. In this region, our calculation is not applicable.  
}
\label{fig:DP_bound_vary}
\end{figure}
Here we discuss the phenomenology of the dark photon dark matter based on nonzero kinetic mixing. 
Prior to the discussion, it is important to note that in our model the kinetic mixing varies during the universe's evolution. Namely, when $T\gtrsim T_\osc$, $\phi$ is at rest with the nonzero field displacement. Afterward, $\phi$ begins the damped oscillation. Hence, the experimental detectability depends on the universe's epoch. 
From Sec.~\ref{sec:cos_his}, we write today's local kinetic mixing $\epsilon_0$ as
\bea
\label{eq:eps0_today+local}
\epsilon_0  \sim \epsilon_{\text{FI}} \left(\frac{T_0}{T_\osc}\right)^{3/2}  \times \left\{ 
\begin{aligned}
& 1 ~ & (m_0 \ll 10^{-25}\eV)\\
& \mathcal{E} & (m_0 \gtrsim 10^{-25}\eV)
\end{aligned},
\right.
\eea
where today's universe temperature is $T_0 \sim 10^{-3}\eV$, the enhancement factor from the structure formation is $\mathcal{E} \sim 600$, and the kinetic mixing for the dark photon dark matter to freeze-in is $\epsilon_\FI \sim 10^{-12}$. Based on \Eq{eq:eps0_today+local}, we divide the discussion into two parts: the $m_0 \ll 10^{-25}\eV$ case and the $m_0 \gtrsim 10^{-25}\eV$ case.

The dark photon parameter space of the $m_0 \ll 10^{-25}\eV$ case is shown in the left panel of \Fig{fig:DP_bound_vary}. One can see that the most relevant constraints come from the dark matter direct detection~(red), the stellar energy loss~(purple), the dark photon decay at CMB~(light blue), and the dark photon decay at the late time~(blue). In the plot, the yellow line is the dark photon dark matter's freeze-in line with the constant kinetic mixing, which is already covered by the current constraints. In contrast, the varying kinetic mixing model opens the parameter space and provides the benchmark values determined by $m_0$. Here, we choose $m_0 \simeq 10^{-30}, 10^{-28}, 10^{-26}\,\eV$ to plot the freeze-in lines on the $m_{A'}-\epsilon_0$ plane. Because $T_\osc \sim \left( m_0 \mpl \right)^{1/2}$, larger $m_0$ makes $\phi$ oscillate earlier, therefore $\epsilon_0$ is smaller. 

In the mass range $m_{A'} \lesssim 0.1 \,\mev$, the most relevant constraints are the dark matter direct detection and the stellar energy loss, as shown in the left panel of \Fig{fig:DP_bound_vary}. For direct detection, the most strict constraint within $\kev$ to $\mev$ mass range comes from  XENONnT~\cite{XENONCollaboration:2022kmb}.\, LUX-ZEPLIN~(LZ), represented by the dashed red line in the plot, can test smaller kinetic mixing by a half order of magnitude~\cite{LZ:2021xov}. We can also expect that future direct detection experiments, such as DarkSide-20k~\cite{DarkSide-20k:2017zyg} and DARWIN~\cite{DARWIN:2016hyl}, with larger detectors and lower backgrounds, can have better detection capabilities. In this mass range, our model is also constrained by the stellar energy loss via $\gamma \rightarrow A'$~\cite{Redondo:2008aa, Pospelov:2008jk, Redondo:2008ec, Redondo:2013lna, An:2013yfc, An:2014twa, Hardy:2016kme}. The direct detection of the non-relativistic dark photons produced by the sun~(solar basin) can also impose comparable constraints~\cite{Lasenby:2020goo}. 


In the mass range $m_{A'} \gtrsim 0.1 \,\mev$, our model is constrained by the dark photon dark matter decay. When $m_{A'} < 2 m_e$, because the two-photon channel is forbidden by the Landau-Yang theorem~\cite{Landau:1948kw, Yang:1950rg}, the dark photon decays through $A' \rightarrow 3 \gamma$ induced by the electron loop. When $m_{A'} \gtrsim 2 m_e$, the dominant channel is $A' \rightarrow e^- e^+$. These two channels are constrained by the CMB and the late-time photon background, which give comparable constraints in the constant kinetic mixing scenario. From \cite{Redondo:2008ec, An:2014twa, Essig:2013goa, Slatyer:2016qyl, Wadekar:2021qae}, we know that $\Gamma_{A' \rightarrow 3 \gamma} \lesssim 10^{-9} H_0$ and $\Gamma_{A' \rightarrow e^- e^+} \lesssim 10^{-7} H_0$ for $m_{A'} \sim \mev$. However, the constraints from the CMB and the late-time photon background are different for the varying kinetic mixing model, because these two physical processes happen in different stages of the universe: Galactic photons are emitted in today's universe, while the CMB epoch~(recombination) is much earlier. From the discussion of $\phi$'s evolution in Sec.~\ref{sec:cos_his}, we know that the kinetic mixing in the CMB epoch is
\bea
\label{eq:eps_CMB_m0_leq10em25}
\epsilon_\CMB \sim \epsilon_0 \times  \left[\frac{\min(T_\CMB, T_\osc)}{T_0} \right]^{3/2}, 
\eea
where $T_\osc \sim (m_0 \mpl)^{1/2}$. Based on \Eq{eq:eps_CMB_m0_leq10em25}, we recast the CMB constraint to the $m_{A'}-\epsilon_0$ diagram. When $m_0 \lesssim 10^{-28}\,\eV$, $T_\osc \lesssim T_\CMB$, so $\epsilon_\CMB \simeq \epsilon_\FI$. For this reason, the dark photon mass region $m_{A'} \gtrsim 0.1\,\mev$ is excluded by CMB, which explains CMB bound's cutting off at $m_{A'} \simeq 0.1\, \mev$ in the left panel of \Fig{fig:DP_bound_vary}. When $m_0 \gtrsim 10^{-28}\,\eV$, $T_\osc \gtrsim T_\CMB$, so the kinetic mixing at $T \sim T_\CMB$ is $\epsilon_\CMB \sim \epsilon_0 \,(T_\CMB/T_0)^{3/2}$, which explains why the CMB constraint is stronger than the late-time photon constraints if $m_{A'} \gtrsim 0.1 \mev$. 

Now we discuss the $m_0 \gtrsim 10^{-25}\eV$ case shown in the right panel of \Fig{fig:DP_bound_vary}. The most relevant constraints come from the dark photon dark matter decay during CMB~(light blue) and BBN~(green). To recast the constraints in the early universe to the $m_{A'}-\epsilon_0$ plane, we use the formula\footnote{From the discussion in Subsec.~\ref{subsec:eta>=1}, we know that thermal effect may change $T_\osc$ for the type-B model when the scalar is heavier than $10^{-16}/\log^2 \eta \, \eV$. Even though the type-B model does not cause qualitative differences, to simplify the discussion, we only discuss the type-A model where $T_\osc \sim (m_0 \mpl)^{1/2}$ as an example.}
\bea
\label{eq:eps_CMB_m0_geq10em25}
\epsilon_{\CMB} \sim \epsilon_0 \times   \left[\frac{T_{\CMB}}{T_0} \right]^{3/2} \frac{1}{\mathcal{E}}, \quad \quad \epsilon_{\BBN} \sim \epsilon_0 \times   \left[\frac{\min(T_{\BBN}, T_\osc)}{T_0} \right]^{3/2} \frac{1}{\mathcal{E}}.
\eea
%
During the BBN, $\epsilon_\BBN$ ranges from $ 10^{-12}$ to $10^{-14}$, which leads to the light elements' disintegration caused by $A' \rightarrow e^- e^+$. Based on this, the BBN constraint on the dark photon freeze-in is imposed~\cite{Forestell:2018txr, Fong:2022cmq}. However, if $m_{A'} \lesssim 5 \,\mev$, the dark photon decay cannot change the light element abundance, because the injected energy is smaller than the deuterium binding energy, which is the smallest among all the relevant light elements~(except ${}^{7}\text{Be}$ whose abundance does not affect the main BBN observables). In the right panel of \Fig{fig:DP_bound_vary}, there is a gray region in the lower left corner, the parameter space where the calculations of the freeze-in lines break because the dark photon freeze-in happens after $\phi$ starts oscillation. At the end of this paragraph, we point out one interesting character: Through the varying kinetic mixing, the dark photon heavier than $1\,\mev$ can be frozen-in and free from $A' \rightarrow e^- e^+$ constraints, because the kinetic mixing portal is closed right after the dark photon production. 

In the end, we want to discuss the warm dark matter bound. Because the dark photon is produced through $\gamma \rightarrow A'$ with the initial momentum $p_{A'} \sim T$, it washes out the dark matter substructure in the late universe~\cite{Irsic:2017ixq, Dvorkin:2020xga, DEramo:2020gpr, Zelko:2022tgf}. To avoid this, we need $m_{A'}\gtrsim \text{few} \times 10\,\kev$. Detailed analysis based on the dark matter phase space distribution is left for future work.


\begin{figure}[t]
\centering
\includegraphics[width=0.499\columnwidth]{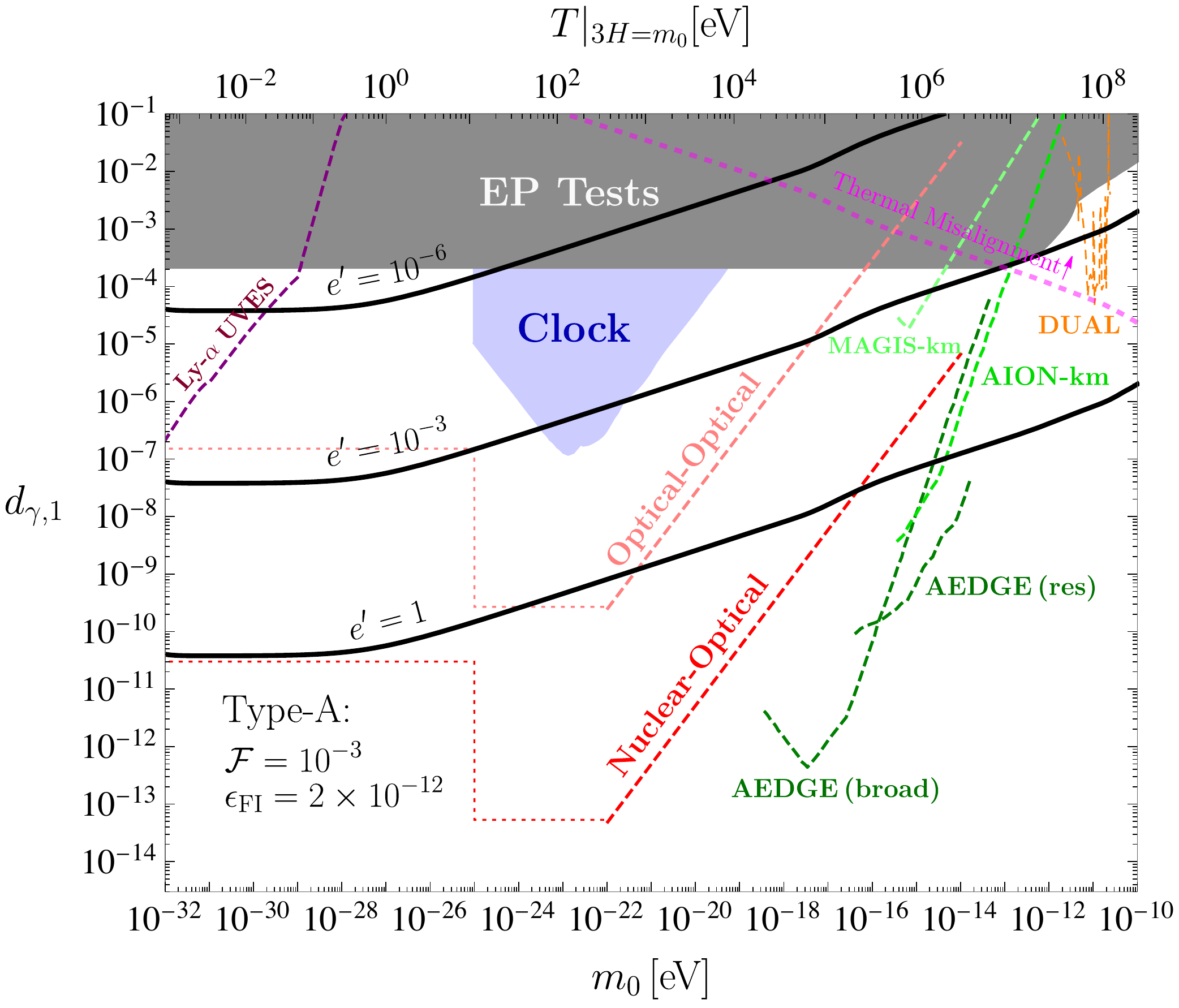}
\hfill
\includegraphics[width=0.492\columnwidth]{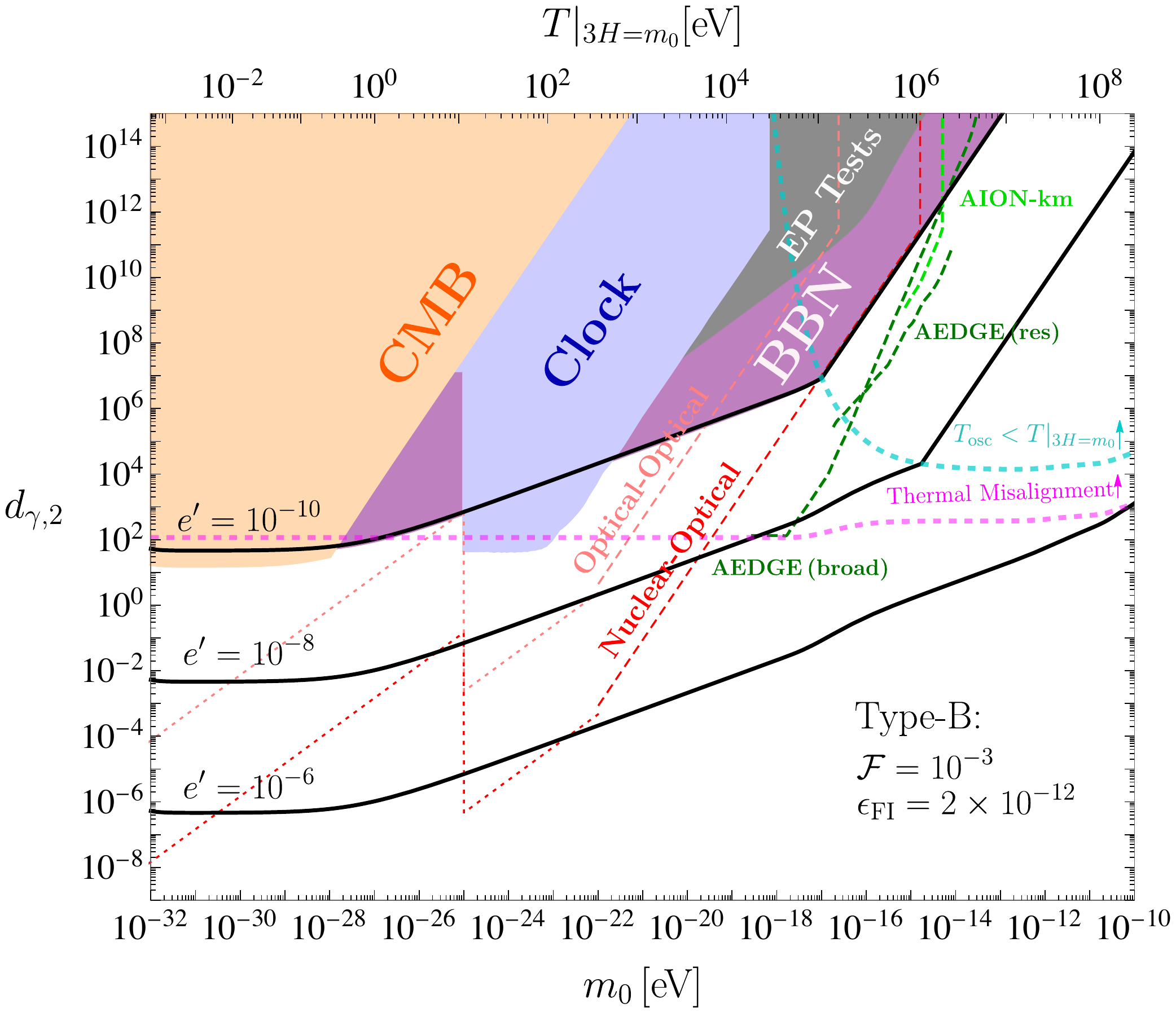}
\caption{The parameter space of the scalar-photon couplings with $\mathcal{F} \simeq 10^{-3}$. The black lines are the experimental targets determined by the dark photon dark matter freeze-in ~($\Omega_{A'} h^2 \simeq 0.12$) with different $e'$\,s.
{\bf Left}: The type-A model~($y'=0$). The thick dashed magenta line is the $\eta \simeq 1$ contour, upon which the scalar's early displacement is set to be $\pi f/2$ by the thermal misalignment. The gray region denotes the constraints from the EP tests~\cite{Smith:1999cr, Schlamminger:2007ht,Berge:2017ovy}. The blue region represents constraints from the clock comparisons~\cite{Arvanitaki:2014faa, VanTilburg:2015oza, Hees:2016gop, Kalaydzhyan:2017jtv, Hees:2018fpg, collaboration2021frequency}. Here, most of the model's parameter space~($e' \lesssim 1$) is inside the projections of the proposed experiments, such as Lyman-$\alpha$ UVES~\cite{Hamaide:2022rwi}, the clock comparison~(optical-optical, optical-nuclear)~\cite{Arvanitaki:2014faa}, the cold-atom interferometer~(AEDGE, AION-km, MAGIS-km)~\cite{Arvanitaki:2016fyj, AEDGE:2019nxb, Badurina:2019hst, MAGIS-100:2021etm}, and the resonant-mass oscillator~(DUAL)~\cite{Arvanitaki:2015iga}. {\bf Right}: The type-B model~($y' = -y$) beginning with the wrong vacuum~($\pi f/2 \lesssim \phi_\rh \lesssim 3 \pi f/2$). In the region beyond the magenta $\eta \simeq 1$ contour, the scalar acquires $\pi f$ via the thermal misalignment. The $T_\osc  \simeq T|_{3H=m_0}$ contour is the thick dashed cyan line, and it envelopes the region where the thermal effect postpones the oscillation. This is the region where $\phi$ does the trapped misalignment. Given the scalar abundance, since the trapped misalignment needs a smaller oscillation amplitude, as discussed in \Eq{eq:typeB_phiosc_rescale}, the power law of the black lines is changed within the upper right corner enveloped by the thick dashed cyan line. In the plot, the constraints from the cosmology~(CMB, BBN)~\cite{Stadnik:2015kia,Hart:2019dxi, Sibiryakov:2020eir, Bouley:2022eer} are comparable with the constraints from clocks and EP tests. }
\label{fig:scalar_plt}
\end{figure}

\subsubsection{Detection of the Ultralight Scalar}

In Sec.~\ref{sec:UV_model}, we reveal that scalar-photon interaction originates from UV physics. Given this foundation, our model can be tested through the observations of the ultralight scalar $\phi$. \Eq{eq:Min_eps_linear} and \Eq{d1_d2} indicate that $d_{\gamma,1}$ and $d_{\gamma,2}$, the dimensionless scalar-photon couplings, are determined by $e'$ and  $\epsilon_\FI$. For the freeze-in model of the dark photon dark matter, there is $\epsilon_\FI \sim 10^{-12}$. This sets the experimental targets of the ultralight scalar experiments. To be more quantitative, we have
\be
\label{eq:d1_d2_FI}
d_{\gamma,i} \sim \left(\frac{\mpl \,\epsilon_\FI}{e' \,\abs{\phi}_\osc }\right)^i \propto \left(\frac{m_0 \,\epsilon_\FI}{e' \, \mathcal{F}^{1/2} \, T_\osc^{3/2} }\right)^i  \,\,\,\,\,\,\,\,\,\,\,\,\,\, (i=1,2),
\ee
where ``$i=1$'' and ``$i=2$'' denote the type-A and type-B models, respectively. From \Eq{eq:d1_d2_FI}, we know that with $\epsilon_\FI$ determined, the smaller $e'$ is, the larger $d_{\gamma, i}$ is, so our model with small $e'$ is more detectable. The reason is that from the UV model in Sec.~\ref{sec:UV_model}, to maintain a determined kinetic mixing,  $y$ and $y'$ should be smaller when $e'$ is larger. Therefore, $d_{\gamma,i}$ gets larger correspondingly. Since the phenomenologies for the linear and quadratic  scalar-photon couplings are quite different, we discuss them separately. 
\\

\paragraph{\bf Type-A Model}

Let us first discuss the type-A model shown in the left panel of \Fig{fig:scalar_plt}. To illustrate how the experimental sensitivities vary with $e'$, we plot the $d_{\gamma,1}$ lines in black choosing $e' = 1, 10^{-3}, 10^{-6}$ as the benchmark values. From \Eq{eq:T_phi_eta<<1} we know that $T_\osc \propto m_0^{2/3}$ when $m_0 \lesssim 10^{-28}\eV$, and $T_\osc \propto m_0^{1/2}$ when $m_0 \gtrsim 10^{-28}\eV$. Given this, we have the analytical form of the black line, which is
\bea
\label{d1_line_TypeA}
d_{\gamma,1} \propto \frac{1}{e' \mathcal{F}^{1/2} } \times 
\left\{
\begin{aligned}
m_0^{1/4} & \,\,\,\quad  (m_0 \gtrsim 10^{-28} \eV)\\
\text{const} & \,\,\,\quad  (10^{-33} \eV \lesssim m_0 \lesssim 10^{-28}\eV)
\end{aligned} 
\right. .
\eea
In \Fig{fig:scalar_plt}, the thick dashed magenta line is the contour of $\eta \simeq 1$. In the upper right corner beyond the dashed magenta line, $\eta \gtrsim \mathcal{O}(1)$, the thermal effect makes $\phi$ converge to $\pi f/2$ in the early time according to Sec.~\ref{subsec:eta>=1}. Because $f\sim \abs{\phi}_\osc$, combining \Eq{eta_from_de} and \Eq{eq:phi_osc_eta<<1} we know that the thick dashed magenta line obeys $d_{\gamma,1} \propto m_0^{-1/4}$. In the region below this thick dashed magenta line, $\eta \ll 1$, so $\phi$ does the standard misalignment. 

One of the strongest constraints for the type-A model comes from the equivalence principle experiments testing the acceleration difference of two objects made by different materials attracted by the same heavy object. In the left panel of \Fig{fig:scalar_plt}, such a constraint is shown as the shaded gray region. Until now, the most stringent constraint is imposed by MICROSCOPE~\cite{Berge:2017ovy} and  E\"{o}t-Wash~\cite{Smith:1999cr, Schlamminger:2007ht}, giving $d_{\gamma,1} \lesssim 10^{-4}$. The clock comparisons of Dy/Dy~\cite{VanTilburg:2015oza}, Rb/Cs~\cite{Hees:2016gop}, and $\text{Al}^+$/$\text{Hg}^+$~\cite{collaboration2021frequency} give stringent constraints based on testing the time-varying $\alphaem$. These constraints are shown as the shaded blue region. 

There are several experiments proposed to go beyond the current constraints. For future clock comparison experiments~\cite{Arvanitaki:2014faa}, the projection of the improved optical-optical clock comparison is shown as the pink line, and the projection of the optical-nuclear clock comparison is shown as the red line. For these projections, the dashed parts denote the projection of the $\alphaem$ oscillation testing, and the dotted parts denote the projection of the $\alphaem$ drift testing. According to \cite{Arvanitaki:2014faa}, the projection has the optimistic assumption that the measurement takes place when the scalar is swiping through the zero such that $\dot{\alpha}_\EM$ is independent of $m_0$. Following this, we extrapolate the projections of the optical-optical and optical-nuclear experiments to $10^{-33}\eV$, considering the homogeneity of the ultralight scalar when the scalar's de Broglie wavelength is much larger than the size of the Milky Way halo. The cold-atom interferometer experiments such as AEDGE~\cite{AEDGE:2019nxb}, AION~\cite{Badurina:2019hst}, and MAGIS~\cite{MAGIS-100:2021etm} have strong detection capability in the mass range $10^{-19}\eV \lesssim m_0 \lesssim 10^{-12}\eV$. In the plot, the projections of AEDGE~(broadband, resonant mode), AION-km, and MAGIS-km are shown in the dashed dark green, dashed green, and dashed light green lines, respectively. The region of the thermal misalignment located in the upper right corner of the left panel of \Fig{fig:scalar_plt} can be tested by the proposed resonant-mass detectors, such as DUAL shown in the dashed orange line~\cite{Arvanitaki:2015iga}. The CP-even scalar in the mass range $10^{-33}\eV \lesssim m_0 \lesssim 10^{-28}\eV$ can be tested via the  Lyman-$\alpha$ UVES observation~\cite{Hamaide:2022rwi} shown in the dashed purple line. 

For the EP tests, because they test the Yukawa interaction mediated by $\phi$, the corresponding constraints are independent of the scalar fraction $\mathcal{F}$. For the $\alphaem$-variation tests, because $\Delta \alphaem \propto d_{\gamma,1} \abs{\phi}$, the $d_{\gamma,1}$ sensitivities are all scaled by $\mathcal{F}^{-1/2}$. We know from \Eq{d1_line_TypeA} that, for the non-EP experiments, the relative position between the experimental targets~(black lines) and the constraints/projections remains the same as $\mathcal{F}$ varies. From \Fig{fig:scalar_plt}, we find that most of the targets are within the detection capabilities of the proposed experiments because $e' \lesssim \mathcal{O}(1)$.
\\

\paragraph{\bf Type-B Model}

Now we discuss the type-B model shown in the right panel of \Fig{fig:scalar_plt}. Here, we discuss the case where $\phi$ has the initial wrong vacuum~($\pi f/2 \lesssim \phi_\rh \lesssim 3 \pi f/2$) as the example. In the plot, the thick dashed magenta line is the contour obeying $\eta \simeq 1$. In the region upon~(below) this line, the ultralight scalar does the thermal~(standard) misalignment. From \Eq{eta_from_de} we have $\eta \sim (\alphaem d_{\gamma,2} )^{1/2}$. Thus, the thick dashed magenta line obeys $d_{\gamma, 2} \sim 10^2$. As discussed in Subsec.~\ref{subsec:eta>=1}, the thermal effect makes $\phi$ converge to $\pi f$ in the early universe in the parameter space upon the thick dashed magenta line. The thick dashed cyan line denotes the contour satisfying $T_\osc  \simeq T|_{3H=m_0}$. In the region enveloped by the thick dashed cyan line, $T_\osc < T|_{3H=m_0}$, meaning that the thermal effect postpones $\phi$'s oscillation. This represents the phase of trapped misalignment. 

In the plot, the thick dashed cyan line is  horizontal when $m_0 \gtrsim 10^{-16}\eV$. This is because $m_T \simeq m_0$ happens when $e^\pm$ are relativistic, which leads to $m_T \sim \eta H$. If $\eta \gg 1$, when $H \sim m_0$, there is $m_T \gg m_0$, which prevents $\phi$ from rolling to the bare minimum. Therefore, $\phi$'s oscillation is postponed by the thermal effect. We also find that the thick dashed cyan line becomes vertical when $m_0$ becomes smaller because $m_T \simeq m_0$ happens when $e^{\pm}$ are non-relativistic. Along the vertical part of the cyan line, $T|_{3H=m_0} \sim m_e$, we have $m_0 \sim m^2_e/\mpl$. 

The experimental benchmarks are shown as the black lines in $m_0-d_{\gamma,2}$ plane. In the region of the standard misalignment~($\eta \ll 1$), the black lines obey
\bea
\label{d1_line_TypeB}
d_{\gamma,2} \propto \frac{1}{e'^2 \mathcal{F} } \times 
\left\{
\begin{aligned}
m_0^{1/2} & \,\,\,\quad  ( 10^{-28} \eV \lesssim m_0 \lesssim m_*)\\
\text{const} & \,\,\,\quad  (10^{-33} \eV \lesssim m_0 \lesssim 10^{-28}\eV)
\end{aligned}
\right. ,
\eea
where $m_*$ is the black and the thick dashed cyan lines' cross point coordinate. To understand $d_{\gamma,2}$ in \Eq{d1_line_TypeB}, we refer to \Eq{eq:T_phi_eta<<1} which shows that $T_\osc \propto m_0^{2/3}$ when $m_0 \lesssim 10^{-28}\eV$, and $T_\osc \propto m_0^{1/2}$ when $m_0 \gtrsim 10^{-28}\eV$. After plugging $T_\osc$ into \Eq{eq:d1_d2_FI}, we have \Eq{d1_line_TypeB}. 

Now let us discuss the behaviors of the black lines in the scalar mass range $m_0\gtrsim m_*$. If the black line intersects with the thick dashed cyan line on its vertical side, i.e., $m_* \lesssim 10^{-16}\eV$, we have $T_\osc \sim m_e$. If the black line crosses with the thick dashed cyan line on its horizontal side, i.e., $m_* \gtrsim 10^{-16}\eV$, we have $T_\osc \sim T|_{3H=m_0}/\eta^{1/2} \propto m_0^{1/2}/d_{\gamma,2}^{1/4}$. Substituting $T_\osc$ into \Eq{eq:d1_d2_FI}, we have
\bea
\text{$d_{\gamma,2} \propto \frac{1}{e'^2 \mathcal{F}} \times m_0^2$ \quad if $m_* \lesssim 10^{-16}\eV$,\,  \quad \quad $d_{\gamma,2} \propto \frac{1}{e'^8 \mathcal{F}^4} \times m_0^2$ \quad if $m_* \gtrsim 10^{-16}\eV$}\,\,\,\quad  ( m_0 \gtrsim m_*).
\eea
This explains the black lines' tilting up in the region enclosed by the thick dashed cyan line. Such enhancement of the signal comes from the postponement of the oscillation, the typical feature of the trapped misalignment~\cite{Nakagawa:2020zjr, DiLuzio:2021gos}.

Similar to the type-A model, the type-B model can be tested through terrestrial experiments. In the plot, the current constraints based on clock comparison~\cite{VanTilburg:2015oza, Hees:2016gop, Hees:2018fpg, collaboration2021frequency, Banerjee:2022sqg} are shown as the shaded blue region, and the constraints from the equivalence principle tests~\cite{Berge:2017ovy, Hees:2018fpg, Banerjee:2022sqg} are shown as the shaded gray region. The projections of the proposed optical-optical and optical-nuclear clock comparisons are shown in orange and red, respectively~\cite{Arvanitaki:2014faa, Banerjee:2022sqg}. Here, the dashed and dotted parts of the projections denote the detection capabilities of the $\alphaem$ oscillation and the $\alphaem$ drift, respectively. Our model can also be tested by the cold-atom interferometers. The projection of AEDGE~\cite{AEDGE:2019nxb} and AION-km~\cite{Badurina:2019hst} are shown in the dashed dark green and dashed green lines, respectively. One may find that the constraints and projections from the ground-based and low-altitude experiments get weakened or cut off in the strong coupling region. This is caused by the scalar's matter effect sourced by the earth, one of the characteristics of the quadratic scalar-SM coupling. Following ~\cite{Hees:2018fpg, Banerjee:2022sqg}, we have
\bea
\label{eq:alphaem_typeB_screen}
d_{\gamma,2,\crit} = \frac{\mpl^2 \Rearth}{3 \Mearth \Qearth} \sim  \text{few}\times 10^{11},
\eea
where $d_{\gamma,2,\crit}$ is the scalar's critical value for the matter effect to appear, $\Rearth \sim 6 \times 10^3 \text{km}$ is the earth's radius, $\Mearth \sim 6 \times 10^{24} \text{kg}$ is the earth's mass, and $\Qearth \sim 10^{-3}$ is the earth's dilaton charge. When $d_{\gamma,2} \gtrsim d_{\gamma,2,\crit}$, the scalar's Compton wavelength is smaller than the earth's radius. In this situation, the scalar easily overcomes the spatial gradient and is pulled toward the origin, so its near-ground and underground oscillation amplitudes are highly suppressed. Even so, if the experiments are carried out in space with altitudes comparable with the earth's radius~(for example, AEDGE), the screen effect sourced by the earth can be largely alleviated. 

Unlike the type-A model, the type-B model has strong constraints from the early universe processes, such as CMB and BBN, which are comparable with the terrestrial constraints. The reason for this character is that when tracing back to the early universe with the $\phi$'s abundance determined, $\Delta \alphaem$ increases much more for the type-B model than the type-A model~ \cite{Stadnik:2015kia, Sibiryakov:2020eir, Bouley:2022eer}. Considering the thermal effect in Subsec.~\ref{subsec:eta>=1}, we impose the constraints on the type-B scalar's parameter space using the current cosmological constraints on $\Delta \alphaem/\alphaem$, which are~\cite{Stadnik:2015kia,Hart:2019dxi,Sibiryakov:2020eir, Bouley:2022eer}\footnote{The results in \cite{Sibiryakov:2020eir, Bouley:2022eer} cannot be recast to our model directly because these works constrain the coupling $\phi^2 F^2$ which can be classified into the case where $\phi$ starts in the range $- \pi f/2\lesssim \phi_\rh \lesssim \pi f/2$. We are discussing the case of the initial wrong vacuum where $\pi f/2\lesssim \phi_\rh \lesssim 3 \pi f/2$. To impose the BBN constraint for our model, we reproduce the CMB and BBN constraints in \cite{Stadnik:2015kia} by using the former CMB bound $\Delta \alphaem/\alphaem \lesssim 10^{-2}$~ \cite{Stadnik:2015kia} and by only including the bare potential effect during BBN. Then we impose the CMB and BBN constraints on the right panel of \Fig{fig:scalar_plt} using \Eq{eq:CMB_BBN_delta_alpha} with the consideration of the scalar's thermal effect as discussed in Subsec.~\ref{subsec:eta>=1}.}
%
\bea
\label{eq:CMB_BBN_delta_alpha}
(\Delta \alphaem/\alphaem)_\CMB \lesssim 2\times 10^{-3}, \quad \quad (\Delta \alphaem/\alphaem)_\BBN \lesssim 6\times 10^{-3} \,.
\eea
In the right panel of \Fig{fig:scalar_plt}, the CMB constraint is shown as the shaded orange region, and the BBN constraint is shown as the shaded purple region. 

 Based on the left panel of \Fig{fig:scalar_plt}, let us discuss the tests of the dark photon dark matter freeze-in given different $\mathcal{F}$'s. Given the scalar fraction $\mathcal{F} \simeq 10^{-3}$, for the model to be tested by the proposed experiments but not excluded by the current constraints, the dark gauge coupling needs to be within $ 10^{-10} \lesssim e' \lesssim 10^{-6}$. When $\mathcal{F}$ varies, the relative position between the black lines and the non-EP experiments does not change qualitatively because all these constraints come from $\Delta \alphaem \propto d_{\gamma,2} {\abs{\phi}^2}$. In this case, their positions are rescaled by $\mathcal{F}^{-1}$. Differently, what the equivalence principle experiments are testing is the acceleration difference of the objects A and B, which obeys $\abs{a_A-a_B}/\abs{a_A+a_B} \propto d_{\gamma,2}^2 \, \phi \abs{\nabla{\phi}}$ according to \cite{Hees:2018fpg, Banerjee:2022sqg}. From this, we know that the constraints on $d_{\gamma,2}$ from the equivalence principle test are rescaled by $\mathcal{F}^{-1/2}$. When $\mathcal{F} \lesssim 10^{-8}$, the equivalence principle constraints are stronger than the current CMB, BBN, and clock constraints. 

In the end, we comment on the constraints from the black hole superradiance. For the $\mathcal{F} \simeq 1$ case, the current constraints from the supermassive black holes exclude the region $10^{-21}\eV \lesssim m_0 \lesssim 10^{-17}\eV$~\cite{Arvanitaki:2014wva, Stott:2018opm, Davoudiasl:2019nlo, Unal:2020jiy, Du:2022trq}, and the constraints from the solar mass black holes exclude the region $10^{-13}\eV \lesssim m_0 \lesssim 10^{-11}\eV$~\cite{Baryakhtar:2020gao}. However, since $\phi$'s self-interaction is $\lambda_\phi \sim m_0^2/f^2$, smaller $\mathcal{F}$ leads to smaller $f$, which increases $\lambda_\phi$. From \cite{Baryakhtar:2020gao, Unal:2020jiy} we know that, for the scalar as the subfraction of the dark matter, the superradiance constraints are alleviated by the scalar's large attractive self-interaction. 

\section{$\mathbb{Z}_N$-Protected Scalar Naturalness} \label{sec:zn_varykm}

Because the Yukawa interaction in \Eq{eq:simpuv} breaks $\Phi$'s global $U(1)$ symmetry, the scalar $\phi$ has quantum correction. Taking the type-B model in Sec.~\ref{sec:UV_model} as an example, we have the mass correction $\Delta m_\phi\sim yM$, and its benchmark value
\bea
\Delta m_{\phi}
& \sim 10^{-15} \text{eV}\left(\frac{\epsilon_\FI }{10^{-12}}\right)\left(\frac{M}{100\,\gev}\right)^2\left(\frac{10^{17}\gev}{ \abs{\phi}_\osc }\right)\left(\frac{1}{e'} \right),
\label{Min_Cancel_phimass}
\eea
which could be larger than the $\phi$'s bare mass in part of the parameter space. For the type-A model, the situation is similar. Even so, by imposing an extra $\mathbb{Z}_N$ symmetry~\cite{Hook:2018jle, Frieman:1995pm}, the global $U(1)$ symmetry can be approximately restored, therefore $\phi$'s quantum correction is exponentially suppressed. To realize this, we introduce $N$ copies of the worlds containing the standard model sector ($\text{SM}_k$), the dark sector ($\text{DS}_k$), and the portal~(${\mathcal{O}_P}_k$) where $k=0,1,\dots,N-1$. Because of the $\zn$ symmetry, the system is invariant under the transformation
\bea
\Phi\rightarrow\Phi\exp\left(i\frac{2\pi}{N}\right)\;,\;(\text{SM}+\mathcal{O}_P+ \text{DS})_k\rightarrow (\text{SM}+ \mathcal{O}_P +\text{DS})_{k+1}\;,
\eea
so the lowest order effective operator is 
\begin{equation}\label{eq:u1break}
	\mathcal{L}_{\cancel{U(1)}}=\const \times \frac{\Phi^N}{\Lambda^{N-4}_1}+\hc, 
\end{equation}
which is invariant under the $\zn$ transformation but not invariant under the $U(1)$ transformation. Such a dimension-N operator shows that even though the global $U(1)$ symmetry is broken, providing that the symmetry under the discrete subgroup $\mathbb{Z}_N$ still exists, the quantum correction is suppressed with the form $\Delta m_\phi^2 \propto f^{N-2}/\Lambda^{N-4}_1$.\footnote{If $N$ is an odd number and the exact $\mathbb{Z}_2$ symmetry under $\CD$~($\Phi \rightarrow - \Phi^\dagger$) is imposed, the lowest order effective operator is $\Phi^{2N}/\Lambda^{2N-4}_2$, so the mass quantum correction is $f^{2N-2}/\Lambda^{2N-4}_2$. Here, we use ``$l$'' for the effective scale in $\mathcal{L} \supset \text{const} \times \Phi^{lN}/\Lambda_l^{lN-4}+\hc$. From \Eq{eq:cwphi}, we have  $\Lambda_l \sim M/y^{lN/(lN-4)}$.} As long as $\Delta m_\phi^2 \lesssim m_0^2$, the scalar's bare mass $m_0$ protected by the $\zn$ symmetry can be naturally small.

\begin{figure}[t]
\centering
\includegraphics[width=0.75\columnwidth]{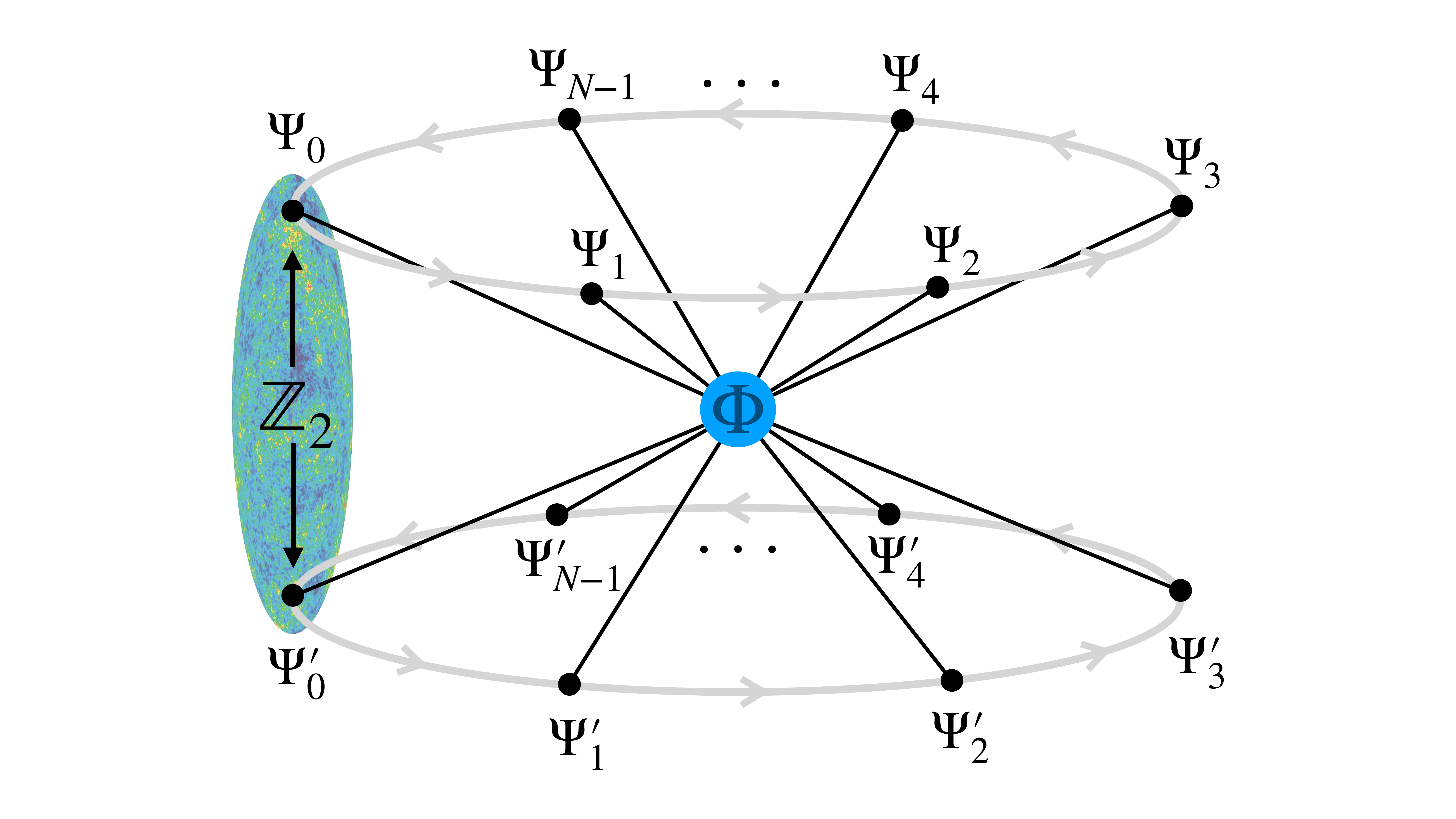}
\caption{The schematic diagram of the $\zn$-invariant UV model. The messenger particles $\F^{(\prime)}_k \, (k=0,1,\cdots,N-1)$ are all coupled with the complex scalar $\Phi$, where ``$k$'' labels the $k$th world. Here, $k=0$ is our world which experienced the reheating. Because of $y$ and $y'$'s suppression, $k$th and $j$th universes~($k \neq j$) do not talk with each other. Here, the $\zn$ rotation $\Phi \rightarrow \Phi\exp(i ~2 \pi/N), \F^{(\prime)}_k \rightarrow \F^{(\prime)}_{k+1}$ is labeled by the gray arrowed circles. The $\mathbb{Z}_2$ symmetry protecting the mass degeneracy between $\F_k$ and $\F_k^{\prime}$ eliminates the time-independent part of the kinetic mixing.}
\label{fig:ZN_Z2}
\end{figure}

\subsection{$\zn$-Protected Model}

To build a concrete varying kinetic mixing model within the $\mathbb{Z}_N$ framework, we embed the minimal model described by \Eq{eq:simpuv} into the  Lagrangian~(See \Fig{fig:ZN_Z2})
\bea
\label{eq:znuv}
\mathcal{L}_\UV \supset \sum_{k=0}^{N-1}\left( y  e^{i \frac{2\pi k}{N}} \Phi \barF_k \F_k + y' e^{i \frac{2 \pi k}{N}} \Phi \barF'_k \F'_k + \hc  \right) - M \sum_{k=0}^{N-1}(\barF_k \F_k + \barF_k' \F_k') - \lambda \left( \left|\Phi\right|^2 - \frac{f^2}{2}\right)^2,
\eea
which is invariant under the $\zn$ transformation $\Phi \rightarrow \Phi \exp(i\frac{2 \pi}{N}), \, \F^{(\prime)}_k \rightarrow \F^{(\prime)}_{k+1}, \, A^{(\prime)}_k \rightarrow A^{(\prime)}_{k+1}$. Here, $\F_k$ and $\F_k'$ are doubly charged messengers in the $k$-th universe carrying the same $k$th-hypercharge but the opposite $k$th-dark charge. Being similar to Sec.~\ref{sec:UV_model}, we introduce the $\mathbb{Z}_2$ symmetry to protect the mass degeneracy between $\F_k$ and $\F'_k$, therefore the allowed bare potential is \Eq{eq:V0}. Given this, the kinetic mixing portal is closed in the late time.  
Representing the complex scalar as $\Phi=i f e^{i\phi/f}/\sqrt{2}$, we write $\F_k^{(\prime)}$'s effective masses as
\bea
\label{eq:M_k_phi}
M_k^{(\prime)}(\phi) = M\left[1+r^{(\prime)}\sin\left(\frac{\phi}{f} + \frac{2 \pi k}{N}\right)\right], \,\,\,\, \text{where $r^{(\prime)} = \frac{\sqrt{2}\, y^{(\prime)} f}{M}$.}
\eea

In the low energy limit, $\F_k$ and $\F_k^{(\prime)}$ are integrated out, so the IR Lagrangian is  
\bea\label{eq:znir}
\mathcal{L}_\IR \supset \sum_{k=0}^{N-1}\left[\frac{1}{2} \epsilon_k {F_k}_{\mu \nu} F'^{\mu \nu}_k + \frac{1}{4}  \left(\frac{\Delta \alphaem}{\alphaem}\right)_k {F_k}_{\mu \nu} F_k^{\mu \nu}\right], 
\eea
where
\bea\label{eq:znir_eps}
\epsilon_k= \frac{\sqrt{2}e e' (y-y')f }{6 \pi^2 M}\sin \left(\frac{\phi}{f} + \frac{2 \pi k}{N}\right)
\eea
and
\bea\label{eq:znir_alpha}
\left(\frac{\Delta \alphaem}{\alphaem}\right)_k = \frac{e^2}{6 \pi^2} \left[-\frac{\sqrt{2}(y+y')f}{M} \sin\left( \frac{\phi}{f} + \frac{2 \pi k}{N} \right) + \frac{(y^2+y'^2)f^2}{M^2}\sin^2\left( \frac{\phi}{f} + \frac{2 \pi k}{N} \right) \right]. 
\eea
From \Eq{eq:znir}, \Eq{eq:znir_eps} and \Eq{eq:znir_alpha}, we know that the IR Lagrangian is invariant under the $\zn$ transformation $\phi\rightarrow \phi + \frac{2 \pi}{N}, \, A^{(\prime)}_{k} \rightarrow A^{(\prime)}_{k+1}$. We find that \Eq{eq:znir_eps} and \Eq{eq:znir_alpha} contain \Eq{eq:Min_eps} and \Eq{alpha_change} when $k=0$, meaning that the minimal model discussed in \Sec{UV_model} is the $k=0$ branch of the $\mathbb{Z}_N$ model. Here, $(\text{SM} + \mathcal{O}_P + \text{DS})_0$ is our universe which experiences the reheating, while the other universes are not reheated. When $m_0 \lesssim H$, $\phi$ does the damped oscillation, so the kinetic mixing between $\text{SM}_0$ and $\text{DS}_0$ gradually decreases, as discussed in Sec.~\ref{sec:UV_model} and Sec.~\ref{sec:cos_his}. 

\subsection{Quantum Correction of $\phi$}

For the ultralight scalar $\phi$ in \Eq{eq:znuv}, the leading order quantum correction is described by the one-loop Coleman-Weinberg potential
\bea
\label{eq:cwphifield}
	V_{\text{cw}}(\phi)=-\frac{1}{16\pi^2 }\sum_{k=0}^{N-1}M_{k}^4(\phi)\left[\log\left(\frac{M_{k}^2(\phi)}{\mu^2}\right)-\frac{3}{2}\right]+ \left(M_{k}\rightarrow M_{k}'\right),
\eea
which has the contributions from $N$ universes with destructive interference. 
According to the calculations in \App{appx:zn_vcw}, we 
 express the Coleman-Weinberg potential as
\bea\label{eq:cwphi}
	V_{\text{cw}}(\phi)
& \simeq\frac{M^4 N}{8\pi^2}\Bigg\{\left(r^N+r'^N + \cdots \right)G(N)\cos\left[N\left(\frac{\phi}{f} + \frac{\pi}{2} \right)\right]\\
& \quad \quad \quad \quad +  \left(r^{2N}+r'^{2N} + \cdots \right)G(2N)\cos\left[2N\left(\frac{\phi}{f} + \frac{\pi}{2} \right)\right]+\cdots\Bigg\}.
	\eea
In \Eq{eq:cwphi}, ``$\cdots$''s in the brackets are $r$ and $r^{\prime}$'s higher order terms, and $G(n)$ is defined as
\bea
\label{eq:Gn}
G(n) \coloneqq  \frac{1}{2^{n-1}} \sum_{j=0}^{4} \binom{4}{j} \frac{(-1)^j }{n-j}  =\frac{3 \, (n-5)!}{2^{n-4} \, n \,(n-1)!}.
\eea

To derive \Eq{eq:cwphi} from \Eq{eq:cwphifield}, we can apply the cosine function sum rules
\begin{equation}\label{cos_sum_rule_brief}
	\sum_{k=1}^N\cos^m\left(\theta+\frac{2\pi k}{N}\right)=\sum_{l=0}^{[m/N]} \mathcal{C}_{lmN} \cos\left(lN\theta\right)+\mathcal{D}_m, \,\,\, \text{where $\mathcal{C}_{l m N}\big|_{m=lN} = \frac{1}{2\,^{lN-1}}$.}
\end{equation}

From \Eq{eq:M_k_phi} we know that $M_k$ contains $\cos(\phi/f+\pi/2+2\pi k/N)$. Therefore, we can expand \Eq{eq:cwphifield} as a polynomial function of the cosine function. From \Eq{cos_sum_rule_brief}, we know that only when the cosine function's power in the effective potential is greater than $N$, the non-constant terms emerge. Since the cosine function appearing in the potential is always accompanied by $r$, based on the cosine sum rules, we find that the lowest order $\mathbb{Z}_N$ potential is proportional to $r^N$ if there is no further cancellation. For the exact calculation of \Eq{eq:cwphifield} to all orders, readers can refer to \App{appx:zn_vcw} containing two different but equivalent derivations, including the cosine sum rules discussed in this section and the Fourier transformation. The lowest order calculation can be found in \cite{Hook:2018jle, Brzeminski:2020uhm}, but the exact calculations listed in \App{appx:zn_vcw} are obtained for the first time as far as we have known. 

Because $V_{\text{cw}}$ receives contributions from both $r$ and $r'$, there is a possible extra cancellation. For even $N$, from \Eq{eq:cwphi} we find that the leading order correction starts from $r^N$. For odd $N$, if $r$ and $r'$  have opposite signs, the quantum correction from $r^N + r'^N$ is reduced. When $r=-r'$, $(r^N+r'^N) \cos[ N (\phi/f+\pi/2)]$ has the exact cancellation, thus the leading order correction starts from $(r^{2N}+r'^{2N})\cos[2N(\phi/f+\pi/2)]$. Such cancellation happens in all orders because the coefficients of $\cos[N(\phi/f+\pi/2)]$ are the series of $r^{(\prime) N+2j}$ according to \Eq{eq:cwphifield_allorders}.

Given that $r \ll 1$, the factor $r^N G(N)$ in  \Eq{eq:cwphi} indicates that the quantum correction of $m_\phi^2$ is suppressed by the $(r/2)^{N-5/2}$ factor in the type-A model, and $(r/2)^{N-2}$ factor in the type-B model. For the freeze-in of dark photon dark matter, $r$'s benchmark value is
\bea
r \sim 10^{-10} \left(\frac{\epsilon_\FI}{10^{-12}}\right) \left( \frac{1}{e'} \right) \left( \frac{f}{\abs{\phi}_\osc} \right).
\eea
In such a case, for $e' \sim \mathcal{O}(1)$, as long as $N\gtrsim 7$, the mass quantum correction is negligible in the whole mass range of $\phi$. For smaller $e'$, as long as $e' \gg 10^{-10}$ which is in the permitted region of the current constraints, we have $r \ll 1$. Therefore, the $\mathbb{Z}_N$ scenario suppressing $\phi$'s quantum correction always works.


\section{Varying Kinetic Mixing From Dirac Gaugino}\label{sec:dirac_gaugino}
Motivated by stabilizing the hierarchy between the light scalars and the heavy fermions, we discuss one of the possible supersymmetric extensions of the varying kinetic mixing in the Dirac gaugino model~\cite{Fox:2002bu, Alves:2015kia} with the superpotential
\begin{equation}\label{eq:dgugino}
\mathcal{W}=\frac{\sqrt{2} \, W_\alpha' W_j^\alpha A_j}{\Lambda_{\DG, j}}.\;
\end{equation} In \Eq{eq:dgugino}, $W_j$ is the gauge field strength of the SM gauge group $G_{\text{SM}, j}$ where the label $j=1,2,3$ denotes the SM gauge groups $U(1)_Y$, $SU(2)_L$ and $SU(3)_c$, respectively, $W'$ is the gauge field strength of $U(1)_d$, and $A_j$ is the chiral multiplet. The operator in \Eq{eq:dgugino} in hidden sector models was firstly introduced by~\cite{Polchinski:1982an} and further understood by~\cite{Fox:2002bu} afterward as the \textit{supersoft} operator such that it provides the Dirac gaugino masses and does not give logarithmic divergent radiative contributions to other soft parameters. Writing $A_j$, $W_j$ and $W'$ in terms of the Taylor expansion of the Grassmann variable $\theta$, we have $A_j\supset \left(S_j+iP_j\right)/\sqrt{2}+\sqrt{2}\theta\tilde{a}_j$, $W_j^\alpha\supset \lambda_j^\alpha+F_j^{\mu\nu}(\sigma_{\mu\nu}\theta)^\alpha + D \,\theta^\alpha$, and $W'_\alpha \supset F'_{\mu\nu}(\sigma^{\mu\nu}\theta)_\alpha+\langle D'\rangle \theta_\alpha$. Plugging these expanded fields into \Eq{eq:dgugino}, and doing the integration over $\theta^2$ firstly and then the auxiliary fields, we have the effective Lagrangian
\bea\label{eq:superW}
	\mathcal{L}
	\supset -\frac{1}{\Lambda_{\DG, j}} \tr\left(S_j F_j^{\mu\nu}\right) F_{\mu\nu}'-\frac{1}{\Lambda_{\DG,j}} \tr\left(P_j F_j^{\mu\nu}\right) \widetilde{F}_{\mu\nu}'-m_{D,j}\tilde{a}_j\lambda_j-2m_{D,j}^2S_j^2+\cdots\;,
\eea
where $S_j$($P_j$) is the CP-even~(CP-odd) scalar, $\lambda_j$ is the gaugino, and $m_{D,j}=\langle D'\rangle/\Lambda_{\DG, j}$ is the gaugino's Dirac mass given by the scale of SUSY breaking. One should note that in \Eq{eq:superW}, $S_j$'s mass term $\mathcal{L} \supset -2 m_{D,j}^2 S_j^2$ comes from integrating out the D-term~(On the contrary, $P_j$ has no extra mass contribution). It is because the supersymmetry is protected that the mass of $S_j$ is correlated with the mass of the gaugino $\lambda_j$. Consequently, $S_j$ is pushed to the heavier mass range given that the gaugino mass is highly constrained: According to \cite{DELPHI:2003uqw,ATLAS:2021yqv,CMS:2020bfa,ATLAS:2020syg}, the LHC has already excluded the electroweakinos and the gluinos masses below $\mathcal{O}(100\gev)$ and $\mathcal{O}(\tev)$ respectively. Unlike the previous discussion, in the Dirac gaugino model, $S_j$ cannot be the ultralight scalar where the misalignment mechanism provides a natural way to open the portal in the early time but gradually close it in the late time. Even so, it is still possible to realize the temporary period of $\langle S_j \rangle\neq 0$ in the early universe through the two-step phase transition, also referred to as the VEV Flip-Flop in some specific dark matter models~\cite{Baker:2016xzo,Baker:2018vos,Baker:2017zwx}. The concrete model building and the phenomenology are beyond the scope of this paper. 

In the $j=1$ case, the first term in \Eq{eq:superW} containing the CP-even scalar $S_1$ corresponds to the kinetic mixing portal between $U(1)_Y$ and $U(1)_d$ which is determined by $S_1$'s VEV. The second term in \Eq{eq:superW} containing $P_1$~(axion) leads to the dark axion portal which is investigated in~\cite{Kaneta:2016wvf, Kaneta:2017wfh, Choi:2018mvk, Choi:2019jwx, Kalashev:2018bra, Hook:2019hdk, Hook:2021ous, Arias:2020tzl, Deniverville:2020rbv, Ge:2021cjz, Domcke:2021yuz, Gutierrez:2021gol, Carenza:2023qxh}. There are several major differences between the kinetic mixing portal $\mathcal{L} \supset -S_1 F^{\mu \nu} F'_{\mu \nu}/\Lambda_{\DG, 1}$ and the dark axion portal $\mathcal{L} \supset -P_1 F^{\mu \nu} \widetilde{F}'_{\mu \nu}/\Lambda_{\DG,1}$: 1.~In the Dirac gaugino model, $S_1$'s mass is correlated with the $\lambda_1^\alpha$ mass which is pushed to $\mathcal{O}(100\gev)$ scale by LHC, while $P_1$'s shift symmetry protests its arbitrarily small bare mass. 2.~The VEV of $P_1$ does not play a direct physical role because it only contributes to the total derivative term in the Lagrangian~($P_1$'s time or spatial derivative still has nontrivial physical effects nonetheless). 

In the $j=2,3$ cases, if $\langle S_j^a \rangle \neq 0$, the first term in \Eq{eq:superW} would mix the non-Abelian gauge field with the dark photon such that the non-Abelian gauge symmetry is broken. Being referred to as the non-Abelian kinetic mixing~\cite{Arkani-Hamed:2008hhe, Arkani-Hamed:2008kxc, Chen:2009ab, Barello:2015bhq, Arguelles:2016ney, Fuyuto:2019vfe, Barello:2015bhq, Gherghetta:2019coi}, the constant mixing models are highly constrained by the collider experiments. However, in the high-temperature environment of the early Universe, the large non-Abelian kinetic mixing can possibly be realized for the non-Abelian vector dark matter production and other intriguing phenomena. We leave the detailed discussion in future work.

\section{Other Cosmologically Varying Portals}\label{sec:other_portal}

Let us begin with the general form of the cosmologically varying portals through which the dark and the visible sectors are connected. To be more generic, we write them as
\begin{equation}\label{eq:genportal}
\mathcal{L} \supset \frac{\phi}{\Lambda^{d-4}} \,\mathcal{O}_\DS  \mathcal{O}_\SM , \,\,\text{where $d = d_\SM +  d_\DS + 1$}.
\end{equation}
In \Eq{eq:genportal}, $\mathcal{O}_\SM$ and $\mathcal{O}_\DS$ are the operators of the visible sector and the dark sector, respectively, $d_\SM$ and $d_\DS$ denote the dimensions of these two operators, and $d$ is the dimension of the time-varying portal. To simplify the notation of \Eq{eq:genportal}, we drop the (spacetime, spin, flavor, \ldots)~indices of $\mathcal{O}_\SM$ and $\mathcal{O}_\DS$ whose contraction makes the varying portal to be a singlet. For simplicity, we only keep the linear form of the CP-even scalar $\phi$, even though in the UV theory, the non-linearity may appear, as we have seen in \Eq{eq:Min_eps}. Based on the EFT, we know that when the effective operator \Eq{eq:genportal} is introduced, the operators merely containing $\phi$ and $\mathcal{O}_\SM$ also appear because the symmetry does not forbid them. The co-appearance of the effective operator shown in \Eq{eq:genportal} and the scalar-SM coupling provides an excellent chance to test these kinds of models from the experiments detecting the portal itself and the ones measuring the scalar-SM coupling. In the rest of this section, we will give some specific examples of the varying portals and show how these minimal extensions illuminate the dark matter model building. 

Let us briefly review the varying kinetic mixing portal in the EFT language. After choosing
\bea
\text{$\mathcal{O}_{\SM,\mu \nu} = F_{\mu \nu}$ \, and \, $\mathcal{O}_{\DS, \mu \nu} = F'_{\mu \nu}$,}
\eea
\Eq{eq:genportal} goes back to the operator $\mathcal{L} \supset \phi \, F_{\mu \nu} F'^{\mu \nu}/\Lambda$ discussed before. Through this operator, the dark photon dark matter can be produced without violating the stringent constraints as shown in Sec.~\ref{sec:dpdm_fi}. Since the spacetime indices of $F_{\mu \nu}$ need to be contracted, the lowest order operator of the scalar-SM coupling is $\phi \,F_{\mu \nu} F^{\mu \nu}$. If there is an exact $\mathbb{Z}_2$ symmetry invariant under the dark charge conjugation $\phi \rightarrow -\phi$, $F'_{\mu \nu} \rightarrow - F'_{\mu \nu}$, $\phi F_{\mu \nu} F^{\mu \nu}$ is forbidden, so the lowest order operator of the scalar-SM coupling becomes $\phi^2 \,F_{\mu \nu} F^{\mu \nu}$. The experiments testing the $\alphaem$-variation and the equivalence principle violation can be used to test $\phi F_{\mu \nu} F^{\mu \nu}$ or $\phi^2 F_{\mu \nu} F^{\mu \nu}$, as discussed in Subsec.~\ref{subsec:DPFI_signal}.

In other situations where $\mathcal{O}_\SM$ is invariant under arbitrary global and gauge transformations, there are no more indices to contract, so the lowest order operator of the scalar-SM coupling is $\phi \,\mathcal{O}_\SM$ or $\phi^2 \,\mathcal{O}_\SM$ depending on whether the exact $\mathbb{Z}_2$ symmetry, i.e., the invariance under $\phi \rightarrow -\phi$, $\mathcal{O}_\DS \rightarrow -\mathcal{O}_\DS$ transformation, exists or not. One typical example is that
\bea
\text{$\mathcal{O}_\SM = \abs{H}^2$ \,\, and \,\, $\mathcal{O}_\DS = s^2$},
\eea
where $s$ is a scalar singlet in the dark sector. Here, $\mathcal{L} \supset \lambda_s \mathcal{O}_\DS \mathcal{O}_\SM = \lambda_s s^2 \abs{H}^2$ is well-known as the singlet-scalar Higgs portal~(SHP) through which the dark matter $s$ reaches today's relic abundance~(The dominant channels are $s s \rightarrow f^- f^+, W^- W^+, ZZ, h h, \cdots$. $f$ refers to the SM fermions.)~\cite{Silveira:1985rk, McDonald:1993ex,Burgess:2000yq}. Besides, because this Lagrangian is invariant under the $\mathbb{Z}_2$ transformation $s \rightarrow -s$, $s$ is stable. Although SHP provides a simple way to realize $\Omega_s h^2 \simeq 0.12$, in the mass range $m_s \lesssim 1\tev$, most of its parameter space except the narrow window of the resonance~($m_s \simeq m_h/2$) is excluded by the Higgs invisible decay $h \rightarrow s s$~\cite{ATLAS:2015gvj, CMS:2016dhk}, the dark matter direct detection, and the indirect detection~(AMS, Fermi)~\cite{Escudero:2016gzx, Casas:2017jjg, Hardy:2018bph, Curtin:2021alk}. By introducing the time-varying SHP
\begin{equation}
	\mathcal{L} \supset \frac{\phi}{\Lambda}s^2|H|^2,
\end{equation}
the parameter space is widely extended. Here, $\phi$'s misalignment supports $s$'s freezeout in the early universe and then starts the damped oscillation such that $\langle \sigma v\rangle_{ss} \propto (T/T_\osc)^3$ when $T \lesssim T_\osc$. For this model, there are two types of experiments: 1.~The future direct detection relying on today's SHP. 2.~The experiments testing $\phi \abs{H}^2$ or $\phi^2 \abs{H}^2$~\cite{Piazza:2010ye, Graham:2015ifn, Arvanitaki:2016fyj, Batell:2022qvr}. 

Another example of the model with singlet $\mathcal{O}_\SM$ is 
\bea
\text{$\mathcal{O}_\SM = \bar{Q}H q_R \,\,\,\text{or}\,\,\, \bar{L}H e_R$ \,\, and \,\, $\mathcal{O}_\DS  = \tilde{s}$}.
\eea
$\tilde{s}$ is the scalar mediator interacting with the dark matter  $\chi$ via the CP-odd coupling $\mathcal{L} \supset i y_{\chi} \tilde{s} \, \chibar \gamma^5 \chi $. Given the constant portal $\mathcal{L} \supset \mathcal{O}_\DS \mathcal{O}_\SM/\Lambda  = y_{f} \tilde{s} \bar{f}f$ where $y_{f} = v_h/\sqrt{2} \Lambda$, $\chi$ reaches today's relic abundance through the freezeout channel $\chi^-\chi^+ \rightarrow f^- f^+$ with $\langle \sigma v \rangle_{\tilde{s}\tilde{s}} \sim y_{f}^2 y_{\chi}^2 m_\chi^2/m^4_{\tilde{s}}$. Here, $v_h$ is the Higgs VEV, and $\Lambda$ is the effective scale of the constant portal. Since $\chi^-\chi^+ \rightarrow f^- f^+$ is s-wave, the region $m_\chi \lesssim 10\,\gev$ is excluded by CMB~\cite{Planck:2018vyg}. To produce lighter but CMB-friendly dark matter, we introduce the operator
\bea
\mathcal{L} \supset \frac{\phi}{\Lambda^2} \tilde{s} \bar{L} H e_R \,\,\, \text{or} \,\,\,\, \frac{\phi}{\Lambda^2} \tilde{s} \bar{Q} H q_R.
\eea
Through the $\phi$-dependent Yukawa coupling $\mathcal{L} \supset $ $y_{f}(\phi) \tilde{s} \bar{f} f$ where $y_{f}(\phi) = \phi v /\sqrt{2} \Lambda^2$ and the aforementioned CP-odd Yukawa coupling $\mathcal{L} \supset i y_{\chi} \tilde{s}\, \chibar \gamma^5 \chi$, the dark matter lighter than $10\,\gev$ can reach today's relic abundance without violating CMB annihilation bound as long as $\phi$'s starts damped oscillation earlier than $T_\text{CMB} \sim \eV$. For this model, there are two kinds of experiments: 1.Direct and indirect detections, such as the next-generation CMB observations~\cite{Madhavacheril:2013cna, Wu:2014hta, Green:2018pmd, Abazajian:2019eic, Dvorkin:2022bsc}. 2.~The tests of the SM fermion mass variations with $\phi \bar{f} f$ or $\phi^2 \bar{f} f$~\cite{Arvanitaki:2014faa, Arvanitaki:2016fyj, Arvanitaki:2015iga, Hees:2016gop, Kalaydzhyan:2017jtv, Hees:2018fpg, Banerjee:2022sqg, Kaplan:2022lmz}.

\section{Conclusion}
\label{sec:conclusion}

In this work, we study the time-dependent kinetic mixing controlled by the ultralight CP-even scalar's cosmological evolution for three reasons: First, to provide a new UV realization of the kinetic mixing. Second, to open the parameter space of the dark photon dark matter freeze-in with $\epsilon_\FI \sim 10^{-12}$, which is experimentally excluded in the time-independent kinetic mixing scenario, as shown in \Fig{fig:DP_bound}. Third, to provide the experimental benchmarks for the ultralight scalar experiments. To realize the model, we introduce the heavy doubly charged messengers coupled with the scalar. To eliminate the time-independent part of the kinetic mixing, we impose the $\mathbb{Z}_2$ symmetry, which is invariant under the dark charge conjugation. Importantly, the scalar-photon coupling also emerges from the UV theory. To categorize the models, we designate the theory with the approximate $\mathbb{Z}_2$ as the type-A model, and the theory with the exact $\mathbb{Z}_2$ as the type-B model. Consequently, the type-A model has the linear scalar-photon coupling, whereas the type-B model has the quadratic scalar-photon coupling.

Through the varying kinetic mixing, the dark photon dark matter ranging from $\kev$ to $\mev$ is frozen-in, free from the late-universe constraints, with the kinetic mixing determined by the scalar mass. Therefore, the target values of the kinetic mixing for $\kev-\mev$ dark photon dark matter experiments are set, as shown in \Fig{fig:DP_bound_vary}. In the meantime, the existence of the nonrelativistic scalar relic with the scalar-photon coupling affects the universe's thermal history, leads to the scalar's thermal misalignment, varies the fine-structure constant, and violates the equivalence principle. These phenomena provide excellent targets to test our model via the ultralight scalar experiments in the mass range $10^{-33}\eV \lesssim m_0 \ll \eV$, as shown in \Fig{fig:scalar_plt}.

We also study the $\mathbb{Z}_N$-protection of the scalar naturalness in the varying kinetic mixing model. We embed the  minimal model into the $\mathbb{Z}_N$ model so that the $U(1)$ shift symmetry is discretely restored. Given that $N \sim 10$, the scalar mass quantum correction can be much lighter than $10^{-33}\eV$. 
Moreover, we provide the analytical methods to expand the $\mathbb{Z}_N$ Coleman-Weinberg potential to all orders. Finally, we briefly discuss the Dirac gaugino realization of the varying mixing and the dark matter models via other varying portals. More generally, the portal controlled by the ultralight scalar can offer a minimal solution. This solves the tension between the portal dark matter's early-time production and late-time constraints.

\section*{Acknowledgments}
 We want to thank Cédric Delaunay, Joshua T. Ruderman, Hyungjin Kim, Raffaele Tito D'Agnolo, Pablo Quílez, Peizhi Du, Huangyu Xiao, Erwin H. Tanin, Xuheng Luo for their helpful discussions and comments on the draft. We also want to thank Neal Weiner, John March-Russell, Ken Van Tilburg, Hongwan Liu, Asher Berlin, Isabel Garcia Garcia, Junwu Huang,  Gustavo Marques Tavares, Andrea Mitridate for useful discussions. DL acknowledges funding from the French Programme d’investissements d’avenir through the Enigmass Labex. XG is supported by James Arthur Graduate Associate~(JAGA) Fellowship.

\appendix
\section{Scalar's Analytical EOM Solutions: High-$T$}\label{appx:analyt_sol}

In this section, we solve $\phi$'s movement analytically in the high-temperature universe where the bare potential's effect is inferior. Taking the joint effects of the thermal mass and the universe's expansion into consideration, we write the equation of motion as
\be
\delta \ddot{\phi} + 3H \delta \dot{\phi} \simeq - m_T^2 \delta \phi, \quad \,\,\text{where $\delta \phi = \phi - \phi_{\min}$}. 
\label{EOM_eta<1_1}
\ee
In \Eq{EOM_eta<1_1}, $\phi_{\min}$ is the minimum of the thermal potential, i.e., $V_T$ in \Eq{TypeAB_VT}, and $\delta \phi$ is the field displacement from such a thermal minimum. As we know in \Eq{eq:TypeAB_VT_min}, within the $2\pi f$ periodicity, $\phi_{\min}=\pi f/2$ for the type-A model, and $\phi_{\min}=0, \pm \pi f$ for the type-B model. In \Eq{EOM_eta<1_1}, we use the linear approximation, because as long as $\phi$ is initially away from the hilltop of the thermal potential, the effect of the nonlinearity is negligible. Applying the equation $\frac{d}{dt} \simeq HT \frac{d}{dT}$, \Eq{EOM_eta<1_1} can be rewritten as 
\be
\label{EOM_eta<1_2}
T^2 \frac{d^2 \delta \phi}{d T^2} \simeq - \frac{\eta^2}{4} \delta \phi,
\ee
which is well-known as the homogeneous linear equation. Utilizing the power-law ansatz, we obtain two independent solutions of \Eq{EOM_eta<1_2}, which are
\bea
\delta \phi \propto T^{\frac{1 \pm \sqrt{1-\eta^2}}{2}} \,\,\,\,\,  (\eta < 1), \quad 
\delta \phi \propto T^{\frac{1}{2}}, T^{\frac{1}{2}} \log\left(T/T_\rh\right) \,\,\,\,\,   (\eta = 1), \quad
\delta \phi \propto T^{\frac{1 \pm i \sqrt{\eta^2-1}}{2}} \,\,\,\,\, (\eta > 1)
\eea
for each case separately. Given the initial condition $\dot{\phi}_\rh = 0$, one can determine the unknown coefficients and write the solution as
\be
\delta \phi \propto T^{\frac{1-\sqrt{1-\eta^2}}{2}} \times \left[ 1- \left(\frac{\sqrt{1-\eta^2}-1}{\eta}\right)^2 \left(\frac{T}{T_\rh}\right)^{\sqrt{1-\eta^2}} \right] \quad \,\,\,\,\,  (\eta < 1),
\label{eq:analyt_delphi_eta<1}
\ee
%
\bea
\delta \phi \propto T^{\frac{1}{2}} \times \left[ 1  - \frac{1}{2}\log\left(\frac{T}{T_\rh}\right)\right] \quad \,\,\,\,\,  (\eta = 1),
\label{eq:analyt_delphi_eta=1}
\eea
and
\be
\delta \phi \propto T^{\frac{1}{2}} \cos\left(\frac{\sqrt{\eta^2-1} \log \frac{T}{T_\rh} }{2}\right) \times \left[ 1 - \frac{1 }{\sqrt{\eta^2 - 1 } } \tan \left(\frac{\sqrt{\eta^2-1} \log \frac{T}{T_\rh}}{2} \right) \right] \,\,\,\,\, \quad \,\,\,\,\,  (\eta > 1).
\label{eq:analyt_delphi_eta>1}
\ee
\begin{figure}[t]
\centering
\includegraphics[width=0.6\columnwidth]{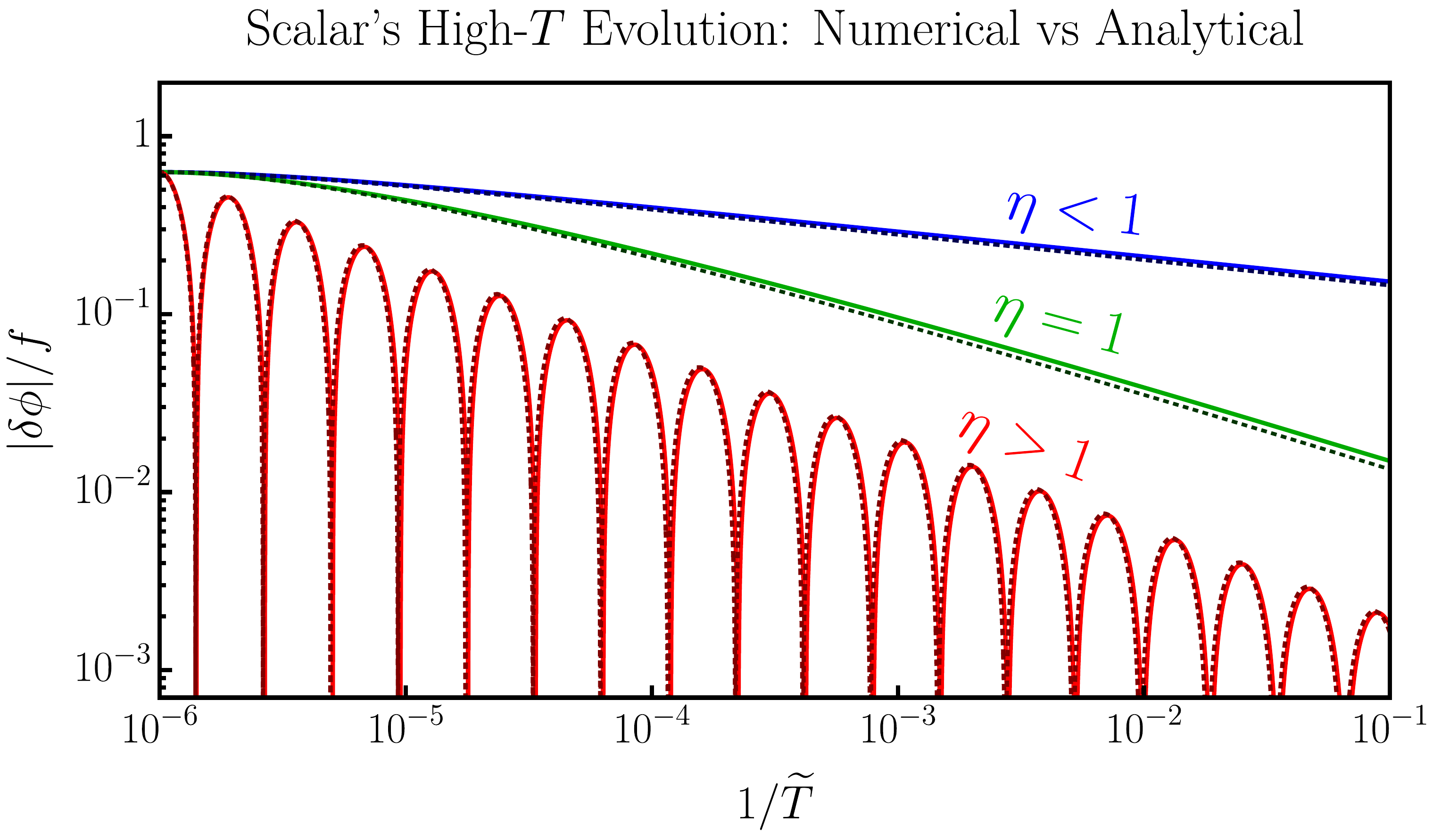}
\caption{ This plot describes the ultralight scalar's evolution in the high-temperature environment and compares the numerical~(solid lines) and analytical~(dotted dark lines) results listed in \Eq{eq:analyt_delphi_eta<1}, \Eq{eq:analyt_delphi_eta=1}, and \Eq{eq:analyt_delphi_eta>1}. In the plot, similar to \Fig{fig:phi_evo}, $\widetilde{T}$ is defined as $\widetilde{T} \coloneqq T/T|_{3H=m_0}$. $\abs{\delta \phi}$ represents the absolute value of the deviation from the thermal potential's minimum. Without loss of generality, we choose the type-A model as an example in this plot, and the discussion of the type-B model is quite similar. Here, the red, green, and blue colors represent the $\eta>1$, $\eta=1$, and $\eta < 1$ cases, respectively, and one can find that the numerical and analytical results for each case match magnificently well. We find that the red and green lines have approximately the same power law, i.e.,  $\abs{\delta \phi} \propto \widetilde{T}^{1/2}$, whose slight deviation can be explained by the $\log$ term in \Eq{eq:analyt_delphi_eta=1}. We can also find that the red line's oscillation period appears the same in the plot with the $\log$-scaled $1/\widetilde{T}$ axis. This is because the only temperature-dependence in \Eq{eq:analyt_delphi_eta>1} comes from $\log\widetilde{T}$, or, more intuitively, is caused by the same temperature power law of $m_T$ and $H$. }
\label{fig:TypeA_phi_highT_evo}
\end{figure}
Let us briefly explain the solution for each case individually. In the $\eta < 1$ case, whose solution is \Eq{eq:analyt_delphi_eta<1}, $\phi$ begins with the initial staticity after the reheating and then slowly slides to the thermal minimum, whose sliding velocity is suppressed by the small $\eta$. In the limit $\eta \ll 1$, one can have $\delta \phi \propto T^{\frac{\eta^2}{4}}$, meaning that $\phi$ is approximately static. The reason for $\phi$'s motionlessness is that the thermal effect is too weak to drive $\phi$ to move under the large Hubble friction. From \Eq{eq:analyt_delphi_eta<1}, we can also find the front coefficient of $T^{\frac{1-\sqrt{1-\eta^2}}{2}}$ dominates over the one of $T^{\frac{1+\sqrt{1-\eta^2}}{2}}$, because for small $\eta$, the latter solution contributes to the nonzero initial velocity, while the former one does not. The $\eta = 1$ case is the critical point in the parameter space which separates the sliding and oscillating phases. In this case, the scalar moves toward the thermal minimum obeying the relation $\delta \phi \propto T^{1/2}$ approximately. When $\eta$ goes beyond the critical value, i.e., in the case where $\eta>1$, the scalar's evolution is in the oscillating phase whose amplitude dwindles like $\abs{\delta \phi} \propto T^{1/2}$ as the universe expands. From \Eq{eq:analyt_delphi_eta>1}, we can find that the front coefficient of the cosine function dominates over the sine function as the result of the initial condition $\dot{\phi}_\rh = 0$. Another feature of the oscillating solution for the $\eta>1$ case worthwhile to be discussed is that the $T$-dependent part is the $\log$ function. Such a $\log$ term comes from the imaginary part of $T$'s power, or, in other words, is the feature of the homogeneous linear equation, which takes its form because $m_T$ and $H$ have the same temperature power law. 

Those who are curious about comparing the analytical and numerical solutions can refer to \Fig{fig:TypeA_phi_highT_evo}, where the solid lines represent the numerical solutions, and the dotted dark lines denote the analytical solutions, as shown in \Eq{eq:analyt_delphi_eta<1}, \Eq{eq:analyt_delphi_eta=1}, and \Eq{eq:analyt_delphi_eta>1}, in the linear approximation. The red, green, and blue colors are related to the $\eta >1$, $\eta = 1$, and $\eta<1$ cases, respectively. Since the type-A and the type-B models have similar behaviors when moving toward the thermal minimum in the high-temperature universe, we take the type-A model as an example to draw the plot, and the numerical-analytical comparison for the type-B model can be identically transplanted. In \Fig{fig:TypeA_phi_highT_evo}, one can easily see that the analytical solutions listed in \Eq{eq:analyt_delphi_eta<1}, \Eq{eq:analyt_delphi_eta=1}, and \Eq{eq:analyt_delphi_eta>1} are perfectly consistent with the numerical results.

\section{Dark Photon Freeze-in}\label{appx:dpfi}

As long as the kinetic mixing is nonzero and the amount of the initial dark photon is negligible, the dark photons are always produced in the late universe through the energy transfer from the visible sector, known as the freeze-in production. Following~\cite{Pospelov:2008jk, Redondo:2008ec}, we give a  pedagogical introduction in this appendix, which is divided into two parts: $m_{A'}<2m_e$ and $m_{A'} \geq 2m_e$. When $m_{A'}<2m_e$, the dark photon production is dominated by the resonant transition $\gamma \rightarrow A'$. When $m_{A'}\geq 2 m_e$, the dark photon mainly comes from the inverse decay $e^- e^+ \rightarrow A'$. Here, we focus on the transverse dark photon, because the longitudinal dark photon production is subdominant~\cite{Redondo:2013lna}.

\begin{figure}[t]
\centering
\includegraphics[width=0.6\columnwidth]{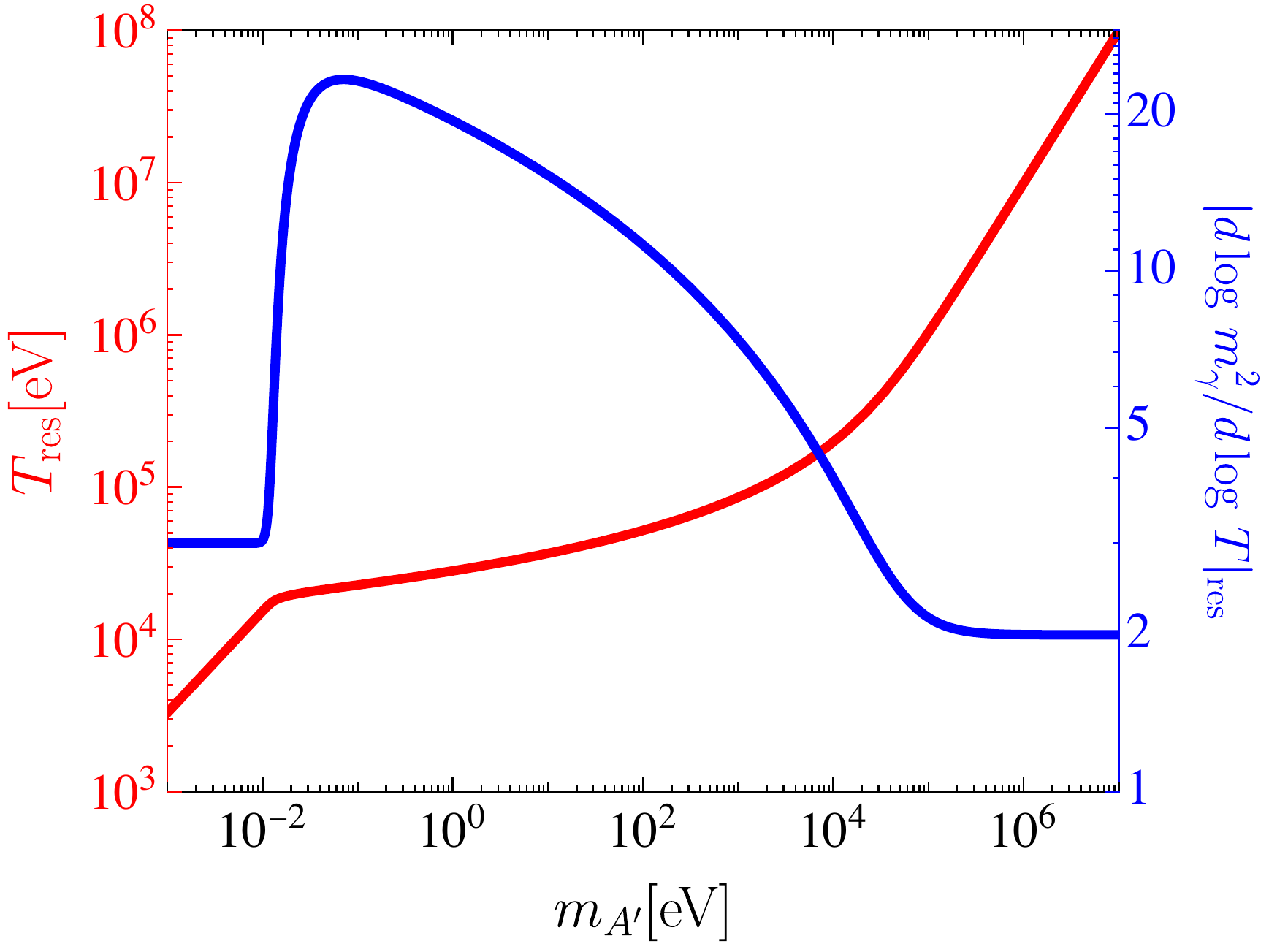}
\caption{The plot showing how $T_\res$ and $\abs{d \log m_\gamma^2/d \log T}_\res$ change in terms of $m_{A'}$. Here, $T_\res$ represented by the red line is the temperature in which the resonant transition $\gamma \rightarrow A'$ happens. Its numerical value can be read from the red vertical axis on the left-hand side of the plot. The dimensionless quantity $\abs{d\log m_\gamma^2/d \log T}_\res$ represents how fast the plasmon mass changes in terms of the temperature at which the resonant transition happens. Readers can refer to the blue axis on the right-hand side for its numerical value. In the mass range $m_{A'} \gtrsim 10^5 \eV$, $T_\res \propto m_{A'}$ and $\abs{d \log m_\gamma^2/d \log T}_\res \simeq 2$ because $m_{\gamma}^2 \propto T^2$ when electrons are relativistic. When $m_{A'} \lesssim 10^{-2}\eV$, $T_\res \propto m_{A'}^{2/3}$ and $\abs{d \log m_\gamma^2/d \log T}_\res \simeq 3$, as a result of the plasmon mass power law $m_\gamma^2 \propto n_{e^-} \propto T^3$ when the symmetric $e^- e^+$ annihilate away but the asymmetric $e^-$ remains. In the middle mass range where $ 10^{-2}\eV \lesssim  m_{A'} \lesssim 10^5 \eV$, the $\gamma \rightarrow A'$ resonance happens when the plasmon mass experiences an exponential drop as the universe temperature goes below the electron mass. In this case, $T_\res$ is in the range $10^{-2}m_e-m_e$ and insensitive to the value of $m_{A'}$. The semi-analytical estimations of $T_\res$ and $\abs{d\log m_\gamma^2/d \log T}_\res$ in such a case can be found in \Eq{Tres_jFactor_NR}.}
\label{fig:DPDM_FI}
\end{figure}

\subsection{$m_{A'} < 2 m_e$}

Knowing that the dominant contribution comes from the resonant transition of the transverse photon when $m_{A'}< 2m_e$, we write down the Boltzmann equation
\bea
\dot{n}_{A'}+3H n_{A'} \simeq n_\gamma \langle \Gamma_{\gamma \rightarrow A'}\rangle, \quad\,\, \text{where\,\, $\langle \Gamma_{\gamma \rightarrow A'}\rangle \simeq \frac{\pi^3}{12 \zeta(3)} \frac{\epsilon^2 m_{A'}^4 }{T} \delta(m_\gamma^2-m_{A'}^2)$ }
\eea
is the thermally-averaged $\gamma \rightarrow A'$ transition rate and $n_\gamma = 2 \zeta(3)T^3/\pi^2$ is the transverse photon number density. The delta function in $\langle \Gamma_{\gamma \rightarrow A'} \rangle$ reveals the feature of the resonant production: Most of the dark photons below the electron mass are produced when $m_{A'}^2 \simeq m_\gamma^2$. Doing the time integration of the Boltzmann equation, we get today's dark photon relic abundance
\bea
\label{OmegaAp_AToAp}
\Omega_{A'} \simeq \left( \frac{3^2 \cdot 5^{3/2}}{2^3 \cdot \pi^{5/2} \cdot g_{*,S} g_*^{1/2}  }  \frac{\epsilon^2 m_{A'}^3 \mpl  s_0}{T^3 \abs{d\log m_\gamma^2/d\log T} \rho_{c} } \right)_{\res},
\eea 
where \enquote{$\text{res}$} means all the temperature-dependent quantities in \Eq{OmegaAp_AToAp} are chosen to be the ones when the plasmon mass and the dark photon mass match with each other. In \Eq{OmegaAp_AToAp}, $s_0$ is today's universe entropy, $\rho_c$ is the critical density, and $\abs{d\log m_\gamma^2/d\log T}_\res$, an order $1-10$ factor, represents how fast $m_\gamma^2$ passes through the vicinity of $m_{A'}^2$. The numerical values of $T_\res$ and $\abs{d\log m_\gamma^2/d\log T}_\res$ are shown in \Fig{fig:DPDM_FI}. 

For $m_{A'} \gtrsim 10^5 \eV$, the resonant transition happens when electrons are relativistic. At this time, the plasmon mass is $m_\gamma^2 \simeq e^2 T^2/6$, based on which we have
\bea
\text{$T_\res \simeq 8 m_{A'}$ \,\,and\,\, $\abs{\frac{d \log m_\gamma^2}{d \log T}}_\res \simeq 2$}.
\label{Tres_jFactor_R}
\eea
Here $T_\res$ is $T_{\gamma \rightarrow A'}$ in \Eq{eq:T_Omega_m_Ap<2me}. We use the lower index ``res'' to emphasize the character of the resonance transition for $\gamma \rightarrow A'$. In \Eq{OmegaAp_AToAp}, $m_{A'}^3$ and $T_\res^3$ cancel with each other, so for fixed dark photon relic abundance, $\epsilon$ is the constant of $m_{A'}$. If all the dark matter is comprised of the dark photon, the needed kinetic mixing is
\bea
\label{eps_approx_1}
\epsilon_\FI \sim 10^{-12} \left( \frac{\Omega_{A'} h^2}{0.12}\right)^{1/2}.
\eea
For $10^{-2} \eV \lesssim m_{A'} \lesssim 10^{5} \eV$, the resonant transition happens when $10\kev \lesssim  T \lesssim m_e$. In this epoch, the plasmon mass is $m_\gamma^2 \simeq e^2 n_e/m_e$, where $n_e\simeq g_{e}(m_e T/2 \pi)^{3/2} e^{-m_e/T}$. By solving the resonant condition, we have
\bea
\text{$T_\res \sim m_e \left[\log\left( \frac{2^{1/2}}{\pi^{3/2}}  \frac{e^2 m_e}{m_{A'}} \right)\right]^{-1}$ \,\,and\,\, $\abs{\frac{d \log m_\gamma^2}{d \log T}}_\res \simeq \frac{3}{2}+\frac{m_e}{T_\res}$}. 
\label{Tres_jFactor_NR}
\eea
Since $T_\res$ is insensitive to $m_{A'}$'s changing, given the dark photon relic abundance, $\epsilon \propto m_{A'}^{-3/2}$. More quantitative representation for the $\epsilon-m_{A'}$ relation is given as
\bea
\label{eps_approx_2}
\epsilon_\FI \sim 10^{-11} \left(\frac{m_{A'}}{1\kev}\right)^{-3/2}\left(\frac{\Omega_{A'} h^2}{0.12} \right)^{1/2}.
\eea
For $m_{A'}\lesssim 10^{-2}\eV$, $\gamma \rightarrow A'$ takes place at the temperature $T \lesssim 10\kev$. Since the symmetric $e^- e^+$ annihilates away, the plasmon mass comes from the asymmetric $e^-$ whose number density is conserved. In this case, the plasmon mass is scaled as $m_\gamma \propto n_e \propto T^{3/2}$, and the resonant temperature is scaled as $T_\res \propto m_{A'}^{2/3}$. The concrete formulas for $T_\res$ and $\abs{d \log m_\gamma^2/d \log T}_\res$ are
\bea
\text{$T_\res \simeq 15 \kev \times \left(\frac{m_{A'}}{10^{-2}\eV}\right)^{2/3}$ \,\,and\,\, $\abs{\frac{d \log m_\gamma^2}{d \log T}}_\res \simeq 3$.}
\eea
Given the dark photon relic abundance, the kinetic mixing is
%
\bea
\label{eps_approx_3}
\epsilon_\FI \sim 10^{-4}  \left(\frac{m_{A'}}{10^{-2}\eV}\right)^{-1/2} \left(\frac{\Omega_{A'} h^2}{0.12}\right)^{1/2}. 
\eea
Even though the warm dark matter bounds exclude the $100\%$ dark photon dark matter via freeze-in when $m_{A'}\gtrsim \text{few} \times 10 \kev$, the subcomponent dark photon dark matter can still be produced. One can rescale $\epsilon_\FI$ by $\Omega_{A'}^{1/2}$. 

\subsection{$m_{A'} \geq 2 m_e$}

For the heavy dark photon which satisfies $m_{A'} \geq 2 m_e$, the inverse decay channel $e^- e^+ \rightarrow A'$ opens up and dominates over the $\gamma \rightarrow A'$ channel. In such a case, the Boltzmann equation is 

\bea
\label{Boltz_EEToAp}
\dot{n}_{A'} + 3H n_{A'} \simeq n_{e^-} n_{e^+} \langle \sigma_{e^- e^+ \rightarrow A'} v \rangle,
\eea
where the collision term is
\bea
\label{Colli_EEToAp_1}
n_{e^-} n_{e^+} \langle \sigma_{e^- e^+ \rightarrow A'} v \rangle \simeq g_{e^-} g_{e^+} \int \frac{d^3 p_{e^-}}{(2\pi)^3} \frac{d^3 p_{e^+}}{(2\pi)^3} f_{e^-} f_{e^+} \sigma_{e^- e^+ \rightarrow A'} v_{\Mol}.
\eea
Then, after calculating the $e^- e^+ \rightarrow A'$ cross section~(We correct $\sigma_{e^- e^+ \rightarrow A'}$ in \cite{Redondo:2008ec})
\bea
\sigma_{e^- e^+ \rightarrow A'} = \pi (\epsilon e)^2 \frac{1+2m_e^2/m_{A'}^2}{\sqrt{1 - 4 m_e^2/m_{A'}^2}} \delta(s-m_{A'}^2),
\eea
substituting it into \Eq{Colli_EEToAp_1}, using the Maxwell-Boltzmann distribution to approximate the electron/positron phase space, and doing the phase space integration  as shown in \cite{Gondolo:1990dk}, we have
\bea
n_{e^-} n_{e^+} \langle \sigma_{e^- e^+ \rightarrow A'} v \rangle \simeq \frac{(\epsilon e)^2}{8 \pi^3}  m_{A'}^3 T \left(1+\frac{2m_e^2}{m_{A'}^2}\right) \sqrt{1- \frac{4 m_e^2}{m_{A'}^2}} K_1\left(\frac{m_{A'}}{T}\right). 
\eea
Finally,  after using the approximation $K_1(m_{A'}/T) \simeq \left( \pi T/2 m_{A'} \right)^{1/2}e^{-m_{A'}/T}$ in the low-temperature limit $T \ll m_{A'}$ and solving the Boltzmann equation \Eq{Boltz_EEToAp}, we get the dark photon relic abundance
\bea
\label{OmegaAp_EEToAp}
\Omega_{A'} \simeq \frac{3^4 \cdot 5^{5/2}}{ 2^{17/2} \cdot \pi^{11/2} }  \frac{ (\epsilon e)^2 }{ g_{*S} g_*^{1/2} } \frac{\mpl s_0}{\rho_{c}} \left( 1 + \frac{2 m_e^2}{m_{A'}^2 } \right) \sqrt{1 - \frac{4 m_e^2}{m_{A'}^2}}.
\eea
From \Eq{OmegaAp_EEToAp}, we know that the $\epsilon$ to reach dark matter's relic abundance when $m_{A'} \geq 2 m_e$ is nearly the constant of $m_{A'}$, which is similar to \Eq{eps_approx_1} but a bit smaller.

\section{$\zn$-Invariant Coleman-Weinberg Potential}\label{appx:zn_vcw}


In this section, we do the calculation to expand \Eq{eq:cwphifield} to all orders. We list the complete form of the 1-loop $\mathbb{Z}_N$ Coleman-Weinberg potential as~($N>4$) 
\bea
\label{eq:cwphifield_allorders}
V_\cw(\theta) = \frac{M^4 N}{8\pi^2} \sum_{l=1}^{+\infty} \left[ \sum_{j=0}^{+\infty} r^{lN+2j} G(lN+2j) \binom{lN+2j}{j} +(r\rightarrow r') \right] \cos(l N \theta) + \const, \quad \theta \coloneqq \frac{\phi}{f} + \frac{\pi}{2}, 
\eea
where $G(n)$ is defined in \Eq{eq:Gn}. In \Eq{eq:cwphifield_allorders}, we find that for the term with $\cos(lN\theta)$, the Fourier coefficient is proportional to $r^{lN}$, which reveals the exponential suppression in the effective operator $\mathcal{L} \supset \const \times (\Phi^{lN}+{\Phi^\dagger}^{lN})/\Lambda^{lN-4}_l$. For the type-A model, the lowest order term is $\mathcal{L} \supset \const \times ({\Phi^{N}+{\Phi^\dagger}^{N}})/\Lambda^{N-4}_1$. For the type-B model, $r=-r'$, so \Eq{eq:znuv} is invariant under the dark charge conjugation $\CD: (A, A',\phi,\F) \leftrightarrow (A, -A',-\phi,\F')$. If $N$ is an odd number, the $\cos(N\theta)$ terms exactly cancel with each other, but the $\cos(2N \theta)$ terms still exist, so the lowest order effective operators are $\mathcal{L} \supset \const \times ({\Phi^{2N}+{\Phi^\dagger}^{2N}})/\Lambda^{2N-4}_2$. In the rest of this appendix, we derive the equation \Eq{eq:cwphifield_allorders} in two different methods: 1.~The Fourier transformation. 2.~The cosine sum rules.

\subsection{Fourier Transformation}

According to \cite{DiLuzio:2021pxd}, the Fourier series of the scalar potential respecting the $\zn$ symmetry only receives contributions from $lN$th modes~($l$ is a positive integer number) so that the Coleman-Weinberg potential can be written as
\begin{equation}\label{eq:fourierV}
	V_\cw(\theta)=N\sum_{\ell =1}^\infty\widetilde{V}_\cw \left(\ell N\right)\cos\left(\ell N\theta\right).
\end{equation}
In \Eq{eq:fourierV}, $\widetilde{V}_\cw\left(\ell N\right)$ denotes the Fourier coefficient of the single-world Coleman-Weinberg potential, and the prefactor $N$ comes from $N$ worlds which have the equal contribution to the Fourier coefficient.

Beginning with \Eq{eq:cwphifield}, we write the Fourier coefficient as
\bea
\label{eq:cwfourier}
	\widetilde{V}_\cw\left(lN\right)=-\frac{M^4}{4\pi^3}\int_0^{\pi}\cos(lN\theta) (1-r\cos\theta)^4\log(1-r\cos\theta) d\theta +(r\rightarrow r')\;.
\eea
%
After expanding $(1-r\cos \theta)^4 \log(1-r\cos\theta)$ in \Eq{eq:cwfourier} in terms of the $r$ and $r'$ powers, we write it as
\bea
\label{eq:fourier_trans_V}
\widetilde{V}_\cw(lN)  = \frac{M^4}{4\pi^3} \sum_{m=1}^{+\infty} r^m 2^{m-1} G(m) \int^\pi_0 d \theta \,\,\cos(l N \theta) \cos^m\theta + (r \rightarrow r').
\eea
Applying the trigonometric identity
\bea
\cos(n\theta) \cos^m\theta = \frac{1}{2^m} \sum_{j=0}^m \binom{m}{j} \cos\left[(n-m+2j)\theta \right],
\eea
we find that for the terms in \Eq{eq:fourier_trans_V}, only the ones satisfying $lN=m-2j$ are picked out. After writing the summation in the Fourier coefficient $\widetilde{V}_\cw(lN)$ as
\bea
\sum_{m=1}^{+\infty} \sum_{j=0}^m \delta_{m,\, lN+2j} = \sum_{\substack{m=lN+2j\\j=0,1,\cdots}}
\eea
and plugging $\widetilde{V}_\cw(lN)$ into \Eq{eq:fourierV}, we get the equation \Eq{eq:cwphifield_allorders}.

We can also derive \Eq{eq:cwphifield_allorders} by only expanding the polynomial $(1-r\cos \theta)^4$ and carrying the integration over $\theta$ using the equation~(See \cite{gradshteyn2014table})
\bea
\int_0^\pi\log[1-r\cos(\theta)]\cos(n\theta)d\theta=-\frac{\pi}{n}[X(r)]^n\;,\quad \text{where $X(r)=\frac{1-\sqrt{1-r^2}}{r}$.}
\eea
After doing this, $\widetilde{V}_\cw\left(lN\right)$ with $N>4$ can be written as
%
\bea\label{eq:cwfourier_exact}
\widetilde{V}_\cw(lN) = \frac{M^4}{4 \pi^2} \sum_{j=0}^4 \sum_{m=0}^j \bmtx4\\4-j,j-m, m\emtx \frac{(-1)^j}{lN - j + 2m} \left( \frac{r}{2} \right)^{lN+2m} \left(\frac{ 1 - \sqrt{1 - r^2}}{r^2/2}\right)^{lN - j + 2m} + (r \rightarrow r').
\eea
Expanding \Eq{eq:cwfourier_exact} in terms of $r$ and $r'$, and then plugging it into \Eq{eq:fourierV}, we can also have \Eq{eq:cwphifield_allorders}.

\subsection{Cosine Sum Rules}

To calculate the Coleman-Weinberg potential in the $\theta$-space directly, we use the cosine sum rules
\bea
\label{cos_sum_rule} \frac{1}{N}\sum_{k=0}^{N-1} \cos^m\left(\theta + \frac{2 \pi k}{N} \right) 
 =\sum_{l=1}^{[m/N]} \mathcal{C}_{lmN}  \cos(\theta l N) +\mathcal{D}_m,
\eea
where
\bea
\label{cos_sum_rule_coefficient}
\left \{
\begin{aligned}
& \mathcal{C}_{lmN} = \frac{1}{2^{m-1}} \binom{m}{\frac{m+lN}{2}}\left( \delta_{l \bmod 2, \,  0} \,\delta_{m \bmod 2, \, 0} + \delta_{l \bmod 2, \, 1} \, \delta_{(m+N) \bmod 2, \, 0} \right)\\
& \mathcal{D}_m = \frac{1}{2^m} \binom{m}{m/2} \, \delta_{m \bmod 2, \, 0}
\end{aligned}.
\right. 
\eea
The cosine sum rules mentioned above can be derived by expanding the cosine functions in terms of the exponential functions and applying the formula $\frac{1}{N} \sum_{k=0}^{N-1}\exp\left( i \frac{2 \pi n k}{N} \right) = \delta_{n \bmod N, \, 0}$. 

Expanding \Eq{eq:cwphifield} in terms of $r$ and $r'$, we have
\bea
\label{eq:Vcw_Expand}
V_\cw(\theta) = \frac{M^4}{8 \pi^2} \sum_{m=1}^{+\infty} r^m 2^{m-1} G(m) \sum_{k=0}^{N-1} \cos^m\left( \theta + \frac{2 \pi k}{N} \right) + (r \rightarrow r')
\eea
for the $N>4$ case. Utilizing \Eq{cos_sum_rule} and \Eq{cos_sum_rule_coefficient} in \Eq{eq:Vcw_Expand} and then reshuffling the summation as
\bea
\left\{
\begin{aligned}
&\sum _{\substack{ m=2j\\ j= 0,1,\cdots} } \sum_{l=1}^{[m/N]} = \sum_{l=1}^{+\infty} \sum_{\substack{m=lN+2j,\\ j= 0,1,\cdots}} & \quad \text{(Even $N$)}\\
& \sum_{\substack{ m=2j+1\\ j= 0,1,\cdots} } \sum^{[m/N]}_{\substack{l=1\\\text{odd $l$}}}+ \sum_{\substack{ m=2j\\j= 0,1,\cdots} } \sum^{[m/N]}_{\substack{l=1\\\text{even $l$}}} = \sum_{l=1}^{+\infty} \sum_{\substack{m=lN+2j,\\j= 0,1,\cdots}} &\quad \text{(Odd $N$)}
\end{aligned},
\right.
\eea
we can also derive the equation \Eq{eq:cwphifield_allorders}.


\bibliographystyle{apsrev4-1}
\bibliography{references}

\end{document}